\documentclass[ALICE,manyauthors]{cernphprep}
\usepackage[comma,square,numbers,sort&compress]{natbib}

\usepackage{lineno}
\usepackage{xspace}
\usepackage{hyperref}
\usepackage{rotating}
\usepackage{multirow}
\usepackage[toc,page]{appendix}
\usepackage{graphicx}
\usepackage[T1]{fontenc}

\usepackage{tikz,colortbl}
\usetikzlibrary{calc}
\usepackage{zref-savepos}
\usepackage{enumitem}

\begin{document}
%%%%%%%%%%%%%%%%%%%%%%%%%%%%%%%%%%%%%%%%%%%%%%%%%%
% These are some new commands that may be useful 
% for paper writing in general. If other newcommands
% are needed for your specific paper, please feel 
% free to add here. 
%
% The currently available commands are organized in: 
% 1) Systems
% 2) Quantities
% 3) Energies and units
% 4) Detectors
% 5) particle species 
%%%%%%%%%%%%%%%%%%%%%%%%%%%%%%%%%%%%%%%%%%%%%%%%%%

% 1) SYSTEMS 
\newcommand{\pp}           {pp\xspace}
\newcommand{\ppbar}        {\mbox{$\mathrm {p\overline{p}}$}\xspace}
\newcommand{\XeXe}         {\mbox{Xe--Xe}\xspace}
\newcommand{\PbPb}         {\mbox{Pb--Pb}\xspace}
\newcommand{\pA}           {\mbox{pA}\xspace}
\newcommand{\pPb}          {\mbox{p--Pb}\xspace}
\newcommand{\AuAu}         {\mbox{Au--Au}\xspace}
\newcommand{\dAu}          {\mbox{d--Au}\xspace}

% 2) QUANTITIES 
\newcommand{\s}            {\ensuremath{\sqrt{s}}\xspace}
\newcommand{\snn}          {\ensuremath{\sqrt{s_{\mathrm{NN}}}}\xspace}
\newcommand{\pt}           {\ensuremath{p_{\rm T}}\xspace}
\newcommand{\meanpt}       {$\langle p_{\mathrm{T}}\rangle$\xspace}
\newcommand{\ycms}         {\ensuremath{y_{\rm CMS}}\xspace}
\newcommand{\ylab}         {\ensuremath{y_{\rm lab}}\xspace}
\newcommand{\etarange}[1]  {\mbox{$\left | \eta \right |~<~#1$}}
\newcommand{\yrange}[1]    {\mbox{$\left | y \right |~<~#1$}}
\newcommand{\acceff}         {\ensuremath{(A \times \epsilon)}\xspace}
\newcommand{\dndy}         {\ensuremath{\mathrm{d}N_\mathrm{ch}/\mathrm{d}y}\xspace}
\newcommand{\dndeta}       {\ensuremath{\mathrm{d}N_\mathrm{ch}/\mathrm{d}\eta}\xspace}
\newcommand{\avdndeta}     {\ensuremath{\langle\dndeta\rangle}\xspace}
\newcommand{\dNdy}         {\ensuremath{\mathrm{d}N_\mathrm{ch}/\mathrm{d}y}\xspace}
\newcommand{\Npart}        {\ensuremath{N_\mathrm{part}}\xspace}
\newcommand{\Ncoll}        {\ensuremath{N_\mathrm{coll}}\xspace}
\newcommand{\dEdx}         {\ensuremath{\textrm{d}E/\textrm{d}x}\xspace}
\newcommand{\RpPb}         {\ensuremath{R_{\rm pPb}}\xspace}
\newcommand{\sigtot}       {\ensuremath{\sigma_{\rm tot}}\xspace}
\newcommand{\dsigdpt}      {\ensuremath{\frac{d\sigma}{d\pt}}\xspace}
\newcommand{\dsigdy}       {\ensuremath{\frac{d\sigma}{dy}}\xspace}
\newcommand{\dsigdptdy}    {\ensuremath{\frac{d^2\sigma}{d\pt dy}}\xspace}
\newcommand{\rapidity}     {\ensuremath{y}\xspace}

% 3) ENERGIES, UNITS
\newcommand{\nineH}        {$\sqrt{s}~=~0.9$~Te\kern-.1emV\xspace}
\newcommand{\seven}        {$\sqrt{s}~=~7$~Te\kern-.1emV\xspace}
\newcommand{\twoH}         {$\sqrt{s}~=~0.2$~Te\kern-.1emV\xspace}
\newcommand{\twosevensix}  {$\sqrt{s}~=~2.76$~Te\kern-.1emV\xspace}
\newcommand{\five}         {$\sqrt{s}~=~5.02$~Te\kern-.1emV\xspace}
\newcommand{\twosevensixnn}{$\sqrt{s_{\mathrm{NN}}}~=~2.76$~Te\kern-.1emV\xspace}
\newcommand{\fivenn}       {$\sqrt{s_{\mathrm{NN}}}~=~5.02$~Te\kern-.1emV\xspace}
\newcommand{\LT}           {L{\'e}vy-Tsallis\xspace}
\newcommand{\GeVc}         {Ge\kern-.1emV/$c$\xspace}
\newcommand{\MeVc}         {Me\kern-.1emV/$c$\xspace}
\newcommand{\TeV}          {Te\kern-.1emV\xspace}
\newcommand{\GeV}          {Ge\kern-.1emV\xspace}
\newcommand{\MeV}          {Me\kern-.1emV\xspace}
\newcommand{\GeVmass}      {Ge\kern-.2emV/$c^2$\xspace}
\newcommand{\MeVmass}      {Me\kern-.2emV/$c^2$\xspace}
\newcommand{\lumi}         {\ensuremath{\mathcal{L}}\xspace}

% 4) DETECTORS 
\newcommand{\ITS}          {\rm{ITS}\xspace}
\newcommand{\TOF}          {\rm{TOF}\xspace}
\newcommand{\ZDC}          {\rm{ZDC}\xspace}
\newcommand{\ZDCs}         {\rm{ZDCs}\xspace}
\newcommand{\ZNA}          {\rm{ZNA}\xspace}
\newcommand{\ZNC}          {\rm{ZNC}\xspace}
\newcommand{\SPD}          {\rm{SPD}\xspace}
\newcommand{\SDD}          {\rm{SDD}\xspace}
\newcommand{\SSD}          {\rm{SSD}\xspace}
\newcommand{\TPC}          {\rm{TPC}\xspace}
\newcommand{\TRD}          {\rm{TRD}\xspace}
\newcommand{\VZERO}        {\rm{V0}\xspace}
\newcommand{\VZEROA}       {\rm{V0A}\xspace}
\newcommand{\VZEROC}       {\rm{V0C}\xspace}
\newcommand{\Vdecay} 	   {\ensuremath{V^{0}}\xspace}

% 4) PARTICLE SPECIES 
\newcommand{\ee}           {\ensuremath{e^{+}e^{-}}} 
\newcommand{\pip}          {\ensuremath{\pi^{+}}\xspace}
\newcommand{\pim}          {\ensuremath{\pi^{-}}\xspace}
\newcommand{\kap}          {\ensuremath{\rm{K}^{+}}\xspace}
\newcommand{\kam}          {\ensuremath{\rm{K}^{-}}\xspace}
\newcommand{\pbar}         {\ensuremath{\rm\overline{p}}\xspace}
\newcommand{\kzero}        {\ensuremath{{\rm K}^{0}_{\rm{S}}}\xspace}
\newcommand{\lmb}          {\ensuremath{\Lambda}\xspace}
\newcommand{\almb}         {\ensuremath{\overline{\Lambda}}\xspace}
\newcommand{\Om}           {\ensuremath{\Omega^-}\xspace}
\newcommand{\Mo}           {\ensuremath{\overline{\Omega}^+}\xspace}
\newcommand{\X}            {\ensuremath{\Xi^-}\xspace}
\newcommand{\Ix}           {\ensuremath{\overline{\Xi}^+}\xspace}
\newcommand{\Xis}          {\ensuremath{\Xi^{\pm}}\xspace}
\newcommand{\Oms}          {\ensuremath{\Omega^{\pm}}\xspace}
\newcommand{\degree}       {\ensuremath{^{\rm o}}\xspace}
\newcommand{\ups}         {\ensuremath{\Upsilon}\xspace}
\newcommand{\upsones}     {\ensuremath{\Upsilon\rm(1S)}\xspace}
\newcommand{\upstwos}     {\ensuremath{\Upsilon{\rm(2S)}}\xspace}
\newcommand{\upsthrees}   {\ensuremath{\Upsilon{\rm(3S)}}\xspace}
\newcommand{\upsns}   {\ensuremath{\Upsilon{\rm(nS)}}\xspace}
\newcommand{\jpsi}         {\ensuremath{\mathrm{J}/\psi}\xspace}
\newcommand{\psitwos}      {\ensuremath{\psi {\rm (2S)}}\xspace}
\newcommand{\chib}         {\ensuremath{\chi_b}\xspace}
\newcommand{\chibone}      {\ensuremath{\chi_{b1}}\xspace}
\newcommand{\chibtwo}      {\ensuremath{\chi_{b2}}\xspace}
\newcommand{\chibonep}     {\ensuremath{\chi_{b}{\rm(1P)}}\xspace}
\newcommand{\chibtwop}     {\ensuremath{\chi_{b}{\rm(2P)}}\xspace}
\newcommand{\chic}         {\ensuremath{\chi_c}\xspace}
\newcommand{\chicone}      {\ensuremath{\chi_{c1}}\xspace}
\newcommand{\chiconep}     {\ensuremath{\chi_{c}{\rm(1P)}}\xspace}
\newcommand{\chictwop}     {\ensuremath{\chi_{c}{\rm(2P)}}\xspace}
\newcommand{\chictwo}      {\ensuremath{\chi_{c2}}\xspace}

%OTHERS
\newcommand{\etal}{{\it et al.}}

%%%%%%%%%%%%%%%  Title page %%%%%%%%%%%%%%%%%%%%%%%%
\begin{titlepage}
% the dates below correspond to CERN approval
% please don't touch: EB chairs will take care
\PHyear{2021}       % required, will be obtained from CERN
\PHnumber{197}      % required, will be obtained from CERN
\PHdate{24 September}  % required, will be obtained from CERN
%%%%%%%%%%%%%%%%%%%%%%%%%%%%%%%%%%%%%%%%%%%%%%%%%%%%

%%% Put your own title + short title here:
\title{Inclusive quarkonium production in \pp collisions at \s = 5.02~TeV}
\ShortTitle{}   % appears on left page headers

%%% Do not change the next lines
\Collaboration{ALICE Collaboration\thanks{See Appendix~\ref{app:collab} for the list of collaboration members}}
\ShortAuthor{ALICE Collaboration} % appears on right page headers, do not change

\begin{abstract}

This article reports on the inclusive production cross section of several quarkonium states, \jpsi, \psitwos, \upsones, \upstwos, and \upsthrees, measured with the ALICE detector at the LHC, in \pp collisions at $\s = 5.02$~TeV. The analysis is performed in the dimuon decay channel at forward rapidity ($2.5 < \rapidity < 4$). The integrated cross sections and transverse-momentum (\pt) and rapidity (\rapidity) differential cross sections for \jpsi, \psitwos, \upsones, and the \psitwos-to-\jpsi cross section ratios are presented. The integrated cross sections, assuming unpolarized quarkonia, are: $\sigma_{\jpsi}$~($\pt<20$~GeV/c) = 5.88 $\pm$ 0.03 $\pm$ 0.34$
~\mu$b, $\sigma_{\psitwos}$~($\pt<12$~GeV/c) = 0.87 $\pm$ 0.06 $\pm$ 0.10$~\mu$b, $\sigma_{\upsones}$~($\pt<15$~GeV/c) = 45.5 $\pm$ 3.9 $\pm$ 3.5~nb, $\sigma_{\upstwos}$~($\pt<15$~GeV/c) = 22.4 $\pm$ 3.2 $\pm$ 2.7~nb, and $\sigma_{\upsthrees}$~($\pt<15$~GeV/c) = 4.9 $\pm$ 2.2 $\pm$ 1.0~nb, where the first (second) uncertainty is the statistical (systematic) one.
For the first time, the cross sections of the three $\Upsilon$ states, as well as the \psitwos one as a function of \pt and \rapidity, are measured at $\s = 5.02$~TeV at forward rapidity. These measurements also significantly extend the \jpsi \pt reach and supersede previously published results.
A comparison with ALICE measurements in \pp collisions at $\s = 2.76$, 7, 8, and 13~TeV is presented and the energy dependence of quarkonium production cross sections is discussed. Finally, the results are compared with the predictions from several production models. 

\end{abstract}
\end{titlepage}

\setcounter{page}{2} %please do not remove this line

%%%%%%%%%%%%%%%%%%%%%%%%%%%%%%%%
% begin main text
%%%%%%%%%%%%%%%%%%%%%%%%%%%%%%%%

\section{Introduction}

Quarkonium production in high-energy hadronic collisions is an important tool to study the perturbative and non-perturbative aspects of quantum chromodynamics (QCD) calculations~\cite{Brambilla:2010cs,Andronic:2015wma}. 
Quarkonia are bound states of either a charm and anti-charm (charmonia) or a bottom and anti-bottom quark pair (bottomonia). 
In hadronic collisions, the scattering process leading to the production of the heavy-quark pair involves momentum transfers at least as large as twice the mass of the considered heavy quark, hence it can be described with perturbative QCD calculations. In contrast, the binding of the heavy-quark pair is a non-perturbative process as it involves long distances and soft momentum scales. Describing quarkonium production measurements in proton--proton (pp) collisions at various colliding energies represents a stringent test for models and, in particular, for the investigation of the non-perturbative aspects that are treated differently in the various approaches. These measurements also provide a crucial reference for the investigation of the properties of the quark--gluon plasma formed in nucleus--nucleus collisions and of the cold nuclear matter effects present in proton--nucleus collisions~\cite{Andronic:2015wma,Rothkopf:2019ipj}. 

Quarkonium production can be described by various approaches that essentially differ in the treatment of the hadronization part. The Color Evaporation Model (CEM)~\cite{Fritzsch:1977ay,Amundson:1996qr} considers that the quantum state of every heavy-quark pair produced with a mass above its production threshold and below twice the open heavy flavor (D or B meson) threshold production evolves into a quarkonium. In this model, the probability to obtain a given quarkonium state from the heavy-quark pair is parametrized by a constant phenomenological factor. The Color Singlet Model (CSM)~\cite{Baier:1981uk} assumes no evolution of the quantum state of the pair from its production to its hadronization. Only color-singlet heavy-quark pairs are thus considered to form quarkonium states. Finally, in the framework of Non-Relativistic QCD (NRQCD)~\cite{Bodwin:1994jh}, both color-singlet and color-octet heavy-quark pairs can evolve towards a bound state. Long Distance Matrix Elements are introduced in order to parametrize the binding probability of the various quantum states of the heavy-quark pairs. They can be constrained from existing measurements and do not depend on the specific production process under study (\pp, electron--proton, etc.).

This article presents measurements of the inclusive production cross section of charmonium (\jpsi and \psitwos) and bottomonium (\upsones, \upstwos, and \upsthrees) states in \pp collisions at a center-of-mass energy $\s = 5.02$~TeV with the ALICE detector. The analysis is performed in the dimuon decay channel at forward rapidity ($2.5 < y < 4$). In this rapidity interval, the total, transverse momentum~(\pt) and rapidity~(\rapidity) differential cross sections for \jpsi as well as the total cross section for \psitwos, were published by the ALICE collaboration based on an earlier data sample~\cite{Adam:2016rdg,Acharya:2017hjh}, corresponding to a factor 12 smaller integrated luminosity. These measurements with improved statistical precision supersede the ones from earlier publication. The \pt and \rapidity differential measurements for the \psitwos and \upsones as well as the total cross sections for all the measured \ups states are presented here for the first time at $\s = 5.02$~TeV and at forward rapidity. The \pt coverage of the \jpsi measurement is extended up to 20~\GeVc.

 The inclusive differential cross sections are obtained as a function of \pt for $\pt < 20$~\GeVc and as a function of \rapidity for $\pt < 12$~\GeVc for \jpsi, for $\pt < 12$~\GeVc for \psitwos, and for $\pt < 15$~\GeVc for \upsones. Only the \pt-integrated cross sections are measured for \upstwos and \upsthrees due to statistical limitations. The inclusive \psitwos-to-\jpsi ratio is also presented as a function of \pt and \rapidity.
The comparison of the \jpsi cross section with recent results from LHCb~\cite{LHCb:2021pyk} is discussed. 
The results are compared with previous ALICE measurements performed at $\s = 2.76$, 7, 8, and 13~TeV~\cite{Abelev:2012kr,Abelev:2014qha,Adam:2015rta,Acharya:2017hjh}. Earlier comparisons with LHCb quarkonium results at $\s = 7$, 8, and 13~TeV~\cite{LHCb:2012aa,LHCb:2013itw,LHCb:2015log,LHCb:2015foc} were performed in~\cite{Abelev:2014qha,Adam:2015rta,Acharya:2017hjh}. Finally, the results are compared with theoretical calculations based on NRQCD and CEM.

The measurements reported here are inclusive and correspond to a superposition of the direct production of quarkonium and of the contribution from the decay of higher-mass excited states (predominantly \psitwos and \chic for \jpsi, \upstwos, \chib, and \upsthrees for  \upsones, \upsthrees and \chib for \upstwos, and \chib for \upsthrees). For \jpsi and \psitwos a non-prompt contribution from beauty hadron decays is also present. 

\clearpage
The article is organized as follows: the ALICE detectors used in the analysis and the data sample are briefly described in Section~\ref{sec:apparatus}, the analysis procedure is presented in Section~\ref{sec:analysis}, and in Section~\ref{sec:results} the results are discussed and compared with theoretical calculations and measurements at other center-of-mass energies from ALICE.

\section{Apparatus and data samples} 
\label{sec:apparatus}

%should contain: \\
%- detector description \\
A detailed description of the ALICE setup and its performance are discussed in Refs.~\cite{Aamodt:2008zz,Abelev:2014ffa}. In this section, the subsystems relevant for this analysis are presented. 

Muons from quarkonium decays are detected in the muon spectrometer within the pseudorapidity range\footnote{In the ALICE coordinate system, detectors located on the muon spectrometer side are defined as being at negative $z$ (and negative pseudorapidity). However, due to the symmetry of pp collisions, the results are presented at positive rapidity.} $-4<\eta<-2.5$~\cite{ALICE:1999aa}. The muon spectrometer consists of a front absorber located along the beam direction ($z$) between $-0.9$ and $-5$~m from the interaction point (IP), five tracking stations (MCH), located between $-5.2$ and $-14.4$~m from the IP, an iron wall at $-14.5$~m, and two triggering stations (MTR), placed at $-16.1$ and $-17.1$~m from the IP. Each station is made of two layers of active detection material, with cathode pad and resistive plate techniques employed for the muon detection in the tracking and triggering devices, respectively. A dipole magnet with a 3~T$\times$m field integral deflects the particles in the vertical direction for the measurement of the muon momentum. The hadronic particle flux originating from the collision vertex is strongly suppressed thanks to the front absorber with a thickness of 10 interaction lengths. Throughout the spectrometer length, a conical absorber at small angle around the $z$ axis reduces the background from secondary particles originating from the interaction of large angle primary particles with the beam pipe. The 1.2~m thick iron wall positioned in front of the triggering stations stops the punch-through hadrons escaping the front absorber, as well as low-momentum muons from pion and kaon decays. In addition, a rear absorber downstream of the trigger stations ensures protection against the background generated by beam--gas interactions. 

Two layers of silicon pixel detectors (SPD) with a cylindrical geometry, covering $|\eta|<2.0$ and $|\eta|<1.4$, respectively, are used for the determination of the collision vertex. They are the two innermost layers of the Inner Tracking System (ITS)~\cite{Aamodt:2010aa} and surround the beam pipe at average radii of 3.9 and 7.6~cm. The T0 quartz Cherenkov counters~\cite{Bondila:2005xy} are made of two arrays positioned on each side of the IP at $-70$~cm and 360~cm. They cover the pseudorapidity ranges $-3.3 < \eta < -3.0$ and $4.6 < \eta < 4.9$, respectively. The T0 is used for luminosity determination and background rejection. 
Similarly, the V0 scintillator arrays~\cite{Abbas:2013taa} are located on both sides of the IP at $-90$ and 340~cm and cover the pseudorapidity ranges $-3.7 < \eta < -1.7$ and $2.8 < \eta < 5.1$, respectively. These are used for triggering, luminosity determination and to reject beam--gas events using offline timing selections together with the T0 detectors.

%- trigger description \\
A minimum bias trigger is issued by the V0 detector~\cite{Abbas:2013taa} when a logical AND of signals from the two V0 arrays on each side of the IP is produced. Single muon, same-sign dimuon, and opposite-sign dimuon triggers are defined by an online estimate of the \pt of the muon tracks using a programmable trigger logic circuit.
A predefined \pt threshold of 0.5~\GeVc is set in order to remove the low-\pt muons, mainly coming from $\pi$ and K decays. The muon trigger efficiency reaches $50\%$ at this threshold value and saturates for $\pt > 1.5$~\GeVc.
%- event selection \\
Events containing an opposite-sign dimuon trigger in coincidence with the minimum bias trigger are selected for the quarkonium analysis. 

The data sample of pp collisions at $\s=5.02$~TeV used for the measurements reported in this article was collected in 2017 with the opposite-sign dimuon trigger, and corresponds to an integrated luminosity $L_{\rm int} = 1229.9~\pm~0.4$~(stat.)~$\pm$ 22.1~(syst.)~nb$^{-1}$~\cite{lumi}. The luminosity determination is based on dedicated van der Meer scans~\cite{vanderMeer:1968zz}, where the cross sections seen by two different minimum bias triggers based on the V0 and T0 signals are derived~\cite{lumi}. 
The number of T0- and dimuon-trigger counts measured with scalers on a run-by-run basis without any data acquisition veto is used along with the T0-trigger cross section to calculate the integrated luminosity of the analyzed data sample. Another method, using reconstructed minimum bias events triggered with the V0 detector only, is used as a cross-check of the first method. In this method, the luminosity is computed as the ratio of the number of equivalent minimum bias events over the V0-trigger cross section. The number of equivalent minimum bias events is evaluated as the product of the total number of dimuon-triggered events with the inverse of the probability of having dimuon-triggered events in a minimum bias triggered data sample recorded with only the V0~\cite{Adam:2015jsa}. The two methods give compatible values and the one based on T0 is used, as it gives a smaller total uncertainty (see section~\ref{subsec:syst}).

\section{Analysis procedure}
\label{sec:analysis}

\subsection{Track selection}
\label{subsec:trig_track_sel}

The number of detected quarkonia
is estimated by pairing muons of opposite charges and by fitting their invariant mass ($m_{\mu^{+}\mu^{-}}$) distribution. Reconstructed tracks must meet several selection criteria. The pseudorapidity of each muon candidate must be within the geometrical acceptance of the muon spectrometer ($-4 < \eta < -2.5$). Muons are identified and selected by applying a matching condition between the tracking system and the trigger stations. 
A selection on the transverse position $R_{\textrm{abs}}$ of the muon at the end of the front absorber ($17.6 < R_{\textrm{abs}} < 89.5$~cm) rejects tracks crossing the thickest sections of the absorber. Finally, the contamination from tracks produced by background events, like beam--gas collisions, 
is reduced by applying a selection on the product of the track momentum and the transverse distance to the primary vertex~\cite{Abelev:2012pi}. Opposite-sign (OS) muon pairs are then formed in the range $2.5 < \rapidity < 4$. The considered \pt~interval varies according to the studied resonance given the available data sample: 
$\pt < 20$~\GeVc for \jpsi; $\pt < 12$~\GeVc for \psitwos; \rapidity-differential and (\pt,\rapidity)-differential \jpsi studies; and $\pt < 15$~\GeVc for \upsns.

\subsection{Signal extraction}
\label{signalext}

A fit to the OS dimuon invariant mass distribution is performed separately for the charmonium and bottomonium mass regions, in each \pt and \rapidity interval considered. In both cases, a maximum log-likelihood fitting method is used. In order to evaluate the systematic uncertainties on the charmonium and bottomonium signal extraction, several fitting functions and ranges are considered, and the parameters that are fixed during the fitting procedure are varied, as described below. 

\begin{figure}[!htb]
\begin{center}
\includegraphics[width=0.52\linewidth]{{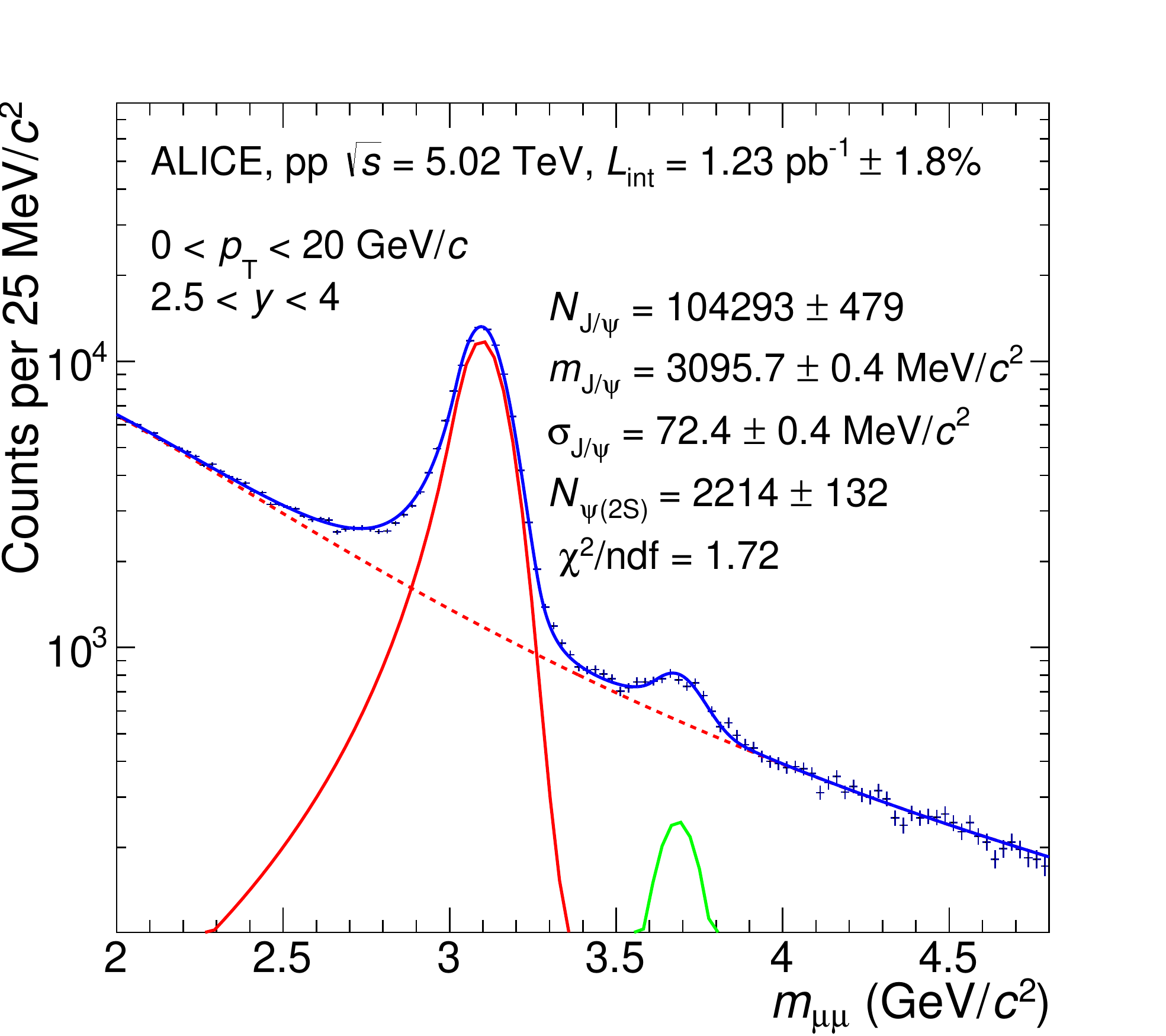}}
\includegraphics[width=0.47\linewidth]{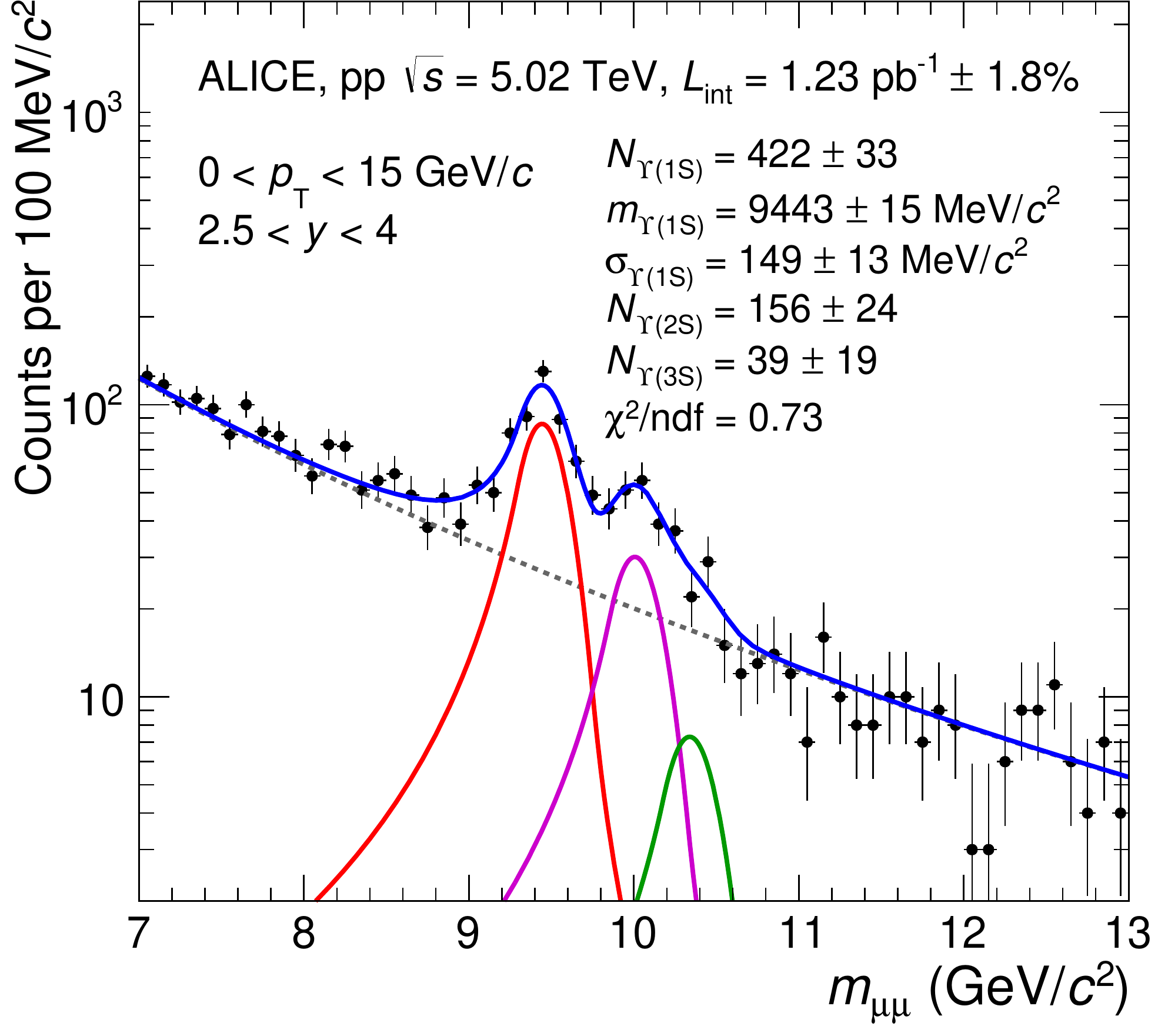}
\caption{Examples of fit to the OS dimuon invariant mass distribution 
%in \pp collisions at \s = 5.02~TeV
 in the mass region $2 < m_{\mu^{+}\mu^{-}} < 5$~GeV/$c^{2}$ for $\pt < 20$~\GeVc (left), and $7< m_{\mu^{+}\mu^{-}} < 13$ GeV/$c^{2}$ for $\pt < 15$~\GeVc (right). The \jpsi, $\psi$(2S) and $\Upsilon$(nS) signals are modelled with extended Crystal Ball functions, while the background is described by a pseudo Gaussian with a width increasing linearly with the invariant mass. The fit is performed on the full data sample. The widths of the \psitwos, \upstwos and \upsthrees, for these examples, are fixed to 73 \MeVc, 156 \MeVc~and 161 \MeVc, respectively.}
\label{fig:InvMass}
\end{center}
\end{figure}

In the charmonium mass region (2 $< m_{\mu^{+}\mu^{-}} <$ 5~GeV/$c^{2}$), the fit is performed using the same functional form to describe the \jpsi and \psitwos signals, on top of an ad-hoc function to describe the background. 
The signal shapes considered are either two extended Crystal Ball functions or two pseudo-Gaussian functions~\cite{ALICEpage}. 
For both functional forms, the \jpsi mass pole and width are left free during the fit procedure, while the \psitwos mass is bound to the \jpsi one by fixing the mass difference between the two states according to the PDG values~\cite{PDG}. The width of the \psitwos signal is also bound to the \jpsi one by means of a scale factor on their ratio. It was obtained via a fit to a large data sample from pp collisions at $\s$~=~13~TeV~\cite{Acharya:2017hjh} which gives 1.01 $\pm$ 0.05. A variation of +5$\%$ of the \psitwos-to-\jpsi width ratio central value, corresponding to the difference observed between data and Monte Carlo (MC) simulation at $\s = 13$~TeV\footnote{It is assumed that the \psitwos-to-\jpsi~width ratio and signal tail parameters do not depend on the collision energy and are the same at $\s = 5.02$~TeV and $\s = 13$~TeV.}, induces a variation of the \jpsi yield at the per mille level and is therefore neglected, while the impact of this variation on the \psitwos yield enters the systematic uncertainty. The parameters describing the left and right signal tails are the same for both resonances and are fixed to the values extracted from either MC simulations at $\s = 5.02$~TeV using the GEANT3~\cite{Brun:1082634} or the GEANT4~\cite{Agostinelli:2002hh} transport codes (see Section~\ref{Sec:AxE}), or from fits to the 13~TeV data sample. While the tail parameters can be extracted in \pt and \rapidity intervals in the MC for both signal shapes, the 13~TeV data sample is only used to constrain the tail parameters of the extended Crystal Ball, when performing a fit to the invariant mass spectrum integrated over \pt and \rapidity. Therefore, when using tail parameters from data, the same set is applied to all the \pt and \rapidity intervals. Various functions successfully model the background in the invariant mass range $2 < m_{\mu^{+}\mu^{-}} < 5$~GeV/$c^{2}$. To extract the \jpsi signal, either a pseudo Gaussian with a width increasing linearly with the invariant mass or the ratio of a first order to second order polynomial is used as a background shape. For the \psitwos signal extraction, either a pseudo Gaussian with a width increasing linearly with the invariant mass or the combination of a fourth order polynomial with an exponential function is used to describe the background. 
In addition to the variation of the background shapes, two different fitting ranges are also used for the evaluation of  the signal extraction systematic uncertainties. For each \pt and \rapidity range, several fits are performed with different combinations of signal shapes, background shapes, fitting ranges, signal tail parameters, and signal width ratios between the two resonances for the \psitwos case. For the charmonium states, the raw yields are computed as the weighted average of the results of all the fits. The statistical uncertainty is the weighted average of the statistical uncertainties of the fits, while the systematic uncertainty is taken as the RMS of the distribution of the results. Given that the choice of the signal tails is the main source of systematic uncertainty, this weight is applied to counterbalance the higher number of fits performed with MC tails with respect to fits with data-driven tails. The raw \jpsi yield is $N_{\jpsi} = 101285 \pm 452$~(stat.)~$\pm$~3012 (syst.) for $\pt~<$~20~\GeVc, and the \psitwos raw yield is  $N_{\psitwos} = 2086 \pm 133$~(stat.) $\pm$ 150 (syst.) for $\pt < 12$~\GeVc. Figure~\ref{fig:InvMass} left shows an example of a fit of the OS dimuon invariant mass distribution in the mass region $2 < m_{\mu^{+}\mu^{-}} < 5$~GeV/$c^{2}$, separately showing the contributions of the two charmonium resonances and the background. In each \pt~and \rapidity~interval, the \psitwos-to-\jpsi yield ratio is evaluated as the weighted average of the \psitwos-to-\jpsi yield ratio values obtained from each individual fits (with a given signal shape, background shape, signal tail choice, fitting range and \psitwos~width) in order to properly account for correlations in the \jpsi and \psitwos signal extraction.
The statistical and systematic uncertainties on the ratio are then evaluated in the same way as for the \jpsi and \psitwos raw yields. 

In the bottomonium mass region ($7 < m_{\mu^{+}\mu^{-}} < 13$~GeV/$c^{2}$), the \upsns shapes are parametrized only with extended Crystal Ball functions, since it was checked that the systematic uncertainty related to the choice of the signal shape is negligible compared to other sources. The \upsones mass and width are left free in the fit, while the \upstwos and \upsthrees masses are bound to the \upsones one by fixing the mass difference between the states according to the PDG values~\cite{PDG}. The width of the \upstwos and \upsthrees signals are also bound to the \upsones one by factors, $\sigma^{\rm MC}_{\upsns} / \sigma^{\rm MC}_{\upsones}$, obtained from MC simulations. 
Two alternative width scalings, namely $\sigma_{\upsns}= \sigma_{\upsones}$ and $\sigma_{\upsns}= \sigma_{\upsones} \times (2 \times \sigma^{\rm MC}_{\upsns} / \sigma^{\rm MC}_{\upsones} - 1)$, are also considered. The \upsns signal tail parameters must also be fixed while fitting. By default, they are fixed to the values extracted in each given \pt and \rapidity range from MC simulations performed with the GEANT3 transport code. The same shapes are used for the three $\Upsilon$~resonances. The systematic uncertainty related to the choice of the tail parameters is evaluated for each resonance on the \pt and \rapidity integrated mass distribution by using several sets of tail parameters that were generated from the fit of the 13 TeV data sample taking into account the correlation among the parameters via the covariance matrix. This uncertainty is then considered to be the same for all \upsones \pt and \rapidity differential intervals. The background shape is described by three empirical functions: an exponential function, a sum of two exponential functions, and a power law function. Additionally, two fit ranges are used. The \upsns raw yields and statistical uncertainties are then computed as the average of all the fit results and statistical uncertainties, respectively. The systematic uncertainty is the RMS of the fit results summed in quadrature with the uncertainty from the choice of signal tails. The main sources of systematic uncertainty come from the choice of the background description and from the choice of the tail parameters. The \upsns raw yields are $N_{\upsones} = 401 \pm 34$ (stat.) $\pm~26$~(syst.), $N_{\upstwos} = 153 \pm 22$~(stat.) $\pm~12$~(syst.), and $N_{\upsthrees} = 38 \pm 17$~(stat.) $\pm 7$~(syst.), for $\pt < 15$~\GeVc. The significance of the \upsthrees signal remains rather limited and amounts to 2.4. Figure~\ref{fig:InvMass} right shows an example of fit to the OS dimuon invariant mass distribution in the mass region $7< m_{\mu^{+}\mu^{-}} < 13$~GeV/$c^{2}$ for $\pt < 15$~\GeVc, showing the contribution of the three $\Upsilon$~resonances. Similarly to the charmonium case, the \upstwos-to-\upsones and \upsthrees-to-\upsones raw yield ratios are extracted on a fit-by-fit basis, in order to account for correlations in the signal extraction.

%likelihood fit

\subsection{Acceptance and efficiency corrections}
\label{Sec:AxE}

The detector acceptance and reconstruction efficiency \acceff corrections 
are applied to the quarkonium raw yields to obtain the corrected yields for the individual resonances. The \acceff values are estimated via MC simulations by computing the ratio between the number of quarkonia reconstructed in the muon spectrometer and the number of generated quarkonia in given \pt and \rapidity intervals. Monte Carlo simulations are performed reproducing on a run-by-run basis the detector conditions during the data taking. 

In the first stage of the simulation procedure, a parametric generator based on phenomenological \pt and $y$ distributions of quarkonia extracted from RHIC, Tevatron, and LHC data~\cite{Bossu:2011qe} is employed, assuming unpolarized resonance production as suggested by the ALICE~\cite{Acharya:2018uww, Abelev:2011md} and LHCb~\cite{Aaij:2013nlm,Aaij:2014qea,Aaij:2017egv} measurements on polarization parameters for quarkonia that are found small or compatible with zero. 
The quarkonium decay to $\mu^+\mu^-$ is implemented using EVTGEN~\cite{Lange:2001uf} and PHOTOS~\cite{Barberio:1990ms} to account for the radiative decay of the quarkonium states.
The decay muons are tracked through a GEANT3~\cite{Brun:1082634} model of the apparatus that includes a realistic description of the detectors and their performance during data taking. An independent test of the detector simulation has also been performed using the GEANT4~\cite{Agostinelli:2002hh} framework. It provides \acceff results compatible with the GEANT3 simulation within a maximum deviation of 2$\%$.

 The \jpsi, \psitwos, and \upsones raw yields are divided by the \acceff correction factors to obtain a first estimate of the \pt and \rapidity distributions. An iterative procedure is performed to tune the quarkonium input \pt and \rapidity MC distributions on the measured data distributions until no significant variation of the input shapes is observed. Because of statistical limitations, the iterative procedure cannot be applied to the \upstwos and \upsthrees as \pt~and \rapidity-differential measurements cannot be performed. Since no significant variation of the \rapidity input shape between the \upsns states is expected~\cite{Aaij:2015awa} and the \upsns \acceff does not strongly depend on the \pt spectrum of the MC input, the \upsones \pt and \rapidity shapes are applied for the \upstwos and \upsthrees. 

\subsection{Systematic uncertainties} \label{subsec:syst}

The main systematic uncertainties on the quarkonium production cross section (see Eq.~\ref{eq:sigma}) come from the following sources: (1) the quarkonium signal extraction, (2) the branching ratio, (3) the determination of the luminosity, and (4) the acceptance and efficiency corrections. The uncertainties on the latter can be broken down into the following contributions: (i) the choice of parametrization for the signal input \pt and \rapidity distributions, (ii) the tracking efficiency in the muon tracking chambers, (iii) the muon trigger efficiency, and (iv) the matching efficiency between the tracks reconstructed in the muon tracker and the track segments measured in the muon trigger systems.

The evaluation of the systematic uncertainty on quarkonium signal extraction is detailed in Section~\ref{signalext}. It amounts to 3$\%$, 7.2$\%$, 6.5$\%$, 7.8$\%$, and 19$\%$ for the integrated \jpsi, \psitwos, \upsones, \upstwos, and \upsthrees signals, respectively. This uncertainty is uncorrelated as a function of \pt and \rapidity, for a given quarkonium state. It is, however, partially correlated between \jpsi and \psitwos, and among the three \upsns resonances. 

The systematic uncertainty on the branching ratio is taken as the current estimate for this quantity according to the PDG~\cite{PDG} and is reported in Tables~\ref{tab:syst1},~\ref{tab:syst2}, and~\ref{tab:syst3} for all the states. This uncertainty is fully correlated versus \pt and \rapidity for a given resonance. 

Regardless of the method used to determine the luminosity, its associated systematic uncertainty has two origins: the uncertainty on the normalization factor between the number of triggered events and the equivalent number of minimum bias events, and the uncertainty on the cross section of the minimum bias trigger evaluated using the van der Meer scan technique~\cite{lumi}. The first source of uncertainty is evaluated by using minimum bias triggers issued either by the V0 or the T0 detectors. The two methods are in agreement within 0.5$\%$. This systematic uncertainty is therefore consistently neglected for all resonances and the method which uses the T0 detector is used as the main one since it gives the result with the smallest statistical uncertainty. The second source of uncertainty is the dominant one and arises from the uncertainty on the T0-trigger cross section. It amounts to 1.8$\%$. This uncertainty is fully correlated as a function of \pt and \rapidity for a given state and also fully correlated among all the quarkonium states.

The systematic uncertainty on \acceff related to the parametrization of the signal input \pt and \rapidity distributions has two components. The first one arises from the fact that the corrected yield used to tune the MC input shape in the iterative procedure is obtained from a data sample with an associated statistical uncertainty. This has a negligible impact on the \jpsi and \psitwos~results, since their reconstructed signals profit from a large sample. For the \upsones state, this uncertainty is not negligible and is evaluated by performing 50 fits to the \pt and \rapidity differential corrected yields after having randomly moved each data point according to a Gaussian smearing within the statistical uncertainty of the data point. The RMS of the resulting distribution of the obtained \acceff values is assigned as the uncertainty. It varies between 1.3$\%$ and 3.5$\%$. 
The second component arises from the fact that the correlations in \pt and \rapidity of the quarkonium input shape are not accounted for in the simulation. It is evaluated by performing several fits to the \rapidity-differential corrected yields in different \pt intervals, and to the \pt-differential corrected yields in different \rapidity intervals. To be conservative, all the possible \pt and \rapidity input shape combinations are then considered, the \acceff values are evaluated and the RMS of the results gives the associated uncertainty, ranging between 0.3$\%$ and 4.9$\%$. Such a study can only be performed for the \jpsi since it requires a large data sample to perform double-differential measurements. For the \psitwos, the uncertainty from the \pt~and \rapidity~double-differential shape variation is assumed to be the same as for the \jpsi. Moreover, additional \pt~and \rapidity~shapes are considered in the systematic uncertainty evaluation. They are obtained by using the measured \jpsi~\pt~and \rapidity-dependent cross sections times the \psitwos-to-\jpsi~cross section ratios. This additional contribution is summed quadratically to the \jpsi~one and is below 1.5$\%$ in all \pt~and \rapidity~intervals. The resulting total MC input shape systematic uncertainty  on the \psitwos ranges between 1.4$\%$ and 5.0$\%$. Given the absence of \pt~and \rapidity~double-differential $\Upsilon$ measurements at $\s = 5.02$~TeV, an estimation of the variation of the \upsones input shape is performed by fitting the \upsones cross sections measured with high statistical precision by LHCb in pp collisions at $\s = 13$~TeV~\cite{Aaij:2018pfp}, as a function of \rapidity~in four \pt~bins and as a function of \pt~in five \rapidity~bins. The combination of these 20 input \pt and \rapidity distributions is used to assess the systematic uncertainty on the \upsones \acceff quantity. Its value ranges between 0.5$\%$ and 1$\%$. For the \upstwos and \upsthrees, the same MC input systematic uncertainty as the \upsones is assumed for the integrated cross section. The two aforementioned sources contributing to the \acceff uncertainty are uncorrelated and are therefore summed in quadrature, when relevant. The total systematic uncertainty on the \acceff related to the parametrization of the signal input \pt and \rapidity distributions is considered uncorrelated as a function of \pt and \rapidity for a given quarkonium state. In addition, it was checked for the \jpsi~that using as MC input shapes the ones obtained from the PYTHIA8 generator~\cite{Bierlich:2022pfr} instead of the parametrization from Ref~\cite{Bossu:2011qe} was leading to similar results within the uncertainties discussed above.

The systematic uncertainty on the tracking efficiency in the muon chambers is obtained by comparing data with MC simulation. The single-muon tracking efficiency can be derived, in both data and MC, from the chamber efficiency, which can be evaluated using the redundancy of the tracking information in each station, since a subset of the detector is sufficient for a track to be reconstructed~\cite{Abelev:2014ffa}. 
The differences between the data and MC tracking efficiencies are taken as systematic uncertainty. A 1$\%$ uncertainty is found at the single muon level, hence a 2$\%$ uncertainty applies at the dimuon level for all the resonances. This uncertainty is assumed uncorrelated versus \pt and \rapidity.

The systematic uncertainty on the trigger efficiency has two origins: the differences in shape of the \pt-dependence of the trigger response function between data and MC in the region close to the trigger threshold, and the intrinsic efficiencies of the muon trigger chambers. The first uncertainty is estimated by comparing the \pt dependence, at the single-muon level, of the trigger response function between data and MC. This difference is then propagated at the dimuon level in the MC to evaluate the effect on the quarkonium \acceff determination. The obtained uncertainty varies, as a function of \pt and \rapidity, between $0.3\%$ and 2.4$\%$ for the \jpsi and \psitwos, and between $0.3\%$ and 1.1\% for the \upsns. The second uncertainty is estimated by comparing the \acceff obtained in the MC, with a second simulation in which the uncertainties on the trigger chamber efficiencies, as measured from data after varying the track selection criteria, are applied at the detector level, taking into account its segmentation, to blur the trigger response. This uncertainty is 1$\%$ for all the quarkonium states. The uncertainty on the trigger efficiency is uncorrelated as a function of \pt and \rapidity. 

The systematic uncertainty associated to the matching efficiency between the tracks reconstructed in the tracking chambers and those reconstructed in the trigger chambers is evaluated from the comparison of the efficiency variation in data and simulation by varying the value of the $\chi^{2}$ selection applied on the matching condition. 
It leads to a systematic uncertainty of 1$\%$ common to all the quarkonium resonances, and uncorrelated versus \pt and \rapidity.  

Tables~\ref{tab:syst1},~\ref{tab:syst2}, and~\ref{tab:syst3} summarize the systematic uncertainties on the evaluation of the \jpsi, \psitwos, and \upsns cross section, respectively. Values marked with an asterisk correspond to uncertainties correlated over \pt~and/or \rapidity. The total systematic uncertainty for a given quarkonium state is the quadratic sum of all the sources listed in the corresponding table. 

The systematic uncertainty on the \psitwos-to-\jpsi, \upstwos-to-\upsones, and \upsthrees-to-\upsones cross section ratios includes the uncertainty on the signal extraction, MC input, and branching ratio of the resonances. 
The systematic uncertainties from MCH and MTR efficiencies, and matching efficiency are similar for the ground and excited states and cancel out in the ratio, as do the luminosity uncertainty. The total systematic uncertainty on the integrated \psitwos-to-\jpsi~ratio is 10$\%$, while this systematic uncertainty varies between $9\%$ and 16$\%$ as a function of \pt~and between $8.9\%$ and 15$\%$ as a function of \rapidity. The total systematic uncertainty on the integrated \upstwos-to-\upsones [\upsthrees-to-\ensuremath{\Upsilon\rm(1S)}] ratio is 12$\%$ [20$\%$] respectively.

\begin{table}[!htb] 
\caption{Summary of the systematic uncertainties on the \jpsi cross section, integrated over \pt, \pt-differential, \rapidity-differential, and double differential in \pt and \rapidity. Values marked with an asterisk correspond to uncertainties correlated over \pt~and/or \rapidity.}

\begin{center}
\vspace{1ex}
\begin{tabular}[t]{c|c|c|c|c}
       
Source & Integrated (\%) & \pt-diff (\%) & \rapidity-diff (\%)& \pt-diff and \rapidity-diff (\%) \\ 
\hline
Branching ratio &  0.6 & 0.6* & 0.6* & 0.6* \\
Luminosity &  1.8 & 1.8* & 1.8* & 1.8* \\
Signal extraction & 3 & 1.9--4.4 & 2.1--4.4 & 0.8--4.4 \\
MC input & 3.2 & 0.3--2.2 & 1.4--4.9 & 0.1--3.3\\ 
MCH efficiency & 2 & 2& 2 & 2\\
MTR efficiency & 2 & 1.0--2.2 & 1.0--2.6 & 1.0--3.1 \\
Matching efficiency & 1 & 1 & 1 & 1 \\
\hline
\end{tabular}
\end{center}

\label{tab:syst1}
\end{table}

\begin{table}[!htb] 
\caption{Summary of the systematic uncertainties on the \psitwos cross section, integrated over \pt and \rapidity, as well as \pt-differential and \rapidity-differential. Values marked with an asterisk correspond to uncertainties correlated over \pt~and/or \rapidity.}

\begin{center}
\vspace{1ex}
\begin{tabular}[t]{c|c|c|c}
       
Source & Integrated (\%) & \pt-diff (\%) & \rapidity-diff (\%) \\ 
\hline
Branching ratio &  7.5 & 7.5* & 7.5*  \\
Luminosity &  1.8 & 1.8* & 1.8*  \\
Signal extraction & 7.2 & 5.8--15.4 & 5.8--13.9  \\
MC input & 3.3 & 1.4--2.4 & 1.4--5.0 \\ 
MCH efficiency & 2 & 2 & 2 \\
MTR efficiency & 2 & 1.4--2.2 & 1.0--2.6 \\
Matching efficiency & 1 & 1 & 1  \\
\hline
\end{tabular}
\end{center}

\label{tab:syst2}
\end{table}

\begin{table}[!htb] 
\caption{Summary of the systematic uncertainties on the \upsns cross section, integrated over \pt and \rapidity, as well as \pt-differential and \rapidity-differential for the \upsones. Values marked with an asterisk correspond to uncertainties correlated over \pt~and/or \rapidity.}

\begin{center}
\vspace{1ex}
\begin{tabular}[t]{c|c|c|c|c|c}
       & \multicolumn{3}{c|}{\upsones} & \upstwos & \upsthrees \\ \hline
Source & Integrated (\%) & \pt-diff (\%) & \rapidity-diff (\%) & Integrated & Integrated \\ 
\hline
Branching ratio &  2.0 & 2.0* & 2.0* & 8.8 & 9.6  \\
Luminosity &  1.8 & 1.8* & 1.8* & 1.8 & 1.8  \\
Signal extraction & 6.5 & 6.1--7.0 & 6.2--7.6 & 7.8 & 19 \\
MC input & 1.7 & 1.5--1.7 & 2.2--3.5 & 1.7 & 1.7 \\ 
MCH efficiency & 2 & 2 & 2 & 2 & 2\\
MTR efficiency & 1.2  & 1.1--1.5  & 1.0--1.2  & 1.2 & 1.2 \\
Matching efficiency & 1 & 1 & 1 & 1 & 1 \\
\hline
\end{tabular}
\end{center}

\label{tab:syst3}
\end{table}

\clearpage

\section{Results and discussion}
\label{sec:results}

The \pt- and \rapidity-differential cross section for inclusive quarkonium production is given by
\begin{equation}
\frac{\text{d}^2\sigma}{\text{d}\pt \text{d}\rapidity} = \frac{N (\Delta y, \Delta \pt ) }{ L_{\rm int} \times {\rm BR} \times \acceff(\Delta y, \Delta \pt )  \times \Delta \pt \times \Delta \rapidity },
\label{eq:sigma}
\end{equation}
where $N (\Delta y, \Delta \pt )$ is the raw quarkonium yield measured in a given \pt and \rapidity interval of width $\Delta \pt$ and $\Delta y$, respectively. 
The dimuon branching ratios BR are $(5.96 \pm 0.03)\%$ for \jpsi, $(0.80 \pm 0.06) \%$ for \psitwos, $(2.48 \pm 0.05)\%$ for \upsones, $(1.93 \pm 0.17)\%$ for \upstwos, and $(2.18 \pm 0.21)\%$ for \upsthrees~\cite{PDG}.

In this section, the results are given with two uncertainties, the first and second being the statistical and systematic ones, respectively. In the figures, the data points are represented with vertical error bars as statistical uncertainties and with boxes as systematic uncertainties. The correlated systematic uncertainties are quoted as text in the legends.

\subsection{Charmonium production} 
\label{sec:results:charmonium}

\subsubsection{\jpsi cross section}
\label{sec:results:jpsi}
\begin{figure}[!htb]
\begin{center}
\includegraphics[width=0.49\linewidth]{{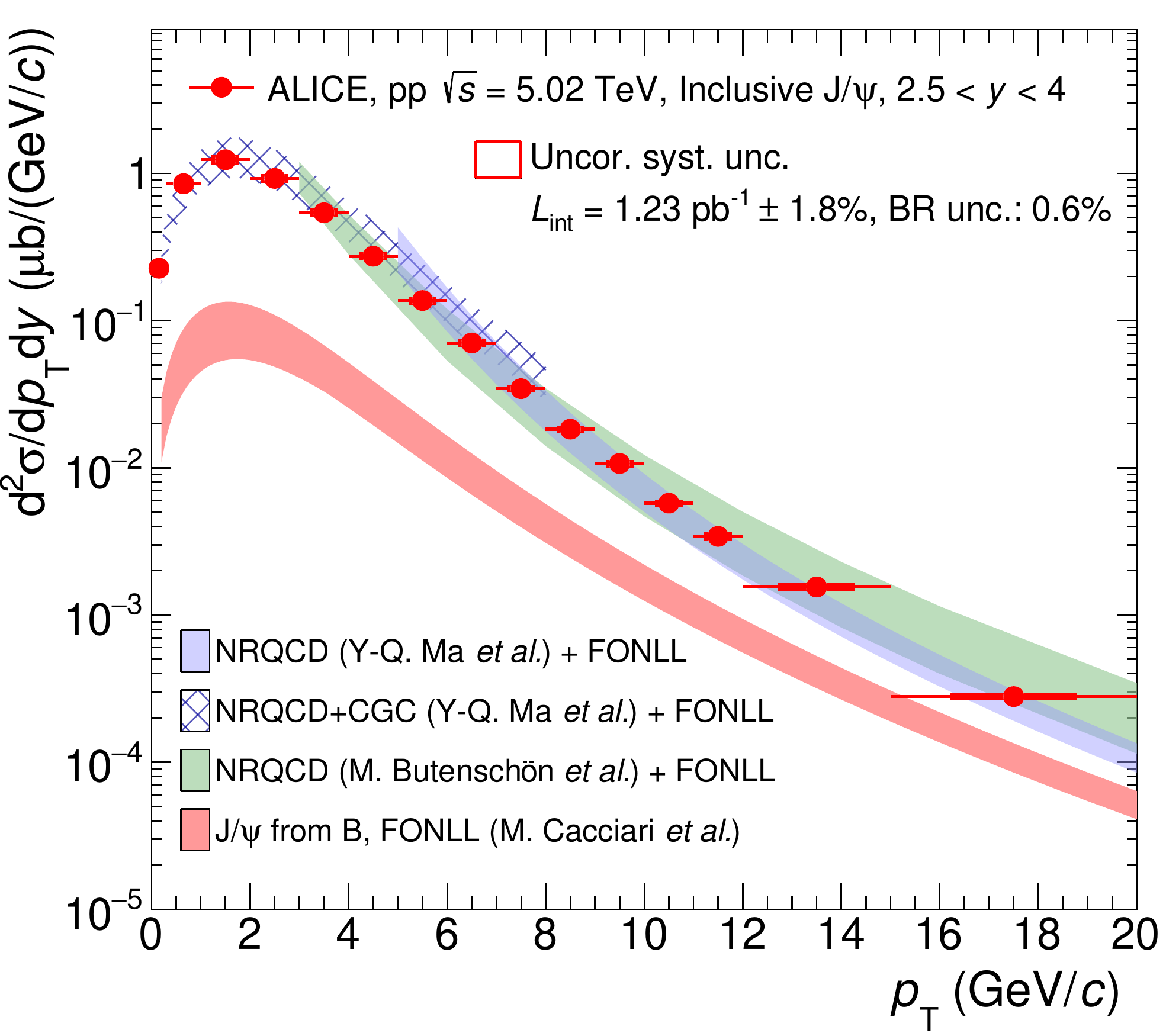}}
\includegraphics[width=0.49\linewidth]{{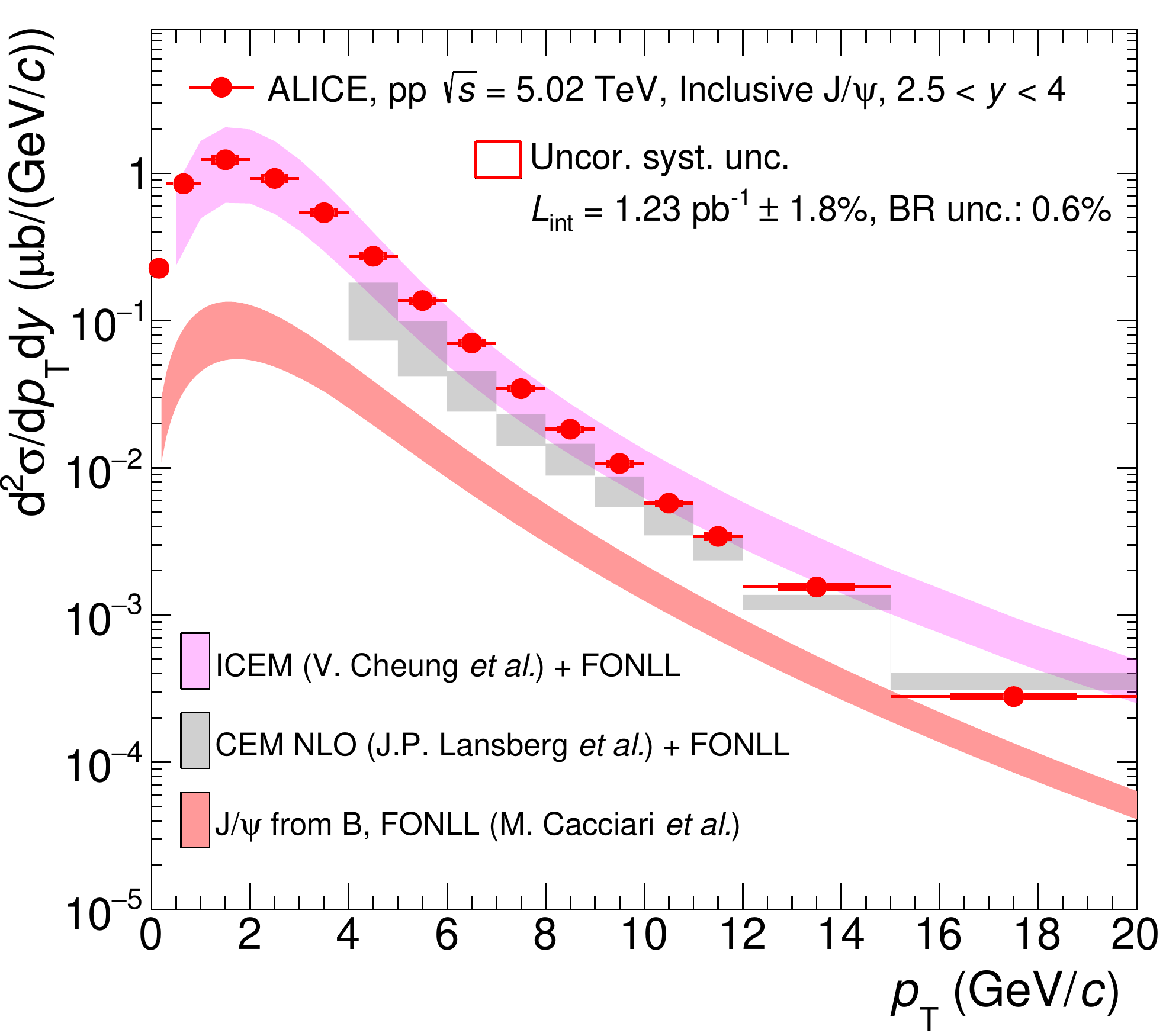}}

\caption{Transverse momentum dependence of the inclusive \jpsi cross section. The measurements are compared to theoretical calculations from Refs.~\cite{Butenschoen:2010rq, Ma:2010yw,Ma:2014mri} (left) and Refs.~\cite{Cheung:2018tvq,Lansberg:2020rft,Cacciari:2012ny} (right). The calculations of the non-prompt contribution~\cite{Cacciari:2012ny} are also shown separately. See text for details.}

\label{fig:JpsiCrosssectionvspt}
\end{center}
\end{figure}

\begin{figure}[!htb]
\begin{center}
\includegraphics[width=0.67\linewidth]{{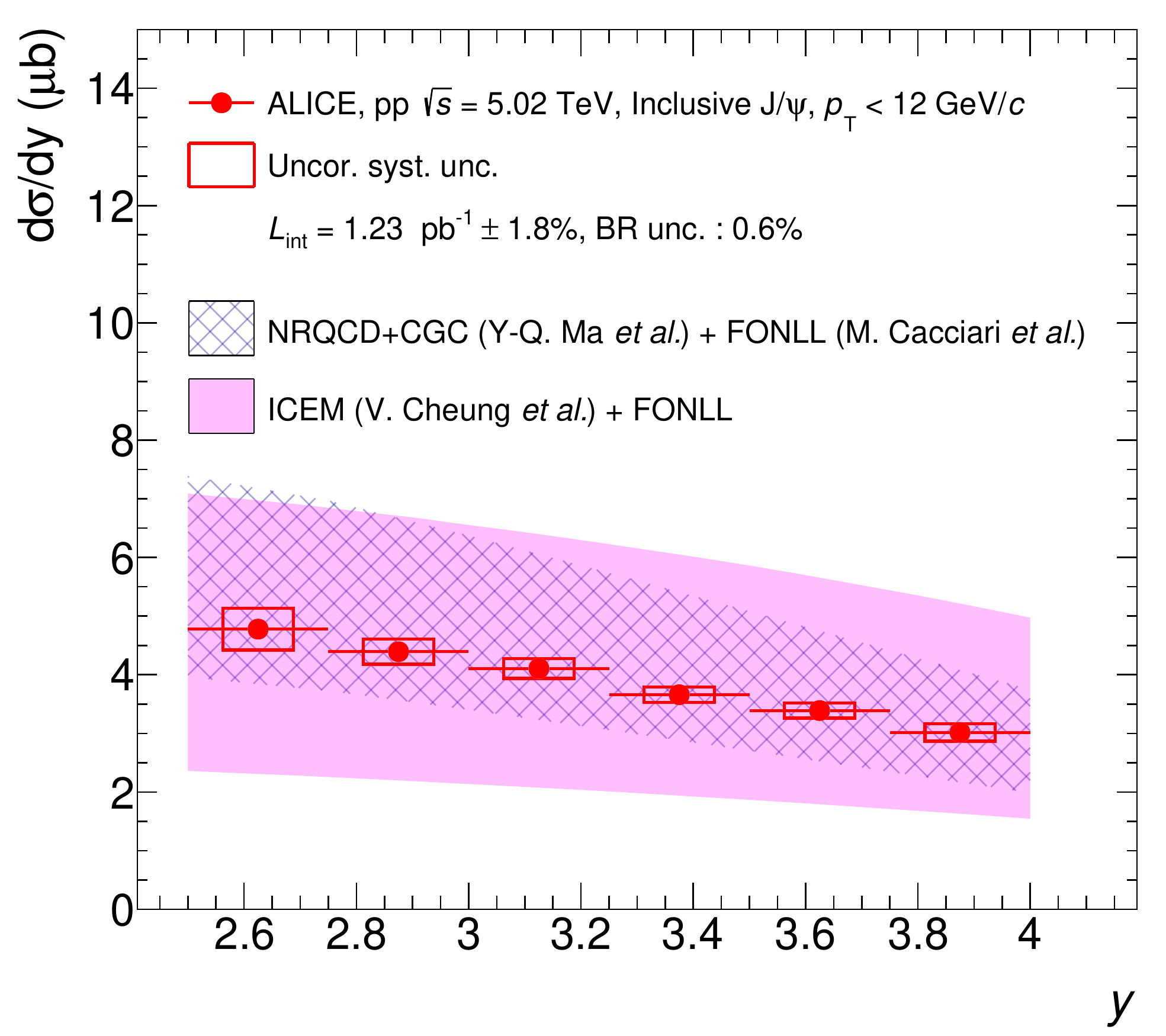}}

\caption{Rapidity dependence of the inclusive \jpsi cross section. The measurements are compared to theoretical calculations from Refs.~\cite{Ma:2014mri,Cheung:2018tvq}.}
\label{fig:JpsiCrosssectionvsy}
\end{center}
\end{figure}

\begin{figure}[!htb]
\begin{center}
\includegraphics[width=0.49\linewidth]{{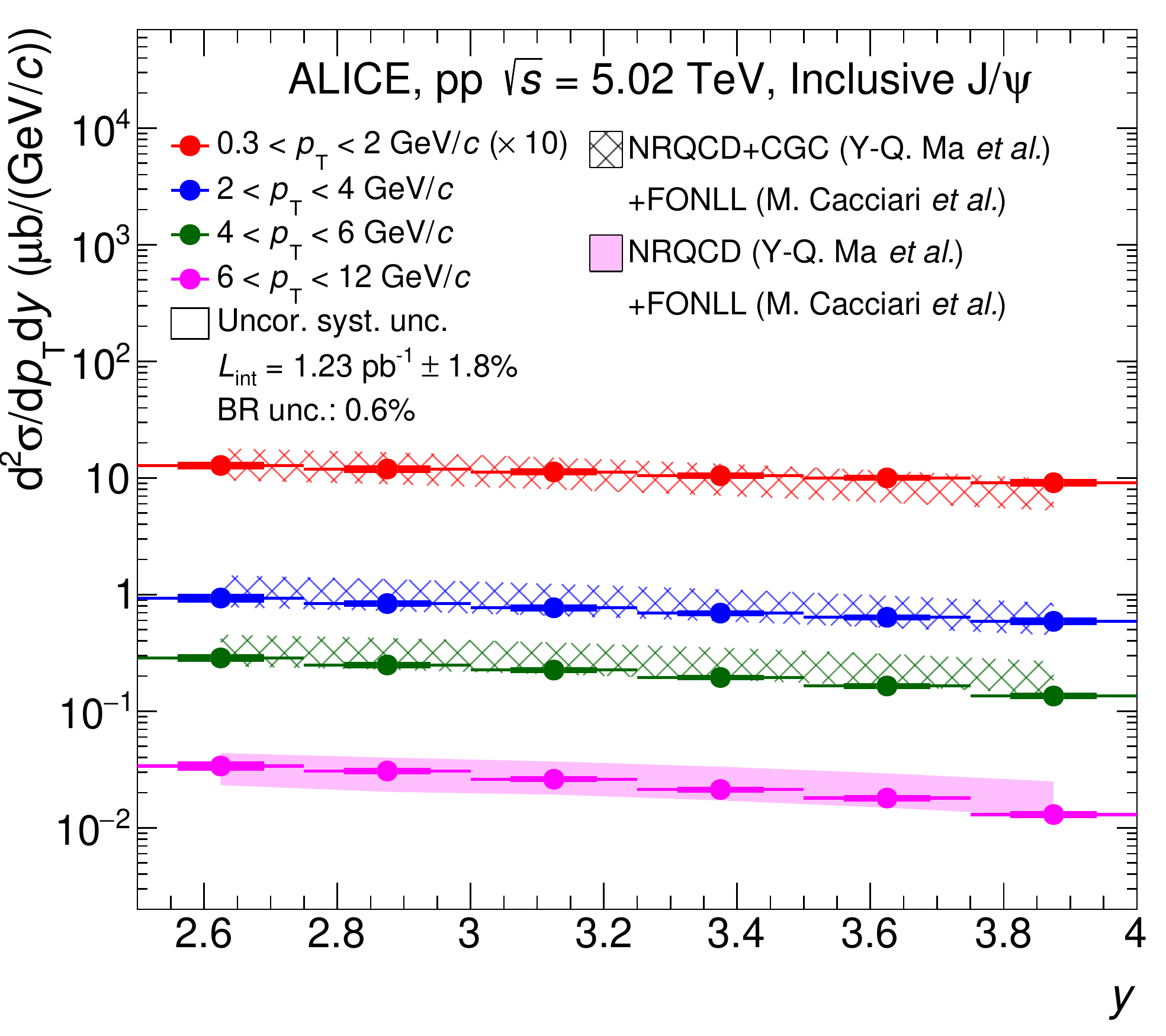}}
\includegraphics[width=0.49\linewidth]{{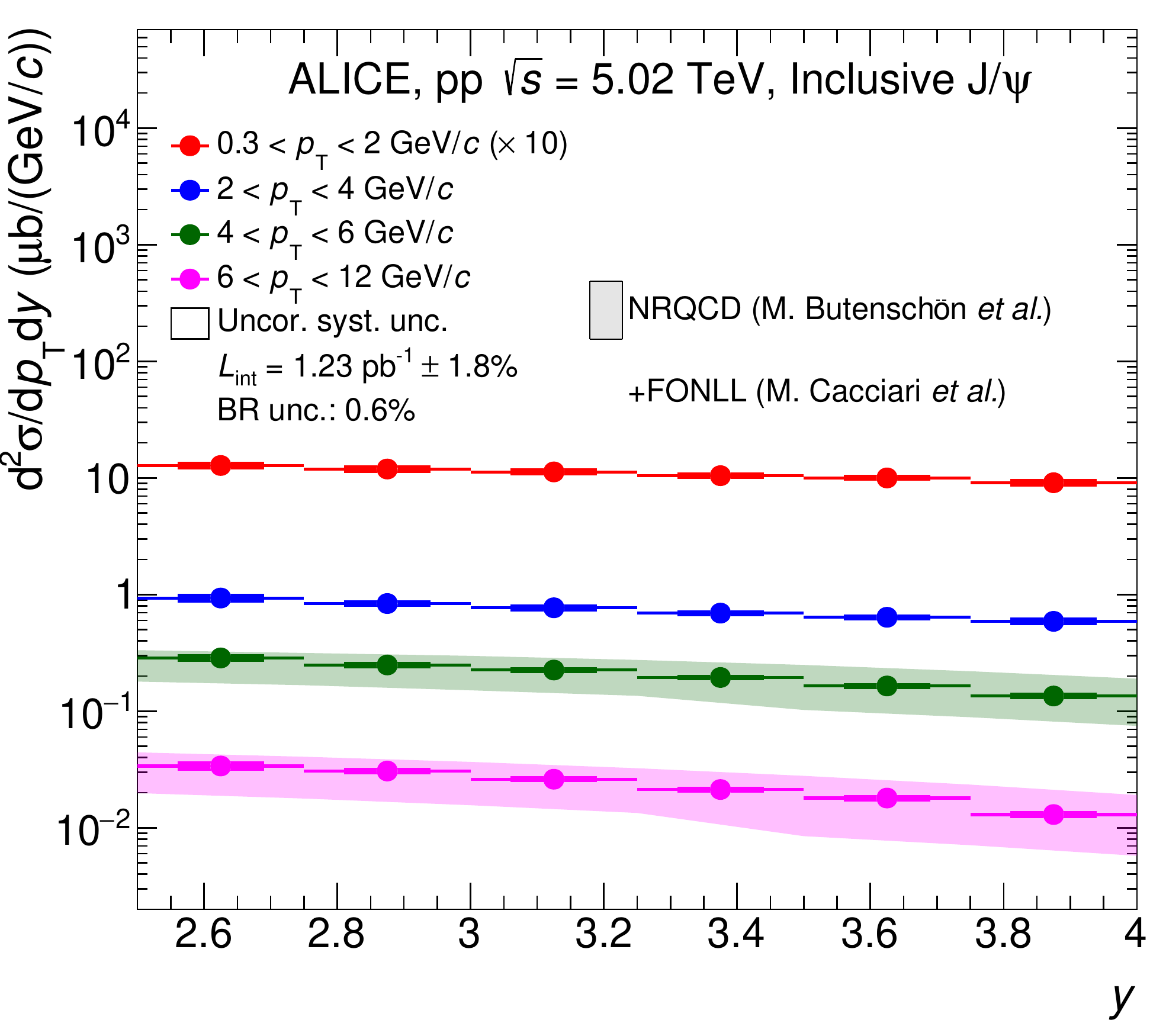}}
\includegraphics[width=0.49\linewidth]{{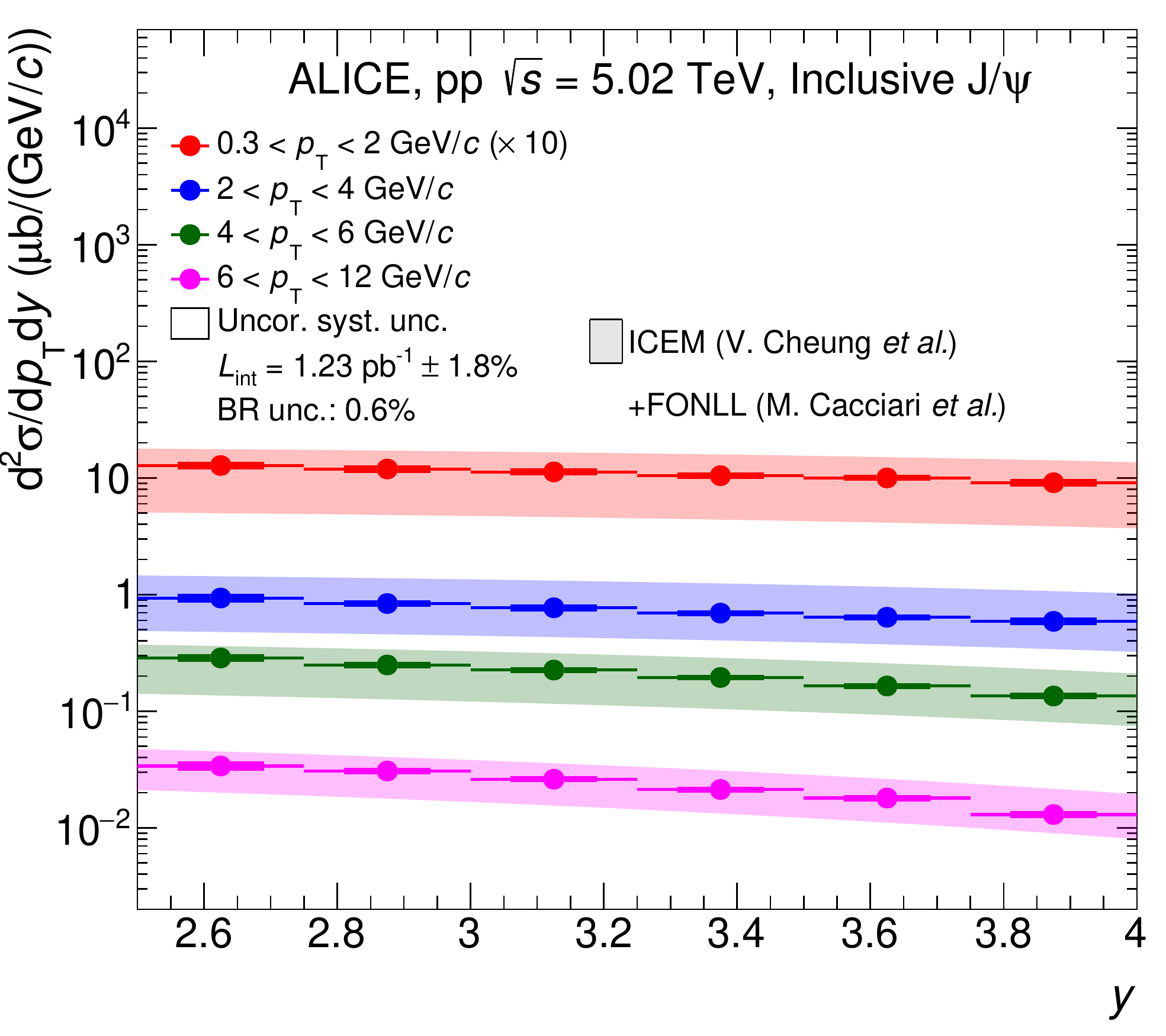}}
\includegraphics[width=0.49\linewidth]{{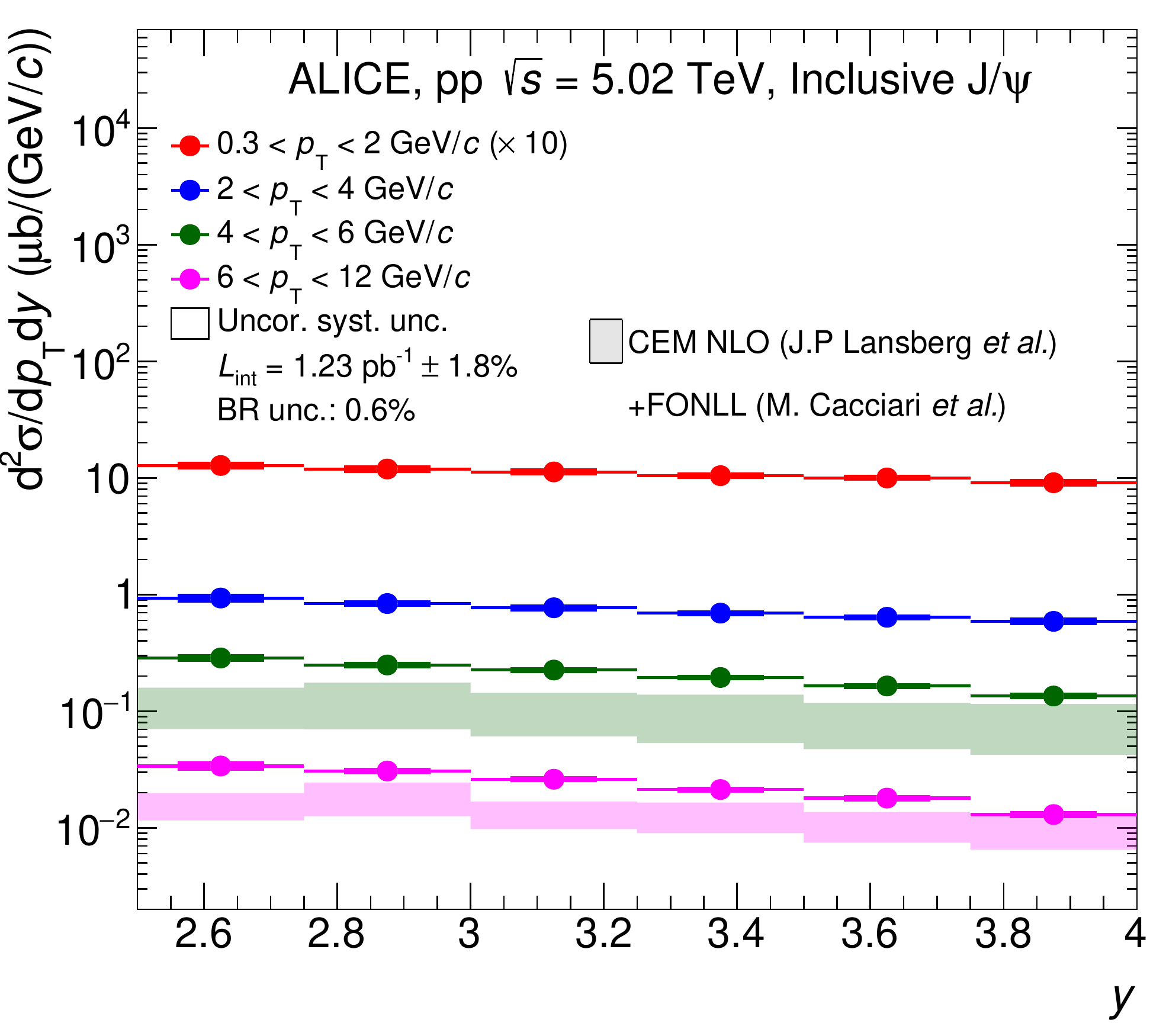}}
\caption{Rapidity dependence of the inclusive \jpsi cross section for various \pt ranges, compared to theoretical calculations~\cite{Butenschoen:2010rq, Ma:2010yw,Ma:2014mri,Cheung:2018tvq,Lansberg:2020rft,Cacciari:2012ny}. The theoretical calculations are scaled as for the data for $0.3<\pt<2$~\GeVc. See text for details.}
\label{fig:JpsiCrosssectionvspty}
\end{center}
\end{figure}

The inclusive \jpsi production cross section in \pp collisions at $\s =5.02$~TeV integrated over $2.5 < y < 4$ and $\pt < 20$~\GeVc is $\sigma_{\jpsi} = 5.88 \pm 0.03$~(stat.)~$\pm~0.34$~(syst.)~$\mu$b. 
The differential cross sections are shown as a function of \pt and \rapidity in Figs.~\ref{fig:JpsiCrosssectionvspt} and~\ref{fig:JpsiCrosssectionvsy}, respectively. The results are in agreement with the previously published ALICE measurements~\cite{Adam:2016rdg,Acharya:2017hjh}. A maximum deviation of $1.8\sigma$ for $4<\pt<5$~GeV/c and $3.75 < y < 4$ is found, where the comparison is performed using the quantity $\sigma_{\jpsi} \times {\rm BR}$ in order to remove the ${\rm BR}$ uncertainty. These new measurements extend the \pt reach from 12~\GeVc to 20~\GeVc. The cross sections are in agreement, within uncertainties, with the recent LHCb results~\cite{LHCb:2021pyk}. The inclusive \jpsi double-differential production cross section is shown as a function of \rapidity for various \pt ranges in the four panels of Fig.~\ref{fig:JpsiCrosssectionvspty}. These measurements will also serve as reference for studying the nuclear modification of \jpsi production in \PbPb collisions. To this purpose, for the \pt and \rapidity double-differential \pp cross sections, \jpsi with $\pt<0.3$~\GeVc were excluded to match a similar selection applied in \PbPb collisions to remove the photoproduction contribution, occurring besides the hadronic one, and relevant at low \pt in peripheral collisions~\cite{Acharya:2019iur}.

The cross sections are compared with three theoretical calculations based on NRQCD: two Next-to-Leading Order (NLO) NRQCD calculations from Butenschön~\etal~\cite{Butenschoen:2010rq} and from Ma~\etal~\cite{ Ma:2010yw}, and a Leading Order (LO) NRQCD calculation coupled to a Color Glass Condensate~(CGC) description of the proton structure for low-$x$ gluons from Ma~\etal~\cite{Ma:2014mri}, labelled as NRQCD+CGC in the following. They are also compared to two theoretical calculations based on CEM: an improved CEM (ICEM) calculation from Cheung~\etal~\cite{Cheung:2018tvq} and a NLO CEM calculation from Lansberg~\etal~\cite{Lansberg:2020rft}. While the NLO calculations presented here are not reliable in the low-\pt region ($\pt \lesssim M_{\rm c\bar{c}}$), the calculations from NRQCD+CGC~\cite{Ma:2014mri} or the semi-hard approach based on $k_{\rm T}$ factorization of the ICEM model~\cite{Cheung:2018tvq} are available also at low \pt. 
 The theoretical calculations are for prompt \jpsi and account therefore for the decay of \psitwos and \chic into \jpsi. 
Since the measurements include as well non-prompt \jpsi, their contribution is estimated from Fixed-Order Next-to-Leading Logarithm (FONLL) calculations from Cacciari~\etal~\cite{Cacciari:2012ny}. The prompt and non-prompt \jpsi calculations are summed in order to obtain inclusive \jpsi calculations to be compared to the measurements. The uncertainties from renormalization and factorization scale and parton distribution function on prompt and non-prompt \jpsi production are considered as uncorrelated.

In Figs.~\ref{fig:JpsiCrosssectionvspt},~\ref{fig:JpsiCrosssectionvsy}, and~\ref{fig:JpsiCrosssectionvspty} the data are compared with the models described above when calculations are available. A good description of the \pt and \rapidity distributions of the data is obtained by the NRQCD models for $\pt > 3$~\GeVc for the model from Butenschön~\etal~and $\pt > 5$~\GeVc for the model from Ma~\etal. The NRQCD+CGC model describes well the data as a function of \pt and \rapidity for $\pt < 8$~\GeVc. The ICEM model also gives a good description of the data as a function of \rapidity and \pt for $\pt< 15$~\GeVc. It overestimates the data for $\pt> 15$~\GeVc. Finally, the CEM NLO calculation underestimates the cross sections for $4 < \pt < 10$~\GeVc and reproduces the data at higher \pt. The non-prompt \jpsi contribution is also shown in Fig.~\ref{fig:JpsiCrosssectionvspt}, indicating that the contribution increases with increasing \pt from 7\% at $\pt \approx 1$~\GeVc to $42\%$ for the largest \pt interval.

%\clearpage

\subsubsection{\psitwos cross section}

\begin{figure}[b]
\begin{center}
\includegraphics[width=0.49\linewidth]{{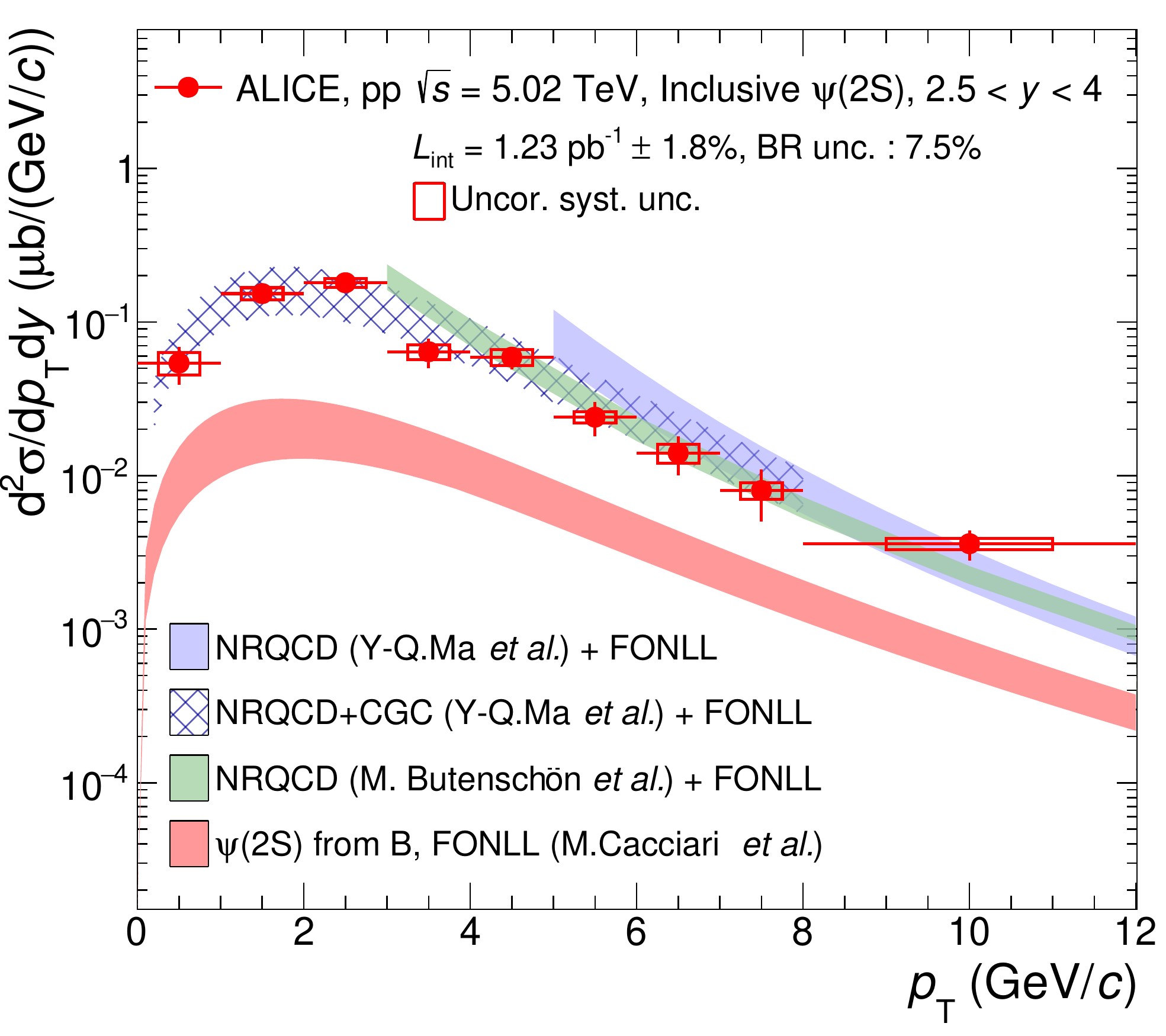}}
\includegraphics[width=0.49\linewidth]{{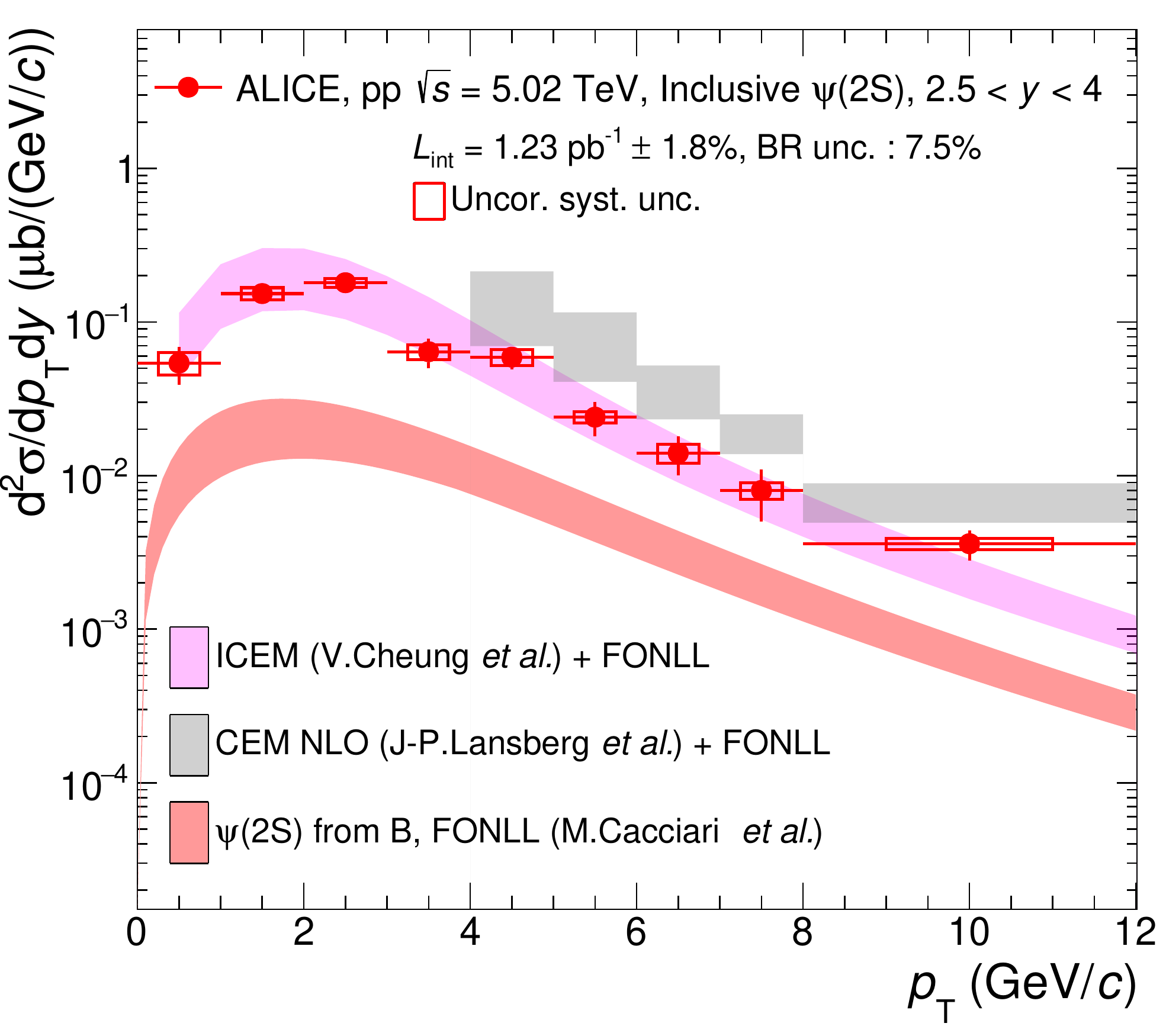}}
\caption{The left and right panels show the \pt dependence for the inclusive \psitwos production cross section in \pp collisions at $\s = 5.02$~TeV. 
The results are compared with the theory predictions based on NRQCD~\cite{Butenschoen:2010rq, Ma:2010yw,Ma:2014mri} (left) and CEM~\cite{Cheung:2018tvq,Lansberg:2020rft} (right) models. The calculation of the non-prompt contribution from FONLL calculations~\cite{Cacciari:2012ny} are also shown separately. See text for details.
}
\label{fig:psitwosCrosssectionvspt}
\end{center}
\end{figure}

\begin{figure}[!htb]
\begin{center}
\includegraphics[width=0.67\linewidth]{{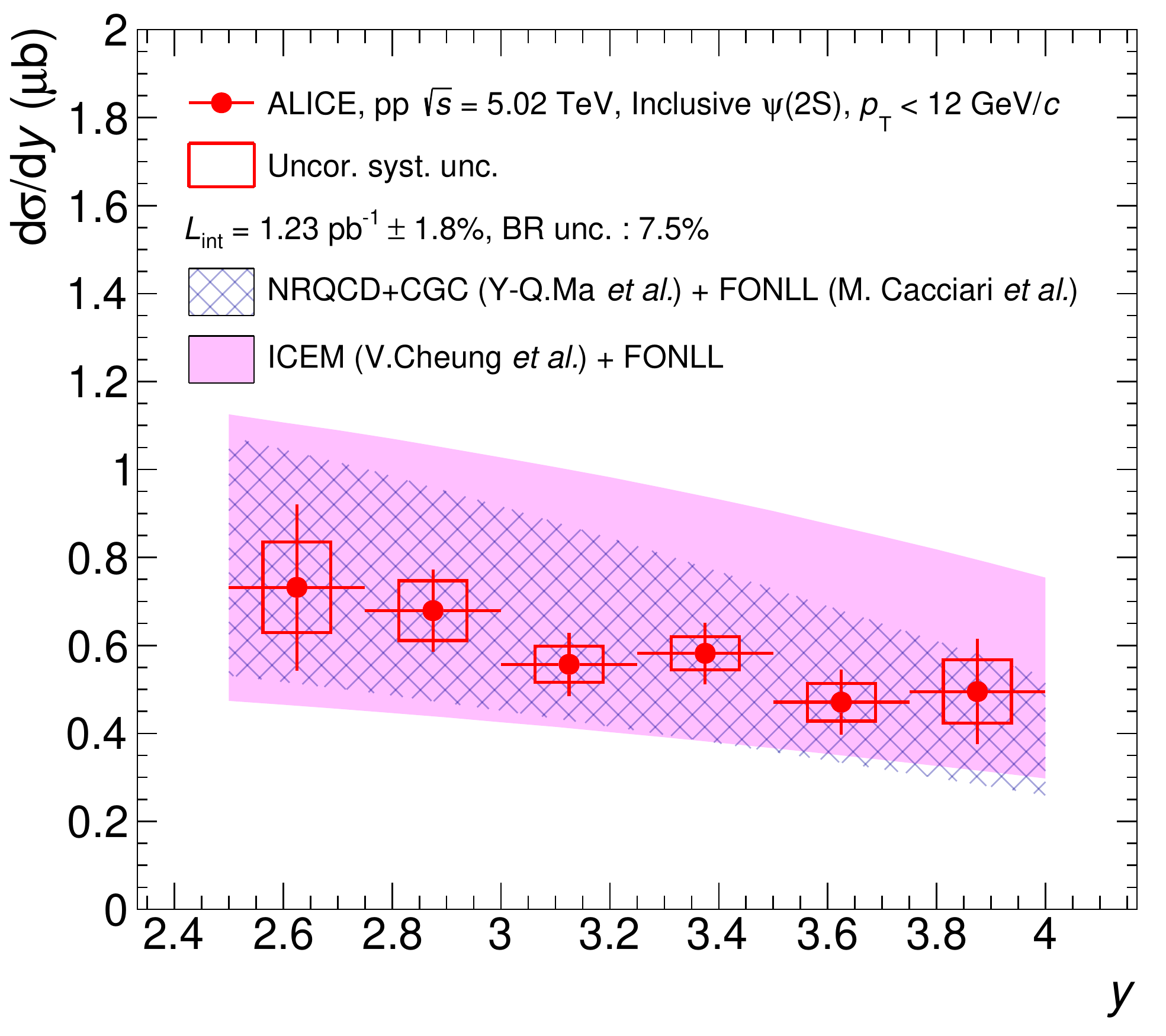}}
\caption{Rapidity dependence for the inclusive \psitwos production cross section in \pp collisions at $\s = 5.02$~TeV. 
The results are compared with the theory predictions based on NRQCD+CGC~\cite{Ma:2014mri} and ICEM~\cite{Cheung:2018tvq} models. See text for details.
}
\label{fig:psitwosCrosssectionvsy}
\end{center}
\end{figure}

The inclusive \psitwos production cross section in pp collisions at $\sqrt{s}=5.02$~TeV integrated over $\pt < 12$~\GeVc and for $2.5 < y < 4$ is $\sigma_{\psitwos} = 0.87 \pm 0.06$~(stat.) $\pm~0.10$~(syst.)~$\mu$b. 
The result is in agreement with the previously published \psitwos cross section~\cite{Acharya:2017hjh} and the deviation is found to be $0.75\sigma$ for the quantity $\sigma_{\psi}(2S) \times {\rm BR}$. 
An improvement of a factor $\sim$3 for the statistical uncertainty is obtained for the most recent data set. The first results on the \pt and \rapidity dependence of the inclusive \psitwos cross section for $2.5 < y < 4$ in \pp collisions at $\s = 5.02$~TeV are shown in Figs.~\ref{fig:psitwosCrosssectionvspt} and~\ref{fig:psitwosCrosssectionvsy}, respectively. 


%\clearpage

Calculations of the same theory models as discussed in Section~\ref{sec:results:jpsi} are compared with the inclusive \psitwos cross section in Figs.~\ref{fig:psitwosCrosssectionvspt} and~\ref{fig:psitwosCrosssectionvsy}.
As for the \jpsi case, the experimental measurements include a prompt and a non-prompt contribution while the model calculations are performed for the former only. 
Therefore, the \psitwos non-prompt contribution, according to FONLL~\cite{Cacciari:2012ny}, is summed to all theoretical predictions.

In the left panel of Fig.~\ref{fig:psitwosCrosssectionvspt}, the NRQCD calculation from Butenschön~\etal~\cite{Butenschoen:2010rq} agrees with the experimental data for $4 < \pt < 12$~\GeVc, and the NRQCD calculation from Ma~\etal~\cite{Ma:2010yw} describes well the data except for $5< \pt < 6$~\GeVc, where it overpredicts them. In addition, in the right panel of Fig.~\ref{fig:psitwosCrosssectionvspt} there are significant deviations between the CEM NLO calculation~\cite{Lansberg:2020rft} and the data at $\pt > 5$~\GeVc.
The NRQCD+CGC~\cite{Ma:2014mri} and ICEM~\cite{Cheung:2018tvq} models provide a good description of the \psitwos cross section as a function of \pt and \rapidity, albeit with large uncertainties for the \rapidity~dependence, as it can be seen in Figs.~\ref{fig:psitwosCrosssectionvspt}~and ~\ref{fig:psitwosCrosssectionvsy}, respectively. 
Finally, the non-prompt \psitwos contribution from FONLL~\cite{Cacciari:2012ny} is also shown in Fig.~\ref{fig:psitwosCrosssectionvspt} and varies from $10\%$ to $25\%$ as a function of \pt. 

%\clearpage

\subsubsection{\psitwos over \jpsi cross section ratio}
\label{sec:psitwostojpsi}

\begin{figure}[!htb]
\begin{center}
\includegraphics[width=0.49\linewidth]{{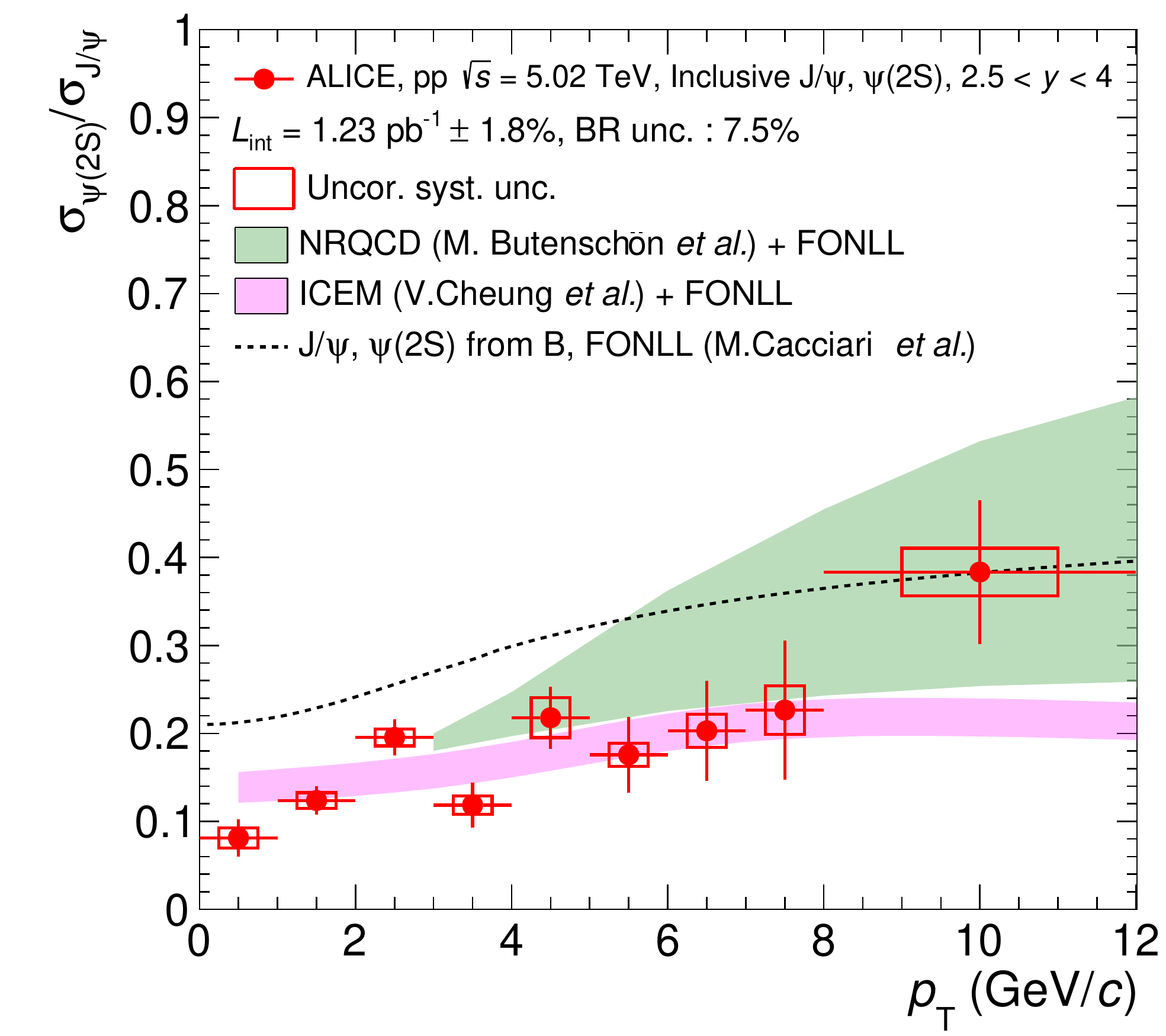}}
\includegraphics[width=0.49\linewidth]{{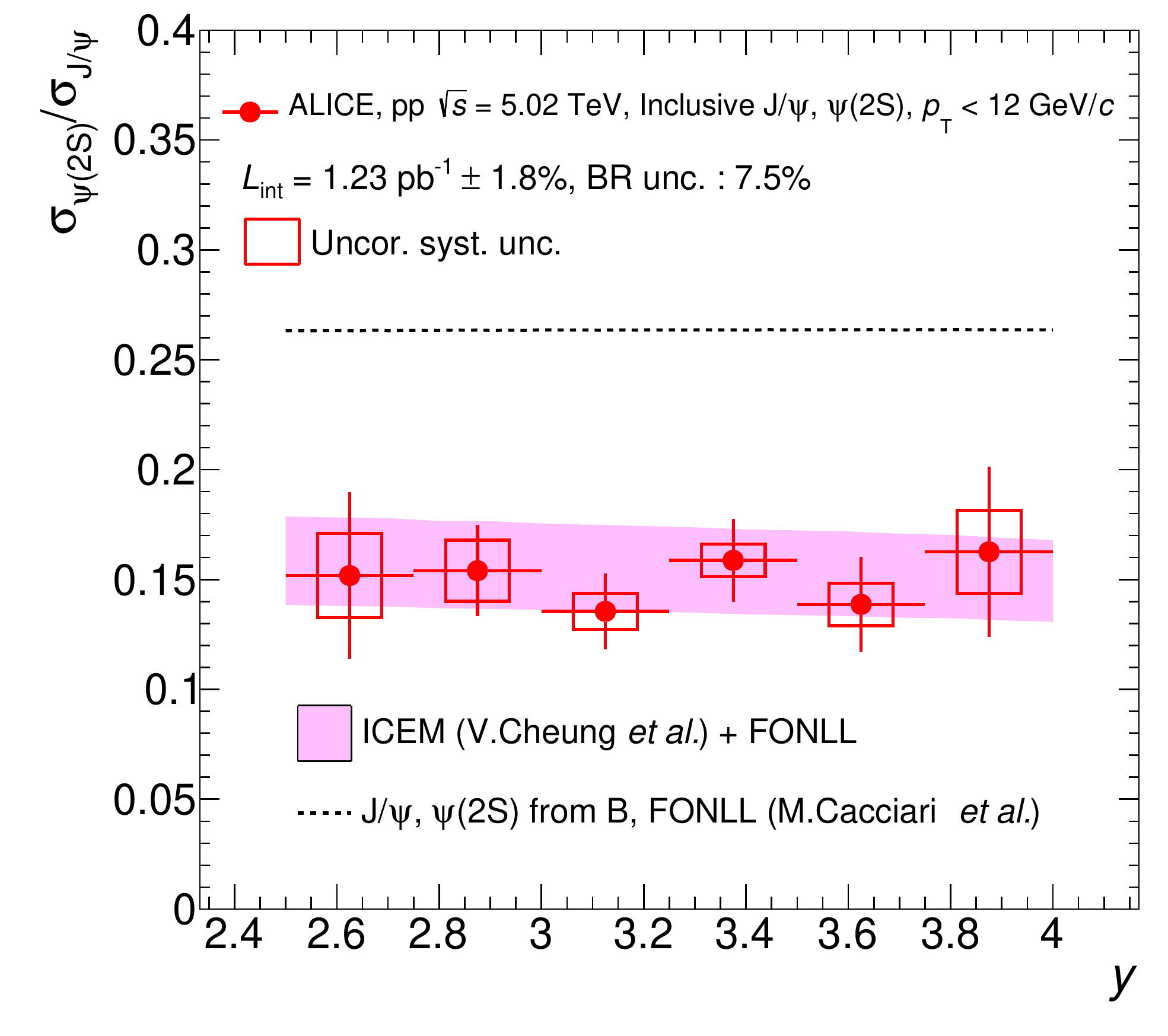}}
\caption{The inclusive \psitwos-to-\jpsi cross section ratio as a function of \pt (left) and \rapidity (right), compared with theoretical calculations~\cite{Butenschoen:2010rq, Cheung:2018tvq,Cacciari:2012ny}. See text for details.}
\label{fig:2S-to-1S_model_comparison}
\end{center}
\end{figure}

The ratio between the inclusive \psitwos and inclusive \jpsi production cross sections integrated over $\pt < 12$~\GeVc and for $2.5 < y < 4$, is $0.15\pm0.01$~(stat.)~$\pm0.02$~(syst.). 
In Fig.~\ref{fig:2S-to-1S_model_comparison}, the \pt and \rapidity dependence of the \psitwos-to-\jpsi cross section ratio in \pp collisions at $\s = 5.02$~TeV are shown in the left and right panel, respectively.   
The boxes represent the uncorrelated systematic uncertainties due to the MC input shapes and the signal extraction. The branching-ratio uncertainties, fully correlated versus \pt~and \rapidity, are reported in the legend of Fig.~\ref{fig:2S-to-1S_model_comparison}. All the other systematic uncertainties are correlated over the two resonances and cancel out in the ratio.

The \psitwos-to-\jpsi production cross section ratio is also compared with theoretical models. 
As in previous sections, the non-prompt contribution from FONLL~\cite{Cacciari:2012ny} is added to all theoretical calculations. Each individual source of theoretical uncertainty is considered as correlated among the two states and partially cancel in the ratio calculation. 
The NRQCD calculations from Butenschön~\etal~\cite{Butenschoen:2010rq} describe well the \pt dependence of the cross section ratio within the large model uncertainties. %,
A good description of the trend of the \psitwos-to-\jpsi cross section ratio as a function of \pt and \rapidity is also provided by the ICEM model~\cite{Cheung:2018tvq}. 
In the left and right panels of Fig.~\ref{fig:2S-to-1S_model_comparison}, the  non-prompt cross section ratios from FONLL~\cite{Cacciari:2012ny} are also shown separately for completeness. 

\subsection{Bottomonium production}

The inclusive production cross sections of the three $\Upsilon$~states are measured for the first time in \pp collisions at $\s = 5.02$~TeV and for $2.5<y<4$. The cross sections, integrated over $\pt < 15$~\GeVc and for $2.5 < y < 4$, are:
\begin{itemize}[label=\textbullet]
\item  $\sigma_{\upsones} = 45.5 \pm 3.9$ {\rm(stat.)} $\pm 3.5$  {\rm(syst.)}~nb,
\item  $\sigma_{\upstwos} = 22.4 \pm 3.2$ {\rm(stat.)} $\pm 2.7$ {\rm(syst.)}~nb,
\item  $\sigma_{\upsthrees} = 4.9 \pm 2.2$ {\rm(stat.)} $\pm 1.0$ {\rm(syst.)}~nb.
\end{itemize}

The corresponding excited to ground-state ratios amount to:
\begin{itemize}[label=\textbullet]
\item  $\sigma_{\upstwos} / \sigma_{\upsones} = 0.50 \pm 0.08$ {\rm(stat.)} $\pm 0.06$ {\rm(syst.)},
\item  $\sigma_{\upsthrees} / \sigma_{\upsones} = 0.10 \pm 0.05$ {\rm(stat.)} $\pm 0.02$ {\rm(syst.)}.
\end{itemize}

\begin{figure}[!htb]
\begin{center}
\includegraphics[width=0.49\linewidth]{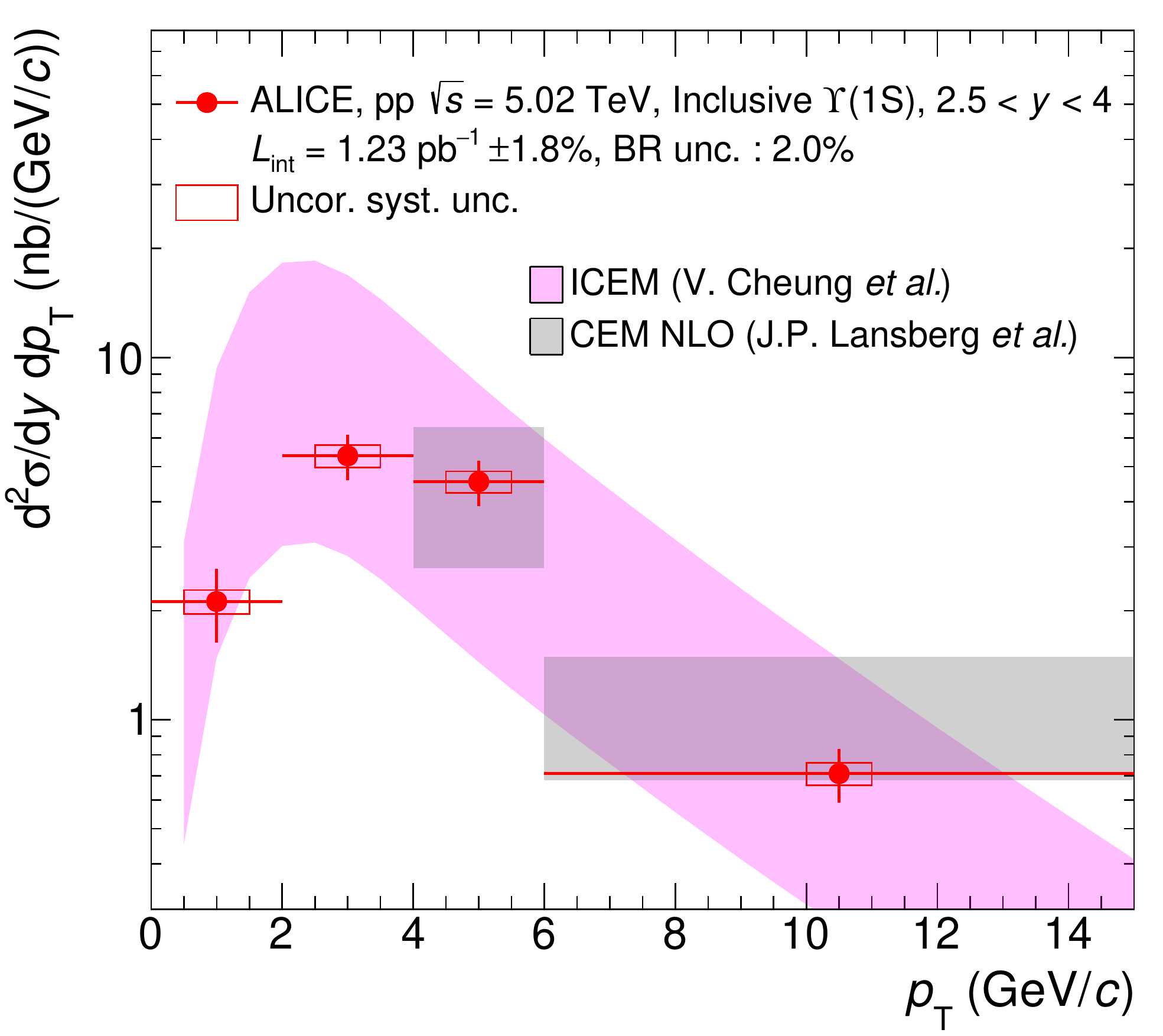}
\includegraphics[width=0.49\linewidth]{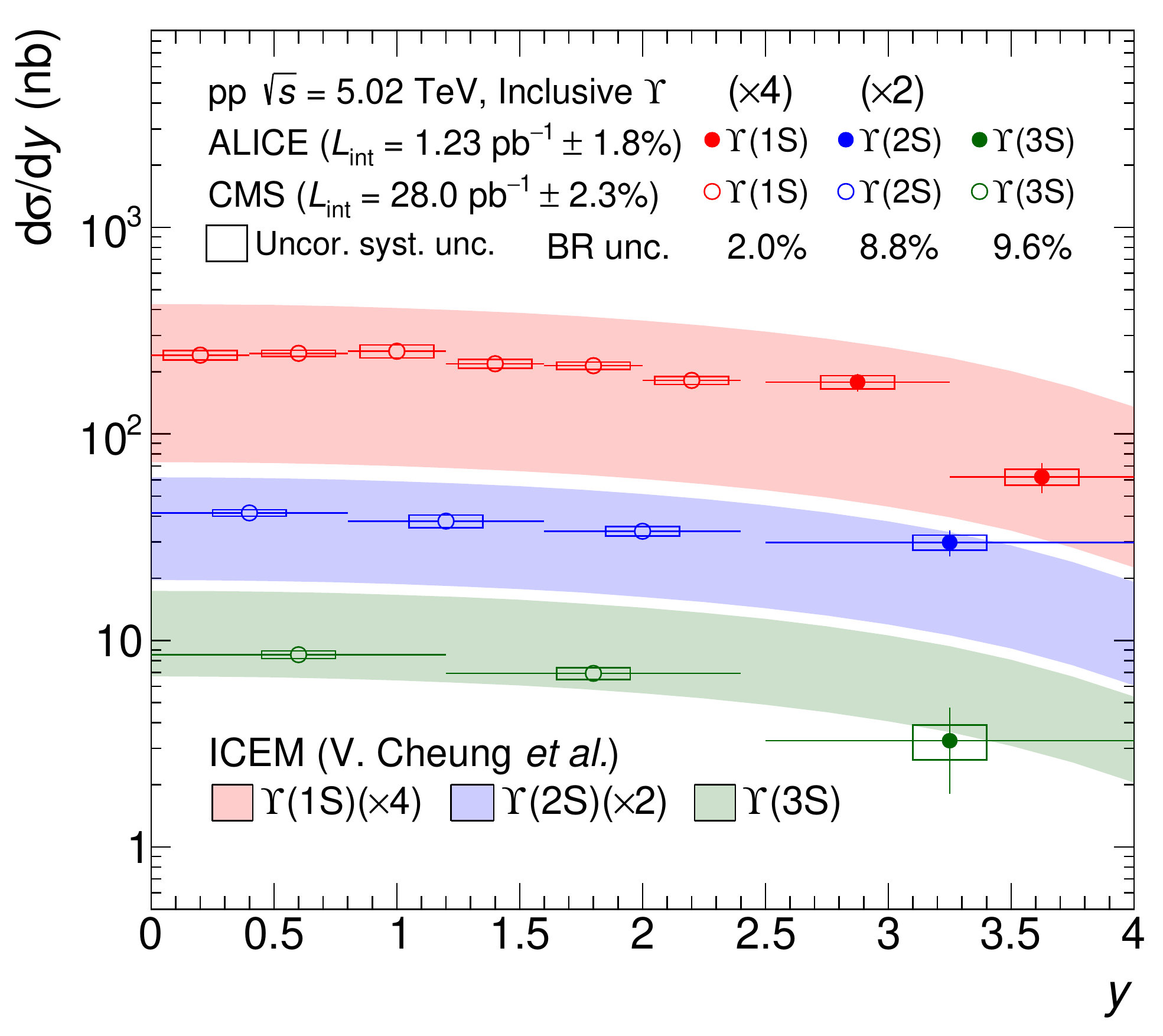}
\caption{Transverse momentum dependence of the \upsones cross section (left) and \rapidity dependence of the \upsones, \upstwos, and \upsthrees (right) measured by ALICE and CMS. The two panels also show theoretical calculations~\cite{Cheung:2018upe,Lansberg:2020rft}. See text for details. 
}
\label{fig:upsilon}
\end{center}
\end{figure}

The cross sections are presented in Fig.~\ref{fig:upsilon} as a function of \pt for the \upsones on the left panel and as a function of \rapidity for the three \ups states, together with the CMS measurements~\cite{Sirunyan:2018nsz}.

The experimental results are compared to ICEM calculations~\cite{Cheung:2018upe} as well as to CEM NLO calculations~\cite{Lansberg:2020rft}. Both approaches account for the feed-down contributions from heavier bottomonium states. The two CEM calculations describe the measured \pt-differential cross section within uncertainties. The \rapidity dependence shows that the forward ALICE acceptance covers the region where the production drops from the midrapidity plateau. This observation is in line with the ICEM expectations. The measured \upstwos production cross section lies in the higher limit of the model while the \upsthrees result lies at the lower edge of the theory band.

\subsection{Energy dependence of quarkonium production} 
\label{sec:results:energy}

\begin{figure}[b]
\begin{center}
\includegraphics[width=0.49\linewidth]{{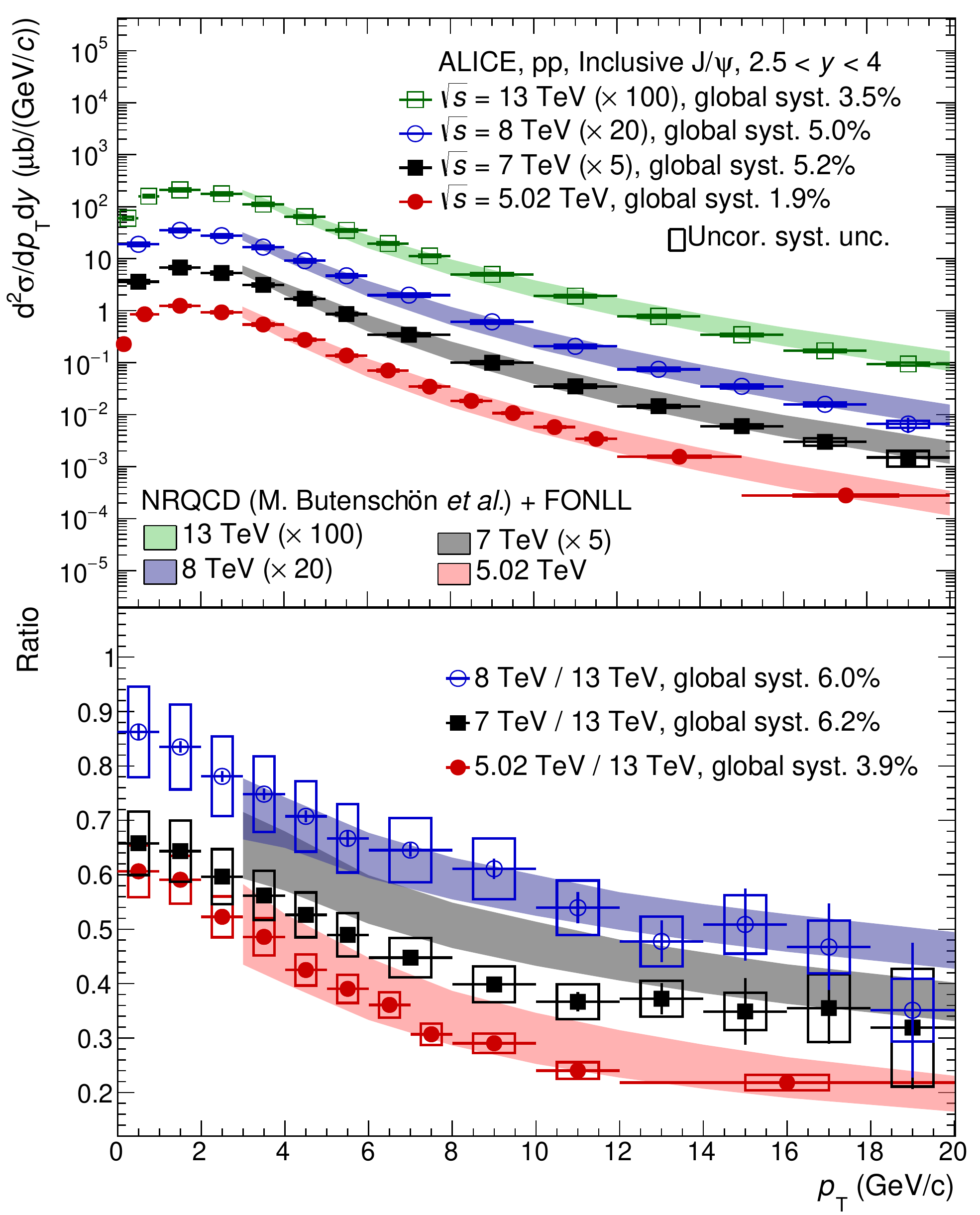}}
\includegraphics[width=0.49\linewidth]{{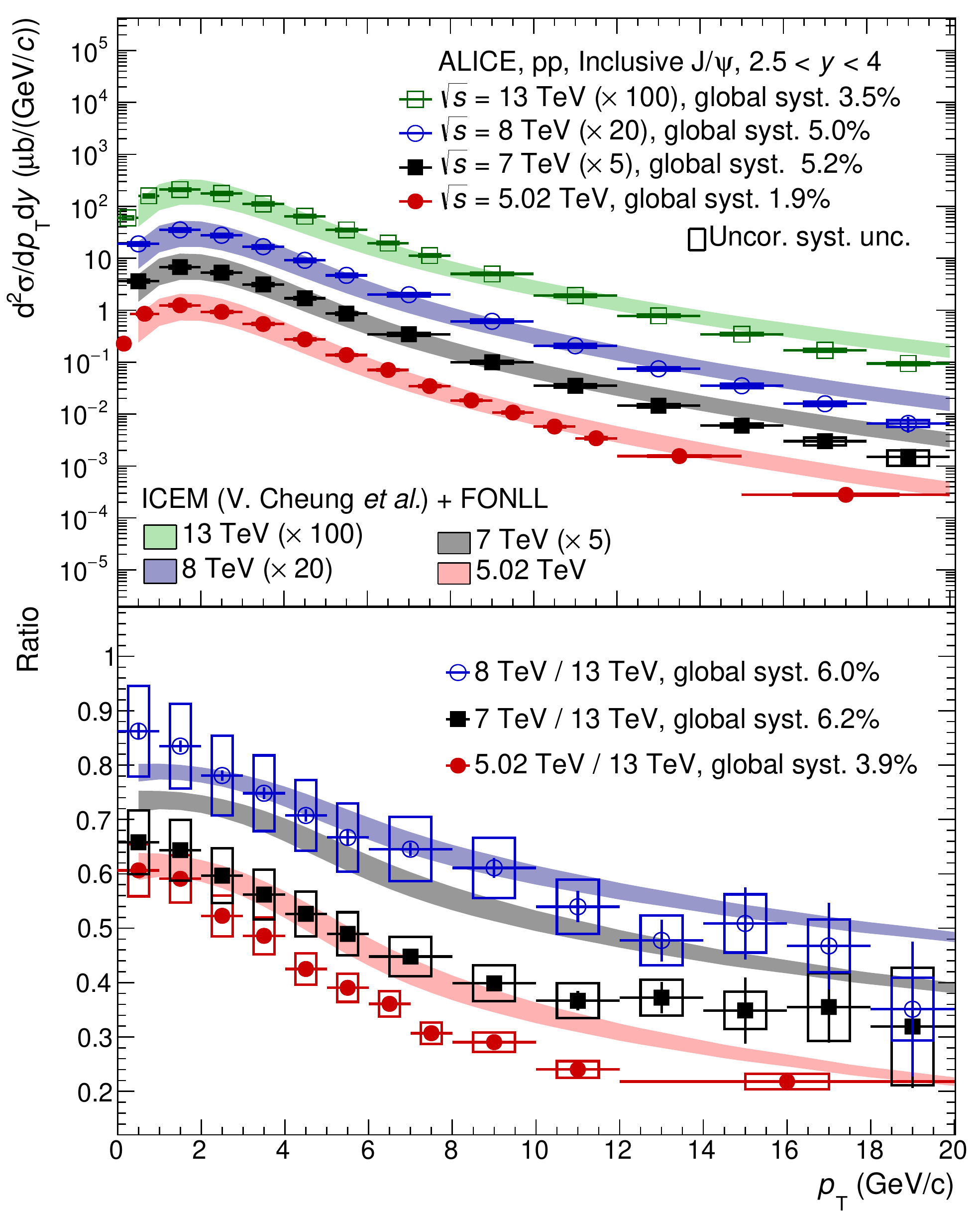}}
\caption{Transverse momentum dependence of the inclusive \jpsi cross section, at forward \rapidity, measured in pp collisions at $\sqrt{s} = 5.02$, 7~\cite{Abelev:2014qha}, 8~\cite{Adam:2015rta}, and 13~\cite{Acharya:2017hjh} TeV (top panels), and ratio of the measurements at 5.02, 7, and 8 TeV to the 13 TeV data (bottom panels). 
The data are compared with the NRQCD theoretical calculations from Butenschön~\etal~+ FONLL (left panels)~\cite{Butenschoen:2010rq,Cacciari:2012ny} and with theoretical calculations from ICEM + FONLL (right panels)~\cite{Cheung:2018tvq,Cacciari:2012ny}.}
\label{fig:jpsienergycomp}
\end{center}
\end{figure}

\begin{figure}[!hb]
\begin{center}
\includegraphics[width=0.49\linewidth]{{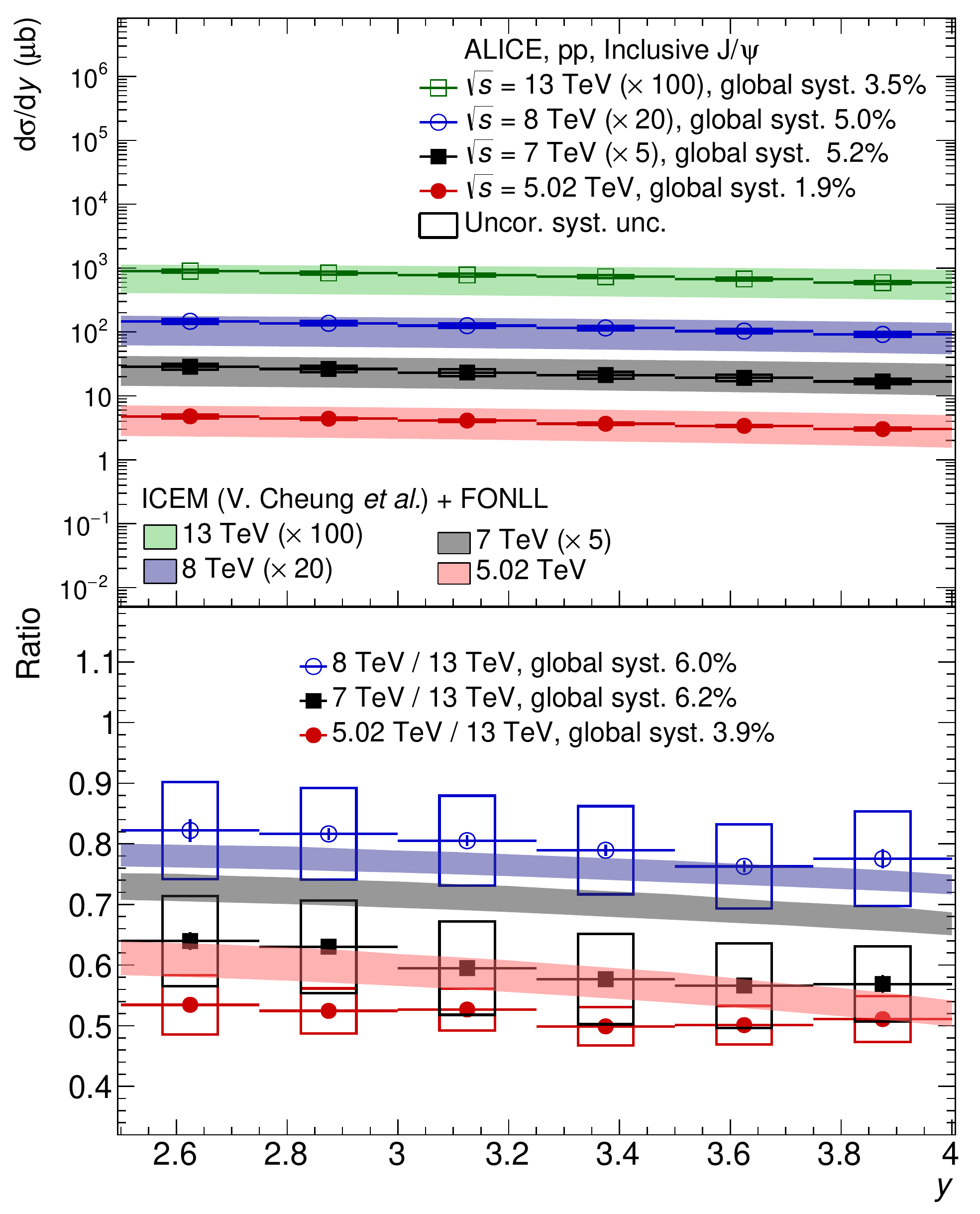}}
\includegraphics[width=0.49\linewidth]{{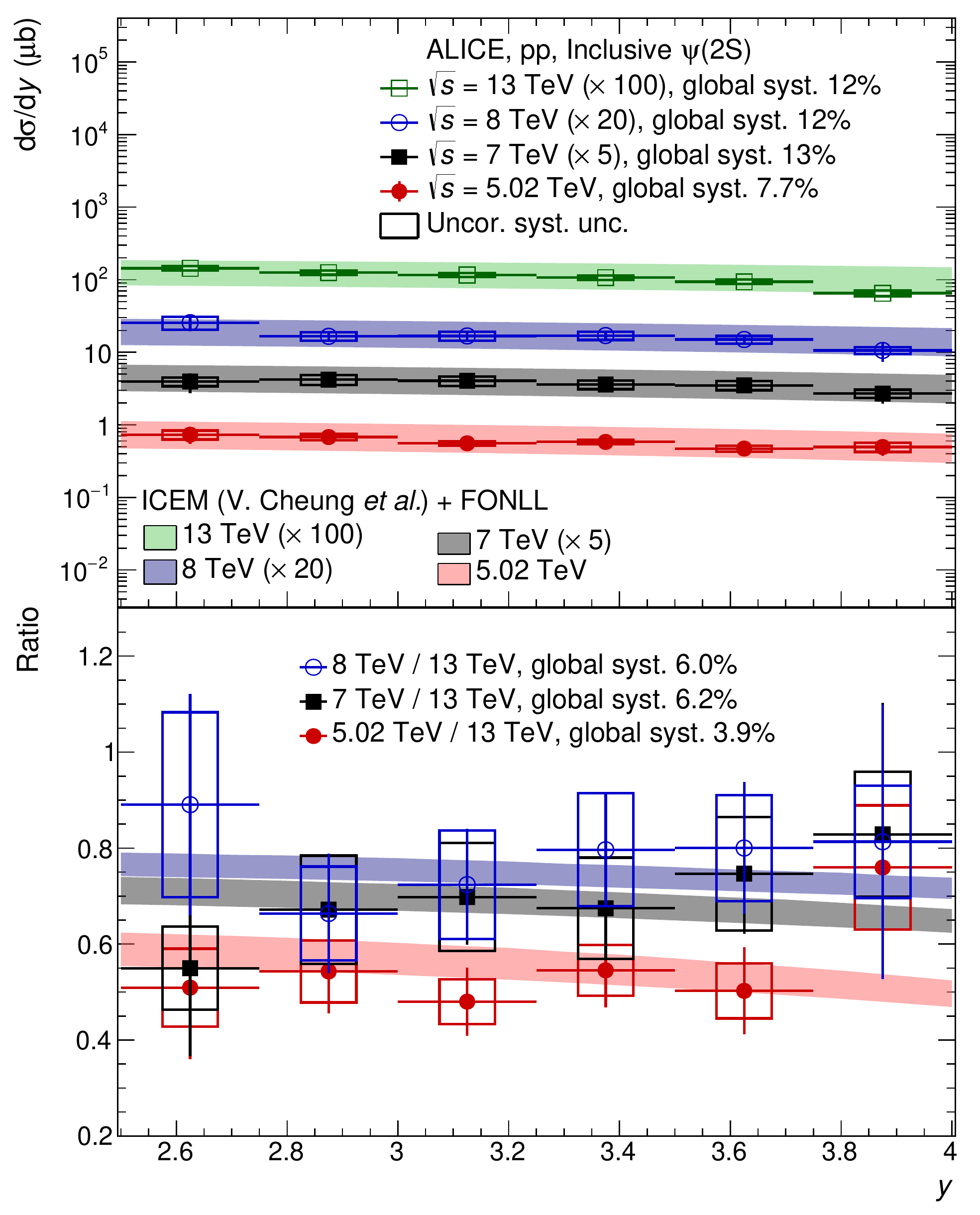}}
\caption{Rapidity dependence of the inclusive \jpsi (left) and \psitwos (right) cross section, at forward \rapidity, measured in pp collisions at $\s = 5.02$, 7~\cite{Abelev:2014qha}, 8~\cite{Adam:2015rta}, and 13~\cite{Acharya:2017hjh} TeV (top panels), and ratio of the measurements at 5.02, 7, and 8 TeV to the 13 TeV data (bottom panels). 
The data are compared with theoretical calculations from ICEM + FONLL~\cite{Cheung:2018tvq,Cacciari:2012ny}.}
\label{fig:jpsienergycomp2}
\end{center}
\end{figure}

\begin{figure}[!b]
\begin{center}
\includegraphics[width=0.49\linewidth]{{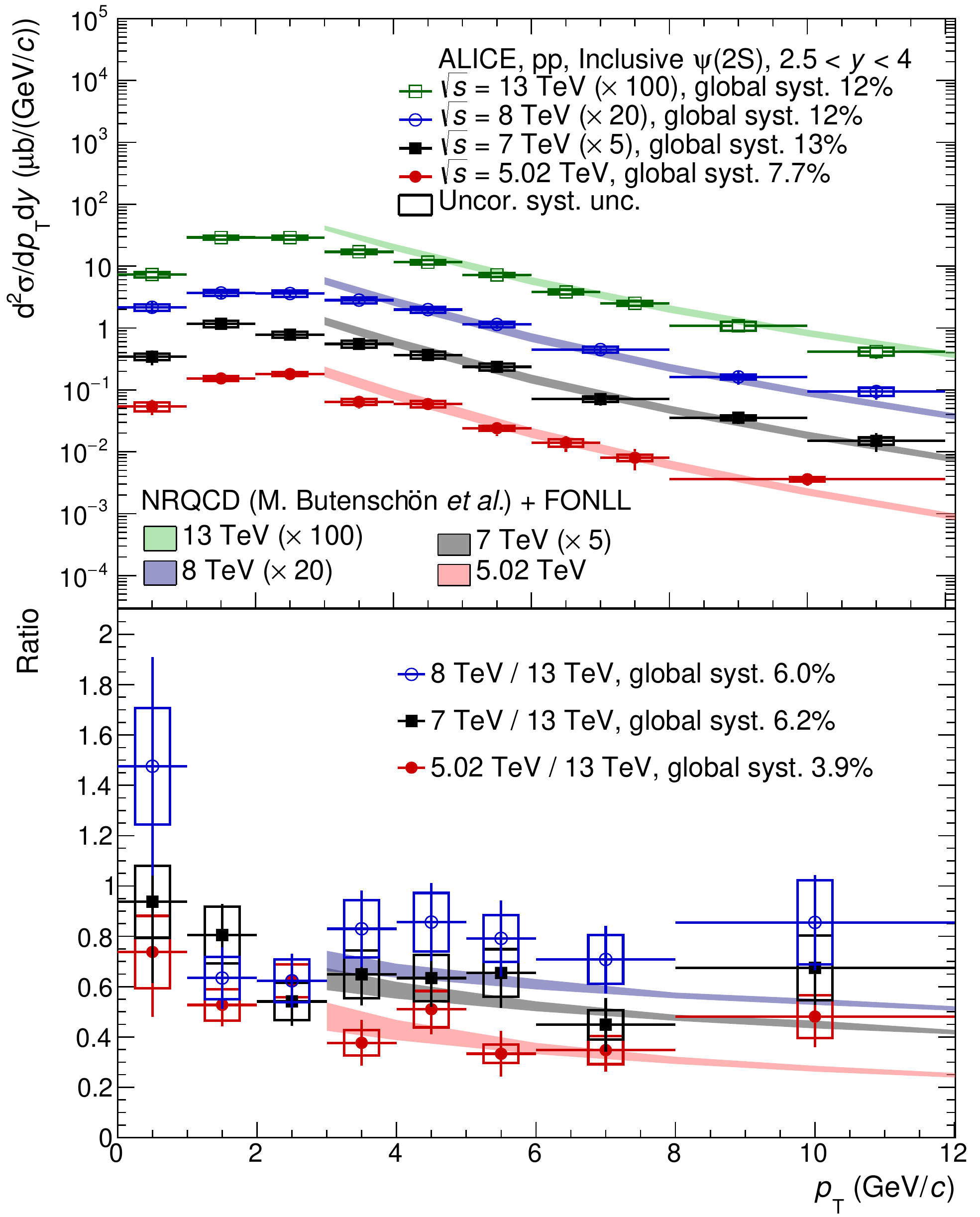}}
\includegraphics[width=0.49\linewidth]{{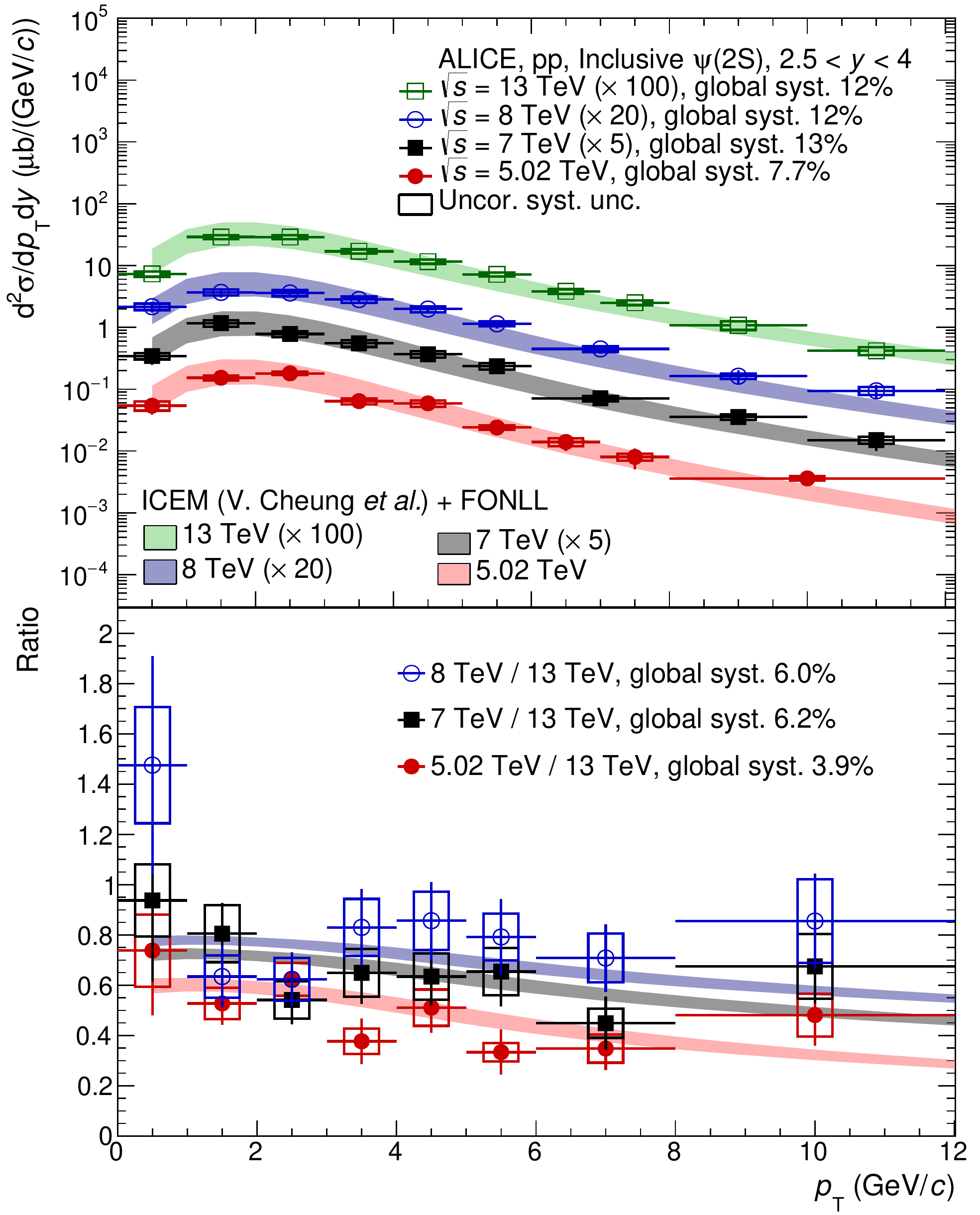}}
\caption{Transverse momentum dependence of the inclusive \psitwos cross section, at forward \rapidity, measured in pp collisions at $\s = 5.02$, 7~\cite{Abelev:2014qha}, 8~\cite{Adam:2015rta}, and 13~\cite{Acharya:2017hjh} TeV (top panels), and ratio of the measurements at 5.02, 7, 8 TeV to the 13 TeV data (bottom panels). 
The data are compared with the NRQCD theoretical calculations from Butenschön~\etal~+ FONLL (left panels)~\cite{Butenschoen:2010rq,Cacciari:2012ny} and with theoretical calculations from ICEM + FONLL (right panels)~\cite{Cheung:2018tvq,Cacciari:2012ny}.}
\label{fig:psitwosenergycomp}
\end{center}
\end{figure}

In Fig.~\ref{fig:jpsienergycomp} and Fig.~\ref{fig:jpsienergycomp2} (left), the \jpsi \pt- and \rapidity-differential cross sections measured at $\s = 5.02$~TeV are compared with previous ALICE measurements at $\s = 7$~\cite{Abelev:2014qha}, 8~\cite{Adam:2015rta}, and 13 TeV~\cite{Acharya:2017hjh}. The ratio of the measurements at 5.02, 7, and 8~TeV to the 13~TeV results are also reported as a function of \pt at the bottom of Fig.~\ref{fig:jpsienergycomp} and as a function of \rapidity at the bottom of the left panel of Fig.~\ref{fig:jpsienergycomp2}. In Fig.~\ref{fig:jpsienergycomp} (and similarly in Fig.~\ref{fig:psitwosenergycomp} for the \psitwos), in order to compute the ratios, the cross sections in some \pt intervals had to be merged. In the merged \pt~intervals, the statistical uncertainty is the quadratic sum of the statistical uncertainties in each \pt~interval, while the systematic uncertainty is the linear sum of the systematic uncertainties in each \pt~interval to conservatively account for possible correlations. In Figs.~\ref{fig:jpsienergycomp},~\ref{fig:jpsienergycomp2} (left), (and similarly in Figs.~\ref{fig:jpsienergycomp2} (right) and~\ref{fig:psitwosenergycomp} for the \psitwos), the global systematic uncertainties quoted as text in the top panel contain the branching ratio and luminosity uncertainty for a given energy, while the global systematic uncertainty quoted as text in the bottom panel contains the combination of the luminosity uncertainties at the two corresponding energies. Both the statistical and systematic uncertainties are assumed to be uncorrelated among different energies when computing the cross section ratios.

Thanks to the large data sample used in this analysis, similar integrated luminosities are now collected at 5.02, 7, and 8~TeV, allowing for a systematic comparison of the \jpsi [\ensuremath{\psi {\rm (2S)}}] differential yields, up to a \pt~of 20~\GeVc [12~GeV/$c$]. The \jpsi \pt- and \rapidity-differential cross section values increase, as expected, with increasing collision energy.  
A stronger hardening of the \pt~spectra is observed in the collisions at 13~TeV with respect to the 5.02, 7, and 8~TeV data, as can be seen in the ratio displayed at the bottom of Fig.~\ref{fig:jpsienergycomp}. This hardening can derive from the increase of the prompt \jpsi~mean \pt~with energy, as well as by the increasing contribution from non-prompt \jpsi~at high \pt. According to FONLL calculations~\cite{Cacciari:2012ny}, the fraction of non-prompt \jpsi~to the inclusive \jpsi~yield, for $\pt>12$~\GeVc, is about 31$\%$ at 5.02~TeV, 37$\%$ at 7 and 8~TeV, and 40$\%$ at 13~TeV. The central values of the 7-to-13 TeV ratio are closer to the 5.02-to-13 than the 8-to-13 TeV ratio at low \pt contrary to the expectation of a smooth increase of the cross section with energy.

The \jpsi \pt-differential cross sections are compared with the NRQCD theoretical calculations from Butenschön~\etal~\cite{Butenschoen:2010rq} (left) and to ICEM calculations~\cite{Cheung:2018tvq} (right). As in Section~\ref{sec:results:charmonium}, a non-prompt contribution from FONLL~\cite{Cacciari:2012ny} is added to all the theoretical calculations for charmonium production. The agreement of both model calculations is rather good for all the energies and covered \pt ranges, although they both tend to slightly overestimate (or are at the upper edge of) the data at $\pt > 12$~\GeVc for ICEM and $\pt > 16$~\GeVc for NRQCD from Butenschön, and this is more pronounced for the ICEM computation. The charmonium \pt- and \rapidity-differential cross section ratios among different energies can provide stronger constraints on the theoretical models. Indeed, similarly as for the \psitwos-to-\jpsi ratio in Section~\ref{sec:psitwostojpsi}, the individual uncertainty sources on prompt charmonium (charm mass, renormalization and factorization scale) and non-prompt charmonium (bottom mass, renormalization and factorization scale, and parton distribution function) are considered as correlated among the considered energies and partially cancel in the ratio calculation. On the bottom panels of Fig.~\ref{fig:jpsienergycomp}, the ratios are compared with theoretical calculations from NRQCD from Butenschön~\etal~\cite{Butenschoen:2010rq} (left) and ICEM~\cite{Cheung:2018tvq} (right) calculations. The NRQCD calculation is able to successfully describe the 5.02-to-13 TeV and 8-to-13 TeV ratios in the whole \pt~range of validity of the model, while it slightly overestimates, or is at the upper edge of data, for the 7-to-13~TeV ratio. The ICEM calculation can only satisfactorily describe the 8-to-13 TeV ratio, while the model calculation is systematically above the 5.02-to-13 and the 7-to-13 TeV data, except in the very-low- and very-high-\pt region.  

In the left panel of Fig.~\ref{fig:jpsienergycomp2}, the \jpsi \rapidity-differential cross section shows a slight decrease with increasing \rapidity at all energies. The ratio of the lower energy data to the 13~TeV data exhibits a flat behaviour within the experimental uncertainties for the three energies. The \rapidity-differential cross sections and cross section ratios between the available energies are also compared to the ICEM model~\cite{Cheung:2018tvq}. The model is able to reproduce the cross sections at all energies, as well as the decreasing trend with increasing \rapidity, but suffers from large theoretical uncertainties. Similarly to what is observed for the \pt-dependent cross section ratios, the ICEM calculation successfully describes the 8-to-13 TeV ratio over the entire \rapidity~range, but overestimates, or is at the edges of the 5.02-to-13 and 7-to-13 TeV cross section ratios. 
The NRQCD model prediction from  Butenschön~\etal~\cite{Butenschoen:2010rq}, being available only for $\pt > 3$~\GeVc, cannot be compared to the \pt-integrated cross section.

The \psitwos \rapidity-differential cross section is presented in the right panel of Fig.~\ref{fig:jpsienergycomp2}. The results at \linebreak $\sqrt{s}~=~13$~TeV, similarly to the \jpsi ones, show a decreasing trend with increasing \rapidity, which is less evident at lower energy because of the larger statistical uncertainties. As shown in Fig.~\ref{fig:jpsienergycomp2} (right), the 5.02-to-13, 7-to-13, and 8-to-13~TeV ratios display no strong \rapidity dependence within the experimental uncertainties. As for the \jpsi, the \rapidity-differential cross section ratios are compared to the ICEM calculation~\cite{Cheung:2018tvq}. The cross sections and their \rapidity~dependence are well reproduced by the model at the various energies. Within the large experimental uncertainties, the ICEM model is able to reproduce consistently the 5.02-to-13, 7-to-13, and 8-to-13~TeV ratios. The inclusive \psitwos \pt-differential cross sections measured at $\s = 5.02$, 7, 8, and 13~TeV are compared in Fig.~\ref{fig:psitwosenergycomp}. The cross section increases with increasing collision energy. 
Contrary to the \jpsi case, the 5.02-to-13, 7-to-13, and 8-to-13~TeV ratios in the bottom panel of Fig.~\ref{fig:psitwosenergycomp} exhibit a flat \pt dependence for $3 \leq \pt < 12$~\GeVc, indicating that no significant hardening of the \pt spectrum is seen, within the data uncertainties, at the highest collision energy with respect to the lower energies. The inclusive \psitwos \pt-differential cross section and cross section ratios among energies are also compared with the NRQCD calculation from Butenschön~\etal~\cite{Butenschoen:2010rq} (left panel of Fig~\ref{fig:psitwosenergycomp}) and with the ICEM model~\cite{Cheung:2018tvq} (right panel of Fig.~\ref{fig:psitwosenergycomp}). Both models are able to satisfactorily describe the \psitwos \pt-differential cross section measurements at all the displayed energies. One can however remark that the NRQCD calculation overestimates systematically the cross sections for 3 $\leq \pt < 4$~\GeVc, and that both the NRQCD and ICEM models are at the lower edges of the measurements for $\pt\geq 8$~\GeVc and for $\s = 5.02$, 7, and 8~TeV. Concerning the cross section ratios, the NRQCD model reproduces the 5.02-to-13 and 7-to-13~TeV data for $3 \leq \pt < 8$~\GeVc, and underestimates them for $\pt \geq 8$~\GeVc and in almost the whole \pt range for the 8-to-13~TeV ratio. Similarly, the ICEM calculation describes successfully the trend versus \pt~of the 5.02-to-13 and 7-to-13~TeV ratios for $\pt < 8$~\GeVc, and additionally it provides a reasonable description of the 8-to-13~TeV ratio for $2 \leq \pt < 8$~\GeVc, given the current experimental uncertainties. Both the NRQCD calculation and ICEM model suggest a weak hardening of the \psitwos \pt~spectrum with the collision energy, which is not observed in data, possibly due to large experimental uncertainties. 

%\clearpage

%\clearpage

\begin{figure}[!htb]
\includegraphics[width=0.49\linewidth]{{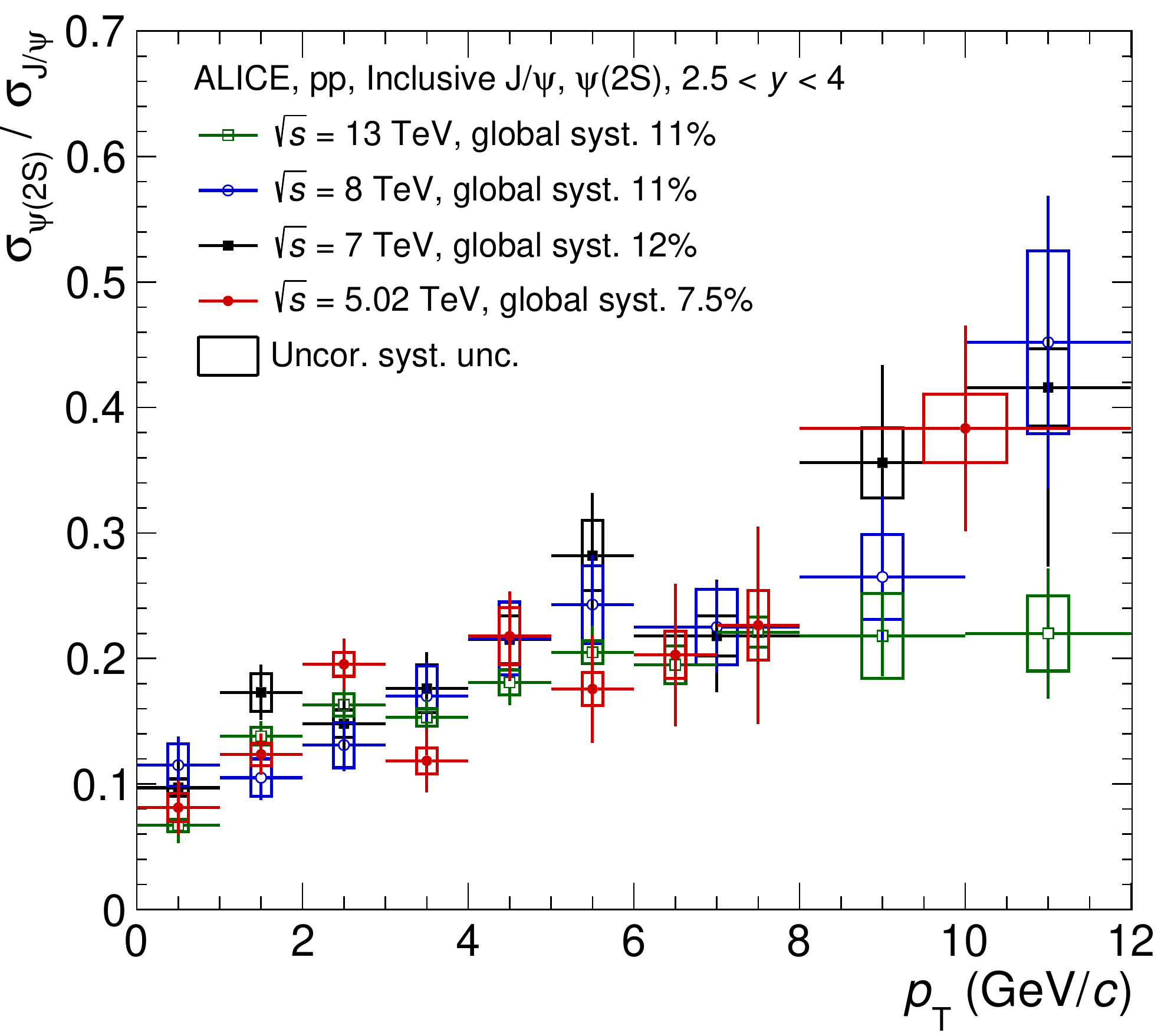}}
\includegraphics[width=0.49\linewidth]{{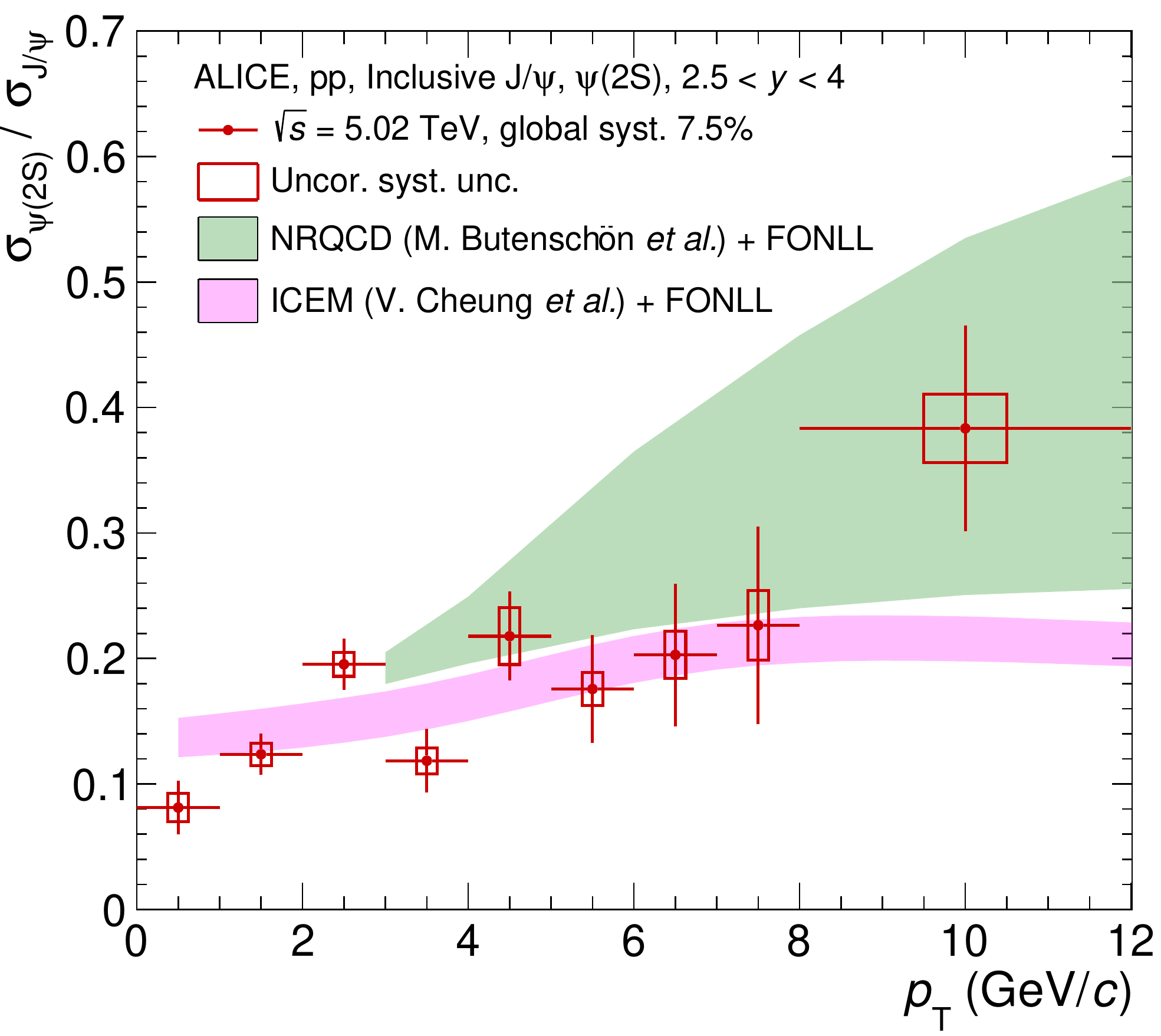}}
\caption{Inclusive \psitwos-to-\jpsi cross section ratio as a function of \pt, at forward \rapidity, in pp collisions at $\s~=~5.02$, 7~\cite{Abelev:2014qha}, 8~\cite{Adam:2015rta}, and 13~\cite{Acharya:2017hjh} TeV (left panel). The data at $\s = 5.02$ TeV are compared with NRQCD theoretical calculations from Butenschön~\etal~+ FONLL~\cite{Butenschoen:2010rq,Cacciari:2012ny} and with theoretical calculations from ICEM + FONLL~\cite{Cheung:2018tvq,Cacciari:2012ny} (right panel).}
\label{fig:psitwosoverjpsienergycomp}
\end{figure}

\begin{figure}[!htb]
\includegraphics[width=0.49\linewidth]{{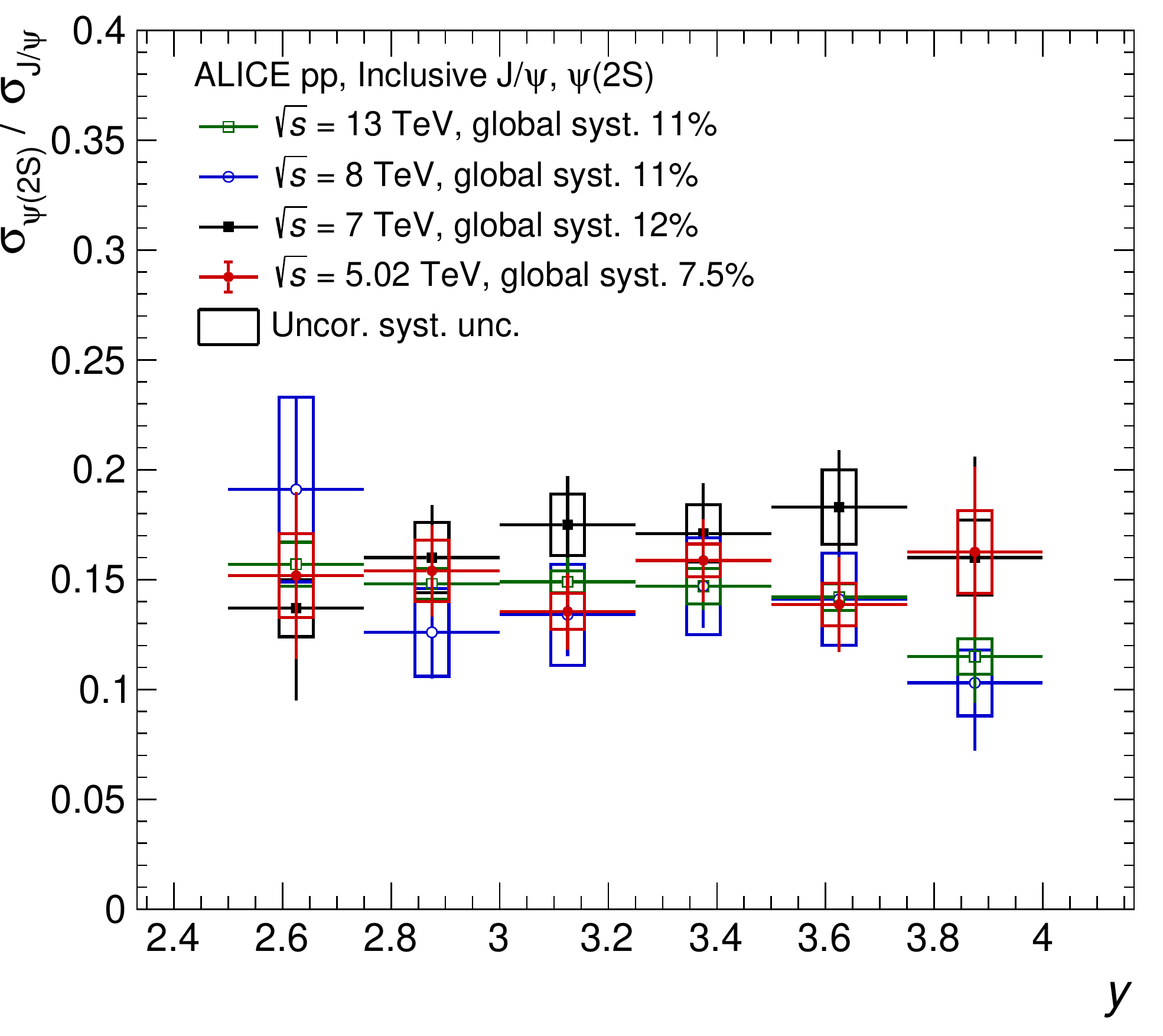}}
\includegraphics[width=0.49\linewidth]{{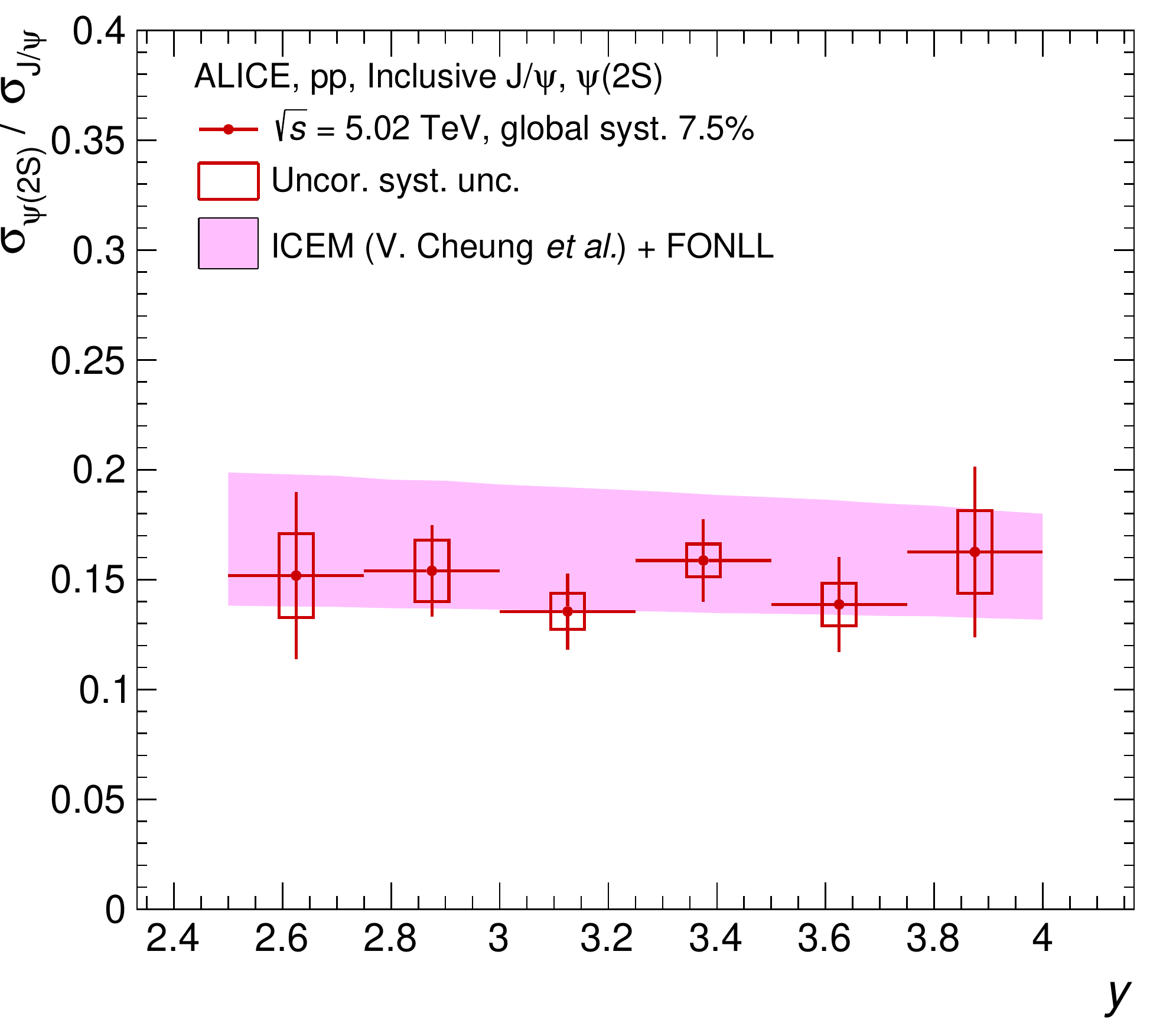}}

\caption{Inclusive \psitwos-to-\jpsi cross section ratio as a function of \rapidity in pp collisions at $\s = 5.02$, 7~\cite{Abelev:2014qha}, 8~\cite{Adam:2015rta}, and 13~\cite{Acharya:2017hjh} TeV (left panel).  
The data at $\s = 5.02$ TeV are compared with theoretical calculations from ICEM + FONLL~\cite{Cheung:2018tvq,Cacciari:2012ny} (right panel).}
\label{fig:psitwosoverjpsienergycomp2}
\end{figure}

\begin{figure}[!t]
\begin{center}
\includegraphics[width=0.49\linewidth]{{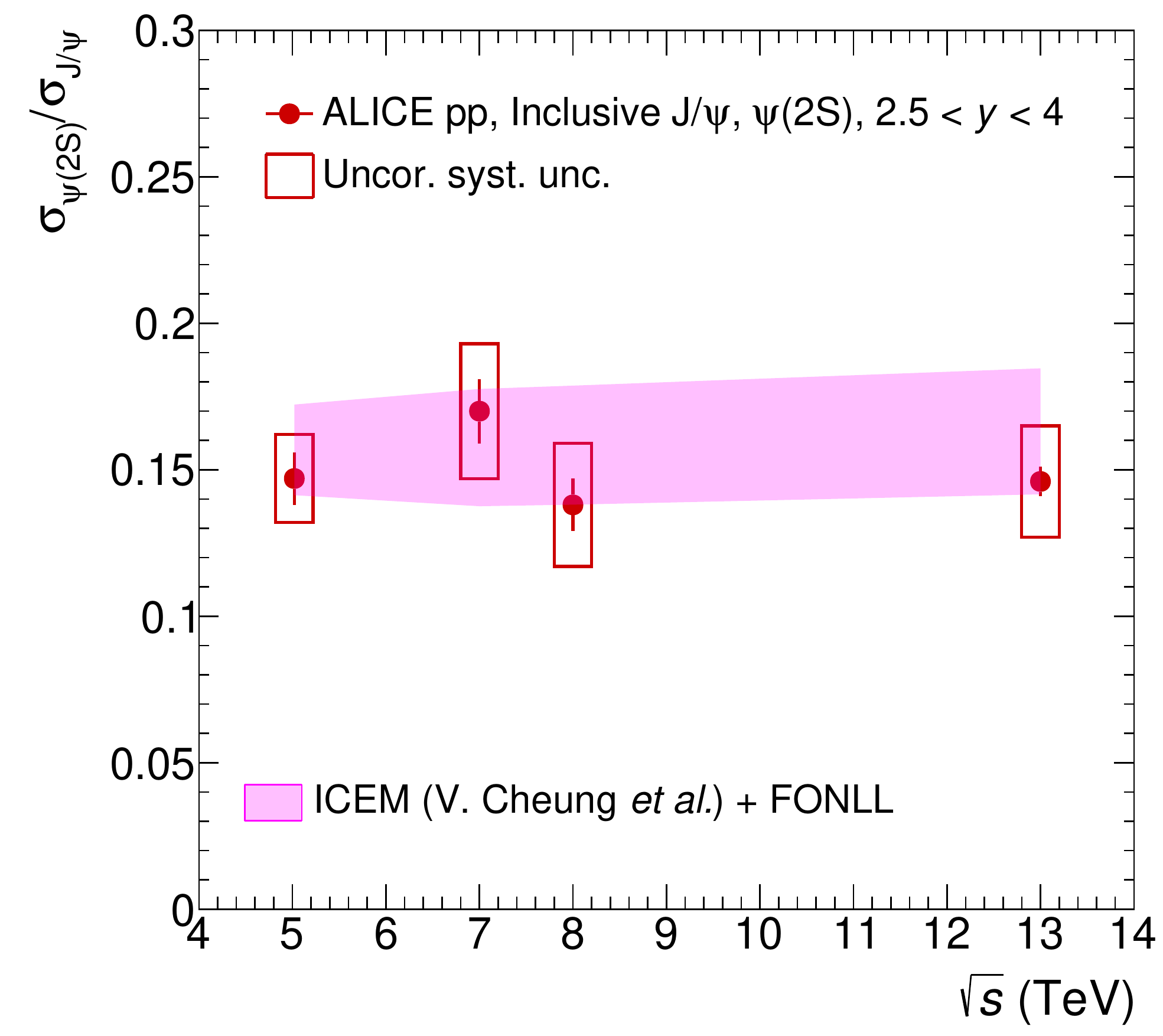}}
\includegraphics[width=0.49\linewidth]{{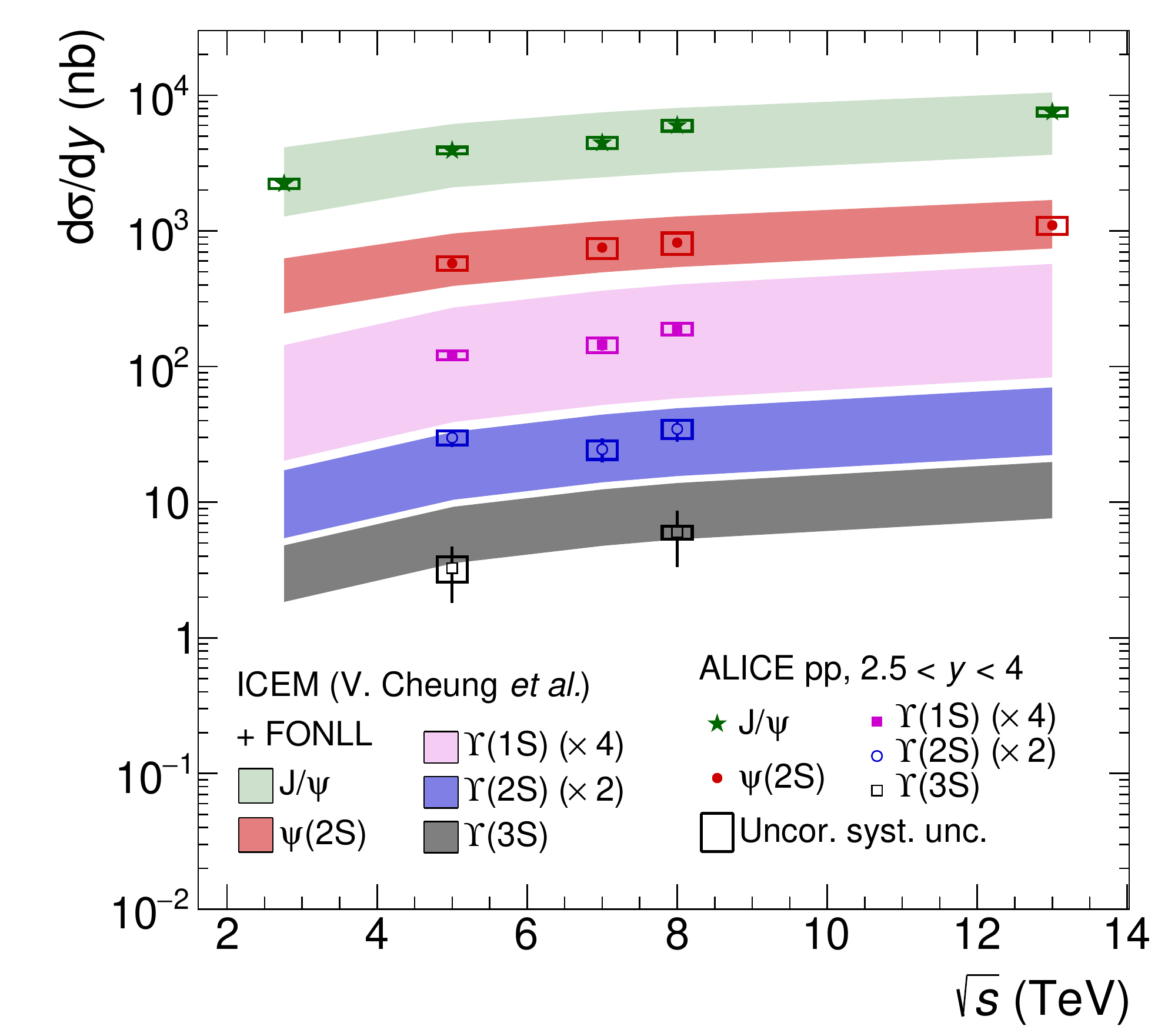}}
\caption{Inclusive \psitwos-to-\jpsi cross section ratio (left) and \jpsi, \psitwos, \upsones, \upstwos, and \upsthrees \pt-integrated cross section per unit of rapidity (right) as a function of the collision energy in pp collisions~\cite{Abelev:2012kr,Abelev:2014qha,Adam:2015rta,Acharya:2017hjh}. In the left panel, the systematic boxes include the BR uncertainties from both resonances, on top of the MC input and signal extraction systematic uncertainties. The 13 TeV data point is computed from the published individual \jpsi and \psitwos \pt-integrated cross sections. The statistical and systematic uncertainties are assumed to be uncorrelated between the resonances when computing the ratio. In the right panel, the luminosity and branching ratio uncertainties are included in the systematic boxes. The data are compared with theoretical calculations from ICEM + FONLL~\cite{Cheung:2018tvq,Cacciari:2012ny}.}
\label{fig:quarkoniacomp}
\end{center}
\end{figure}

The \psitwos-to-\jpsi cross section ratio is displayed as a function of \pt, \rapidity, and integrated over \pt and y for 2.5 $< \rapidity <$ 4 for the different \pp colliding energies in Figs.~\ref{fig:psitwosoverjpsienergycomp} (left),~\ref{fig:psitwosoverjpsienergycomp2} (left), and~\ref{fig:quarkoniacomp} (left), respectively. The \pt-differential \psitwos-to-\jpsi ratio increases with increasing \pt and does not exhibit any energy dependence within the current uncertainties. Similarly, no significant change in shape nor in magnitude is observed in the \rapidity-dependent \psitwos-to-\jpsi cross section ratio, which follows a flat trend with \rapidity. The \rapidity~and \pt-integrated \psitwos-to-\jpsi ratio for 2.5 $< \rapidity <$ 4 is also compatible with no energy dependence within the measurement uncertainties. The \psitwos-to-\jpsi ratio as a function of \pt is also compared to the NRQCD from Butenschön~\etal~\cite{Butenschoen:2010rq} and ICEM~\cite{Cheung:2018tvq} models in the right panel of Fig.~\ref{fig:psitwosoverjpsienergycomp} for $\s$~=~5.02~TeV and in Fig.~\ref{fig:psitwosoverjpsienergycomp_appendix} of the appendix for $\s$ = 7, 8 and 13 TeV. In both models the \psitwos-to-\jpsi ratio does not exhibit a strong energy dependence, as in data. The NRQCD model describes within uncertainties the \psitwos-to-\jpsi ratio at 5.02, 7, and 8 TeV for $\pt \geq 3$~\GeVc, but it tends to overestimate it at 13 TeV, where the uncertainties are smaller. The ICEM calculation qualitatively describes the \pt~dependence of the~\psitwos-to-\jpsi ratio at the four energies for $\pt < 8$~\GeVc, and suggests a flat behaviour for \pt $\geq$ 8~\GeVc in agreement with the 13 TeV data which are the most precise ones. In Fig.~\ref{fig:psitwosoverjpsienergycomp2} right, the \rapidity-differential \psitwos-to-\jpsi cross section ratio at $\s$~=~5.02~TeV is compared with the ICEM calculation. Similar data to theory comparison can be found in Fig.~\ref{fig:psitwosoverjpsienergycomp2_appendix} of the appendix for pp collisions at $\s$ = 7, 8 and 13 TeV. The model predicts a flat \rapidity~dependence and properly describes the data at the four energies within the experimental and theoretical uncertainties.

%\clearpage

The energy dependence of the \psitwos-to-\jpsi ratio integrated in \pt and \rapidity for 2.5 $< \rapidity < 4$ is also compared with the ICEM model in Fig.~\ref{fig:quarkoniacomp} (left). The charmonium cross section ratio does not exhibit a significant energy dependence and is well reproduced by the ICEM model. Finally, the cross section per unit of rapidity for 2.5 $< \rapidity <$ 4 and integrated over \pt is displayed as a function of the collision energy in the right panel of Fig.~\ref{fig:quarkoniacomp}, for all available ALICE quarkonium measurements. A steady increase of the cross section is observed with increasing \s for all the states. ALICE data are compared with theoretical calculations from ICEM~\cite{Cheung:2018tvq}. The model, within its large uncertainties, is able to consistently reproduce the energy dependence of the cross section for all the quarkonium states. However, the \upsthrees results lie on the lower edge of the theoretical calculation band.

%\clearpage

\section{Conclusion}

The inclusive production cross sections of \jpsi, \psitwos, \upsones, \upstwos, and \upsthrees have been measured with the ALICE detector at forward rapidity ($2.5 < \rapidity < 4$) in \pp collisions at \s~=~5.02~TeV. The \jpsi and \psitwos results are in agreement with earlier measurements at the same energy. Thanks to the larger integrated luminosity by a factor 12 of these new measurements, a \pt reach up to 20~\GeVc has been achieved for the \jpsi, and the double-differential cross section as a function of \pt and \rapidity could also be extracted. The \psitwos and \upsones production cross section and the \psitwos-to-\jpsi cross section ratio have been measured for the first time as a function of \pt and \rapidity at forward rapidity, as well as the \pt-integrated \upsones, \upstwos, and \upsthrees cross sections. 
The collision energy dependence has been discussed for the five quarkonium states and the ratios of the cross sections at \s~=~5.02, 7, and 8~TeV to the one obtained at \s~=~13~TeV have been presented as a function of \pt and \rapidity. 
Calculations based on CEM or NRQCD describe well the charmonium and bottomonium cross sections at all collision energies, as well as the \psitwos-to-\jpsi ratio, in the kinematic range they cover. The charmonium cross sections and their ratios relative to the values at $\sqrt{s} = 13$~TeV can be described by a NRQCD model within uncertainties.
These combined measurements provide additional experimental constraints to quarkonium production models. This is particularly evident for the determination of the cross section calculations, where a reduction in the size of the theory should now be pursued in order to match the experimental precision.
Moreover, the \s~=~5.02~TeV \pp measurements represent a more accurate reference for the measurement of the quarkonium nuclear modification factor in \PbPb collisions collected during the LHC Run 2 at the same nucleon--nucleon center-of-mass energy.

%%%%%%%%%%%%%%%%%%%%%%%%%%%%%%%%
% end main text 
%%%%%%%%%%%%%%%%%%%%%%%%%%%%%%%%

%\clearpage
%%%%% acknowledgements - handled by EB chairs 
\newenvironment{acknowledgement}{\relax}{\relax}
\begin{acknowledgement}
\section*{Acknowledgements}
% add specific acknowledgements here 
% ...but please don't remove the line below: funding agencies
% will be acknowledged with a custom tex file handled by EB chairs after Collab Round 2
% Version: 2021-09-21

The ALICE Collaboration would like to thank all its engineers and technicians for their invaluable contributions to the construction of the experiment and the CERN accelerator teams for the outstanding performance of the LHC complex.
The ALICE Collaboration gratefully acknowledges the resources and support provided by all Grid centres and the Worldwide LHC Computing Grid (WLCG) collaboration.
The ALICE Collaboration acknowledges the following funding agencies for their support in building and running the ALICE detector:
A. I. Alikhanyan National Science Laboratory (Yerevan Physics Institute) Foundation (ANSL), State Committee of Science and World Federation of Scientists (WFS), Armenia;
Austrian Academy of Sciences, Austrian Science Fund (FWF): [M 2467-N36] and Nationalstiftung f\"{u}r Forschung, Technologie und Entwicklung, Austria;
Ministry of Communications and High Technologies, National Nuclear Research Center, Azerbaijan;
Conselho Nacional de Desenvolvimento Cient\'{\i}fico e Tecnol\'{o}gico (CNPq), Financiadora de Estudos e Projetos (Finep), Funda\c{c}\~{a}o de Amparo \`{a} Pesquisa do Estado de S\~{a}o Paulo (FAPESP) and Universidade Federal do Rio Grande do Sul (UFRGS), Brazil;
Ministry of Education of China (MOEC) , Ministry of Science \& Technology of China (MSTC) and National Natural Science Foundation of China (NSFC), China;
Ministry of Science and Education and Croatian Science Foundation, Croatia;
Centro de Aplicaciones Tecnol\'{o}gicas y Desarrollo Nuclear (CEADEN), Cubaenerg\'{\i}a, Cuba;
Ministry of Education, Youth and Sports of the Czech Republic, Czech Republic;
The Danish Council for Independent Research | Natural Sciences, the VILLUM FONDEN and Danish National Research Foundation (DNRF), Denmark;
Helsinki Institute of Physics (HIP), Finland;
Commissariat \`{a} l'Energie Atomique (CEA) and Institut National de Physique Nucl\'{e}aire et de Physique des Particules (IN2P3) and Centre National de la Recherche Scientifique (CNRS), France;
Bundesministerium f\"{u}r Bildung und Forschung (BMBF) and GSI Helmholtzzentrum f\"{u}r Schwerionenforschung GmbH, Germany;
General Secretariat for Research and Technology, Ministry of Education, Research and Religions, Greece;
National Research, Development and Innovation Office, Hungary;
Department of Atomic Energy Government of India (DAE), Department of Science and Technology, Government of India (DST), University Grants Commission, Government of India (UGC) and Council of Scientific and Industrial Research (CSIR), India;
Indonesian Institute of Science, Indonesia;
Istituto Nazionale di Fisica Nucleare (INFN), Italy;
Japanese Ministry of Education, Culture, Sports, Science and Technology (MEXT), Japan Society for the Promotion of Science (JSPS) KAKENHI and Japanese Ministry of Education, Culture, Sports, Science and Technology (MEXT)of Applied Science (IIST), Japan;
Consejo Nacional de Ciencia (CONACYT) y Tecnolog\'{i}a, through Fondo de Cooperaci\'{o}n Internacional en Ciencia y Tecnolog\'{i}a (FONCICYT) and Direcci\'{o}n General de Asuntos del Personal Academico (DGAPA), Mexico;
Nederlandse Organisatie voor Wetenschappelijk Onderzoek (NWO), Netherlands;
The Research Council of Norway, Norway;
Commission on Science and Technology for Sustainable Development in the South (COMSATS), Pakistan;
Pontificia Universidad Cat\'{o}lica del Per\'{u}, Peru;
Ministry of Education and Science, National Science Centre and WUT ID-UB, Poland;
Korea Institute of Science and Technology Information and National Research Foundation of Korea (NRF), Republic of Korea;
Ministry of Education and Scientific Research, Institute of Atomic Physics and Ministry of Research and Innovation and Institute of Atomic Physics, Romania;
Joint Institute for Nuclear Research (JINR), Ministry of Education and Science of the Russian Federation, National Research Centre Kurchatov Institute, Russian Science Foundation and Russian Foundation for Basic Research, Russia;
Ministry of Education, Science, Research and Sport of the Slovak Republic, Slovakia;
National Research Foundation of South Africa, South Africa;
Swedish Research Council (VR) and Knut \& Alice Wallenberg Foundation (KAW), Sweden;
European Organization for Nuclear Research, Switzerland;
Suranaree University of Technology (SUT), National Science and Technology Development Agency (NSDTA) and Office of the Higher Education Commission under NRU project of Thailand, Thailand;
Turkish Energy, Nuclear and Mineral Research Agency (TENMAK), Turkey;
National Academy of  Sciences of Ukraine, Ukraine;
Science and Technology Facilities Council (STFC), United Kingdom;
National Science Foundation of the United States of America (NSF) and United States Department of Energy, Office of Nuclear Physics (DOE NP), United States of America.
\end{acknowledgement}

%%%%%%%% Bibliography 
\bibliographystyle{utphys}   % Remember we use title in the biblio
\bibliography{bibliography}

\providecommand{\href}[2]{#2}\begingroup\raggedright\begin{thebibliography}{10}

\bibitem{Brambilla:2010cs}
N.~Brambilla {\em et~al.}, ``{Heavy Quarkonium: Progress, Puzzles, and
  Opportunities}'',
  \href{http://dx.doi.org/10.1140/epjc/s10052-010-1534-9}{{\em Eur. Phys. J.}
  {\bfseries C71} (2011) 1534},
\href{http://arxiv.org/abs/1010.5827}{{\ttfamily arXiv:1010.5827 [hep-ph]}}.
%%CITATION = ARXIV:1010.5827;%%.

\bibitem{Andronic:2015wma}
A.~Andronic {\em et~al.}, ``{Heavy-flavour and quarkonium production in the LHC
  era: from proton-proton to heavy-ion collisions}'',
  \href{http://dx.doi.org/10.1140/epjc/s10052-015-3819-5}{{\em Eur. Phys. J.}
  {\bfseries C76} (2016) 107},
\href{http://arxiv.org/abs/1506.03981}{{\ttfamily arXiv:1506.03981 [nucl-ex]}}.
%%CITATION = ARXIV:1506.03981;%%.

\bibitem{Rothkopf:2019ipj}
A.~Rothkopf, ``{Heavy Quarkonium in Extreme Conditions}'',
  \href{http://dx.doi.org/10.1016/j.physrep.2020.02.006}{{\em Phys. Rept.}
  {\bfseries 858} (2020) 1--117},
  \href{http://arxiv.org/abs/1912.02253}{{\ttfamily arXiv:1912.02253
  [hep-ph]}}.

\bibitem{Fritzsch:1977ay}
H.~Fritzsch, ``{Producing Heavy Quark Flavors in Hadronic Collisions: A Test of
  Quantum Chromodynamics}'',
  \href{http://dx.doi.org/10.1016/0370-2693(77)90108-3}{{\em Phys. Lett. B}
  {\bfseries 67} (1977) 217--221}.

\bibitem{Amundson:1996qr}
J.~Amundson, O.~J. Eboli, E.~Gregores, and F.~Halzen, ``{Quantitative tests of
  color evaporation: Charmonium production}'',
  \href{http://dx.doi.org/10.1016/S0370-2693(96)01417-7}{{\em Phys. Lett. B}
  {\bfseries 390} (1997) 323--328},
  \href{http://arxiv.org/abs/hep-ph/9605295}{{\ttfamily arXiv:hep-ph/9605295}}.

\bibitem{Baier:1981uk}
R.~Baier and R.~Ruckl, ``{Hadronic Production of J/psi and Upsilon: Transverse
  Momentum Distributions}'',
  \href{http://dx.doi.org/10.1016/0370-2693(81)90636-5}{{\em Phys. Lett. B}
  {\bfseries 102} (1981) 364--370}.

\bibitem{Bodwin:1994jh}
G.~T. Bodwin, E.~Braaten, and G.~Lepage, ``{Rigorous QCD analysis of inclusive
  annihilation and production of heavy quarkonium}'',
  \href{http://dx.doi.org/10.1103/PhysRevD.55.5853}{{\em Phys. Rev. D}
  {\bfseries 51} (1995) 1125--1171},
  \href{http://arxiv.org/abs/hep-ph/9407339}{{\ttfamily arXiv:hep-ph/9407339}}.
  [Erratum: Phys.Rev.D 55, 5853 (1997)].

\bibitem{Adam:2016rdg}
{\bfseries ALICE} Collaboration, J.~Adam {\em et~al.}, ``{J/$\psi$ suppression
  at forward rapidity in Pb-Pb collisions at $\sqrt{s_{\rm NN}} = 5.02$ TeV}'',
  \href{http://dx.doi.org/10.1016/j.physletb.2016.12.064}{{\em Phys. Lett.}
  {\bfseries B766} (2017) 212--224},
\href{http://arxiv.org/abs/1606.08197}{{\ttfamily arXiv:1606.08197 [nucl-ex]}}.
%%CITATION = ARXIV:1606.08197;%%.

\bibitem{Acharya:2017hjh}
{\bfseries ALICE} Collaboration, S.~Acharya {\em et~al.}, ``{Energy dependence
  of forward-rapidity $\mathrm {J}/\psi $ and $\psi \mathrm {(2S)}$ production
  in pp collisions at the LHC}'',
  \href{http://dx.doi.org/10.1140/epjc/s10052-017-4940-4}{{\em Eur. Phys. J.}
  {\bfseries C77} (2017) 392},
\href{http://arxiv.org/abs/1702.00557}{{\ttfamily arXiv:1702.00557 [hep-ex]}}.
%%CITATION = ARXIV:1702.00557;%%.

\bibitem{LHCb:2021pyk}
{\bfseries LHCb} Collaboration, R.~Aaij {\em et~al.}, ``{Measurement of
  $J/\psi$ production cross-sections in $pp$ collisions at $\sqrt{s}=5$ TeV}'',
  \href{http://arxiv.org/abs/2109.00220}{{\ttfamily arXiv:2109.00220
  [hep-ex]}}.

\bibitem{Abelev:2012kr}
{\bfseries ALICE} Collaboration, B.~Abelev {\em et~al.}, ``{Inclusive $J/\psi$
  production in pp collisions at $\sqrt{s} = 2.76$ TeV}'',
  \href{http://dx.doi.org/10.1016/j.physletb.2012.10.078}{{\em Phys. Lett. B}
  {\bfseries 718} (2012) 295--306},
  \href{http://arxiv.org/abs/1203.3641}{{\ttfamily arXiv:1203.3641 [hep-ex]}}.
  [Erratum: Phys. Lett. B748 (2015) 472].

\bibitem{Abelev:2014qha}
{\bfseries ALICE} Collaboration, B.~Abelev {\em et~al.}, ``{Measurement of
  quarkonium production at forward rapidity in pp collisions at $\sqrt{s} = 7$
  TeV}'', \href{http://dx.doi.org/10.1140/epjc/s10052-014-2974-4}{{\em Eur.
  Phys. J.} {\bfseries C74} (2014) 2974},
\href{http://arxiv.org/abs/1403.3648}{{\ttfamily arXiv:1403.3648 [nucl-ex]}}.
%%CITATION = ARXIV:1403.3648;%%.

\bibitem{Adam:2015rta}
{\bfseries ALICE} Collaboration, J.~Adam {\em et~al.}, ``{Inclusive quarkonium
  production at forward rapidity in pp collisions at $\sqrt{s}=8$ TeV}'',
  \href{http://dx.doi.org/10.1140/epjc/s10052-016-3987-y}{{\em Eur. Phys. J.}
  {\bfseries C76} (2016) 184},
\href{http://arxiv.org/abs/1509.08258}{{\ttfamily arXiv:1509.08258 [hep-ex]}}.
%%CITATION = ARXIV:1509.08258;%%.

\bibitem{LHCb:2012aa}
{\bfseries LHCb} Collaboration, R.~Aaij {\em et~al.}, ``{Measurement of Upsilon
  production in pp collisions at $\sqrt{s}$ = 7 TeV}'',
  \href{http://dx.doi.org/10.1140/epjc/s10052-012-2025-y}{{\em Eur. Phys. J. C}
  {\bfseries 72} (2012) 2025}, \href{http://arxiv.org/abs/1202.6579}{{\ttfamily
  arXiv:1202.6579 [hep-ex]}}.

\bibitem{LHCb:2013itw}
{\bfseries LHCb} Collaboration, R.~Aaij {\em et~al.}, ``{Production of J/psi
  and Upsilon mesons in pp collisions at sqrt(s) = 8 TeV}'',
  \href{http://dx.doi.org/10.1007/JHEP06(2013)064}{{\em JHEP} {\bfseries 06}
  (2013) 064}, \href{http://arxiv.org/abs/1304.6977}{{\ttfamily arXiv:1304.6977
  [hep-ex]}}.

\bibitem{LHCb:2015log}
{\bfseries LHCb} Collaboration, R.~Aaij {\em et~al.}, ``{Forward production of
  $\Upsilon$ mesons in $pp$ collisions at $\sqrt{s}=7$ and 8TeV}'',
  \href{http://dx.doi.org/10.1007/JHEP11(2015)103}{{\em JHEP} {\bfseries 11}
  (2015) 103}, \href{http://arxiv.org/abs/1509.02372}{{\ttfamily
  arXiv:1509.02372 [hep-ex]}}.

\bibitem{LHCb:2015foc}
{\bfseries LHCb} Collaboration, R.~Aaij {\em et~al.}, ``{Measurement of forward
  $J/\psi$ production cross-sections in $pp$ collisions at $\sqrt{s}=13$
  TeV}'', \href{http://dx.doi.org/10.1007/JHEP10(2015)172}{{\em JHEP}
  {\bfseries 10} (2015) 172}, \href{http://arxiv.org/abs/1509.00771}{{\ttfamily
  arXiv:1509.00771 [hep-ex]}}. [Erratum: JHEP 05, 063 (2017)].

\bibitem{Aamodt:2008zz}
{\bfseries ALICE} Collaboration, K.~Aamodt {\em et~al.}, ``{The ALICE
  experiment at the CERN LHC}'',
\href{http://dx.doi.org/10.1088/1748-0221/3/08/S08002}{{\em JINST} {\bfseries
  3} (2008) S08002}.
%%CITATION = JINST,3,S08002;%%.

\bibitem{Abelev:2014ffa}
{\bfseries ALICE} Collaboration, B.~Abelev {\em et~al.}, ``{Performance of the
  ALICE Experiment at the CERN LHC}'',
  \href{http://dx.doi.org/10.1142/S0217751X14300440}{{\em Int. J. Mod. Phys.}
  {\bfseries A29} (2014) 1430044},
\href{http://arxiv.org/abs/1402.4476}{{\ttfamily arXiv:1402.4476 [nucl-ex]}}.
%%CITATION = ARXIV:1402.4476;%%.

\bibitem{ALICE:1999aa}
{\bfseries ALICE} Collaboration, ``{ALICE dimuon forward spectrometer:
  Technical Design Report}'', CERN-LHCC-99-022.
  \url{http://cds.cern.ch/record/401974}.

\bibitem{Aamodt:2010aa}
{\bfseries ALICE} Collaboration, K.~Aamodt {\em et~al.}, ``{Alignment of the
  ALICE Inner Tracking System with cosmic-ray tracks}'',
  \href{http://dx.doi.org/10.1088/1748-0221/5/03/P03003}{{\em JINST} {\bfseries
  5} (2010) P03003},
\href{http://arxiv.org/abs/1001.0502}{{\ttfamily arXiv:1001.0502
  [physics.ins-det]}}.
%%CITATION = ARXIV:1001.0502;%%.

\bibitem{Bondila:2005xy}
M.~Bondila {\em et~al.}, ``{ALICE T0 detector}'',
  \href{http://dx.doi.org/10.1109/TNS.2005.856900}{{\em IEEE Trans.\ Nucl.\
  Sci.} {\bfseries 52} (2005) 1705--1711}.

\bibitem{Abbas:2013taa}
{\bfseries ALICE} Collaboration, E.~Abbas {\em et~al.}, ``{Performance of the
  ALICE VZERO system}'',
  \href{http://dx.doi.org/10.1088/1748-0221/8/10/P10016}{{\em JINST} {\bfseries
  8} (2013) P10016},
\href{http://arxiv.org/abs/1306.3130}{{\ttfamily arXiv:1306.3130 [nucl-ex]}}.
%%CITATION = ARXIV:1306.3130;%%.

\bibitem{lumi}
{\bfseries ALICE} Collaboration, ``{ALICE 2017 luminosity determination for pp
  collisions at \s = 5 TeV}'', ALICE-PUBLIC-2018-014.
  \url{http://cds.cern.ch/record/2648933}.

\bibitem{vanderMeer:1968zz}
S.~van~der Meer, ``{Calibration of the Effective Beam Height in the ISR}'',
  CERN-ISR-PO-68-31. \url{https://cds.cern.ch/record/296752}.

\bibitem{Adam:2015jsa}
{\bfseries ALICE} Collaboration, J.~Adam {\em et~al.}, ``{Centrality dependence
  of inclusive J/$\psi$ production in p-Pb collisions at $
  \sqrt{s_{\mathrm{NN}}}=5.02 $ TeV}'',
  \href{http://dx.doi.org/10.1007/JHEP11(2015)127}{{\em JHEP} {\bfseries 11}
  (2015) 127}, \href{http://arxiv.org/abs/1506.08808}{{\ttfamily
  arXiv:1506.08808 [nucl-ex]}}.

\bibitem{Abelev:2012pi}
{\bfseries ALICE} Collaboration, B.~Abelev {\em et~al.}, ``{Heavy flavour decay
  muon production at forward rapidity in proton\textendash{}proton collisions
  at $\sqrt{s} =$ 7 TeV}'',
  \href{http://dx.doi.org/10.1016/j.physletb.2012.01.063}{{\em Phys. Lett. B}
  {\bfseries 708} (2012) 265--275},
  \href{http://arxiv.org/abs/1201.3791}{{\ttfamily arXiv:1201.3791 [hep-ex]}}.

\bibitem{ALICEpage}
{\bfseries ALICE} Collaboration, ``{Quarkonium signal extraction in ALICE}'',
  ALICE-PUBLIC-2015-006. \url{https://cds.cern.ch/record/2060096}.

\bibitem{PDG}
{\bfseries Particle Data Group} Collaboration, P.~A. Zyla {\em et~al.},
  ``{Review of Particle Physics}'',
  \href{http://dx.doi.org/10.1093/ptep/ptaa104}{{\em PTEP} {\bfseries 2020}
  (2020) 083C01}.

\bibitem{Brun:1082634}
R.~Brun, F.~Bruyant, F.~Carminati, S.~Giani, M.~Maire, A.~McPherson,
  G.~Patrick, and L.~Urban,
  \href{http://dx.doi.org/10.17181/CERN.MUHF.DMJ1}{{\em {GEANT: Detector
  Description and Simulation Tool; Oct 1994}}}.
\newblock CERN Program Library. CERN, Geneva, 1993.
\newblock \url{https://cds.cern.ch/record/1082634}.
\newblock Long Writeup W5013.

\bibitem{Agostinelli:2002hh}
{\bfseries GEANT4} Collaboration, S.~Agostinelli {\em et~al.}, ``{GEANT4: A
  Simulation toolkit}'',
\href{http://dx.doi.org/10.1016/S0168-9002(03)01368-8}{{\em Nucl. Instrum.
  Meth.} {\bfseries A506} (2003) 250--303}.
%%CITATION = NUIMA,A506,250;%%.

\bibitem{Bossu:2011qe}
F.~Bossù, Z.~Conesa~del Valle, A.~de~Falco, M.~Gagliardi, S.~Grigoryan, and
  G.~Martinez~Garcia, ``{Phenomenological interpolation of the inclusive J/psi
  cross section to proton-proton collisions at 2.76 TeV and 5.5 TeV}'',
  \href{http://arxiv.org/abs/1103.2394}{{\ttfamily arXiv:1103.2394 [nucl-ex]}}.

\bibitem{Acharya:2018uww}
{\bfseries ALICE} Collaboration, S.~Acharya {\em et~al.}, ``{Measurement of the
  inclusive J/ $\psi $ polarization at forward rapidity in pp collisions at
  $\mathbf {\sqrt{s} = 8}$ TeV}'',
  \href{http://dx.doi.org/10.1140/epjc/s10052-018-6027-2}{{\em Eur. Phys. J. C}
  {\bfseries 78} (2018) 562}, \href{http://arxiv.org/abs/1805.04374}{{\ttfamily
  arXiv:1805.04374 [hep-ex]}}.

\bibitem{Abelev:2011md}
{\bfseries ALICE} Collaboration, B.~Abelev {\em et~al.}, ``{$J/\psi$
  polarization in pp collisions at $\sqrt{s}=7$ TeV}'',
  \href{http://dx.doi.org/10.1103/PhysRevLett.108.082001}{{\em Phys. Rev.
  Lett.} {\bfseries 108} (2012) 082001},
  \href{http://arxiv.org/abs/1111.1630}{{\ttfamily arXiv:1111.1630 [hep-ex]}}.

\bibitem{Aaij:2013nlm}
{\bfseries LHCb} Collaboration, R.~Aaij {\em et~al.}, ``{Measurement of
  $J/\psi$ polarization in $pp$ collisions at $\sqrt{s}=7$ TeV}'',
  \href{http://dx.doi.org/10.1140/epjc/s10052-013-2631-3}{{\em Eur. Phys. J. C}
  {\bfseries 73} (2013) 2631}, \href{http://arxiv.org/abs/1307.6379}{{\ttfamily
  arXiv:1307.6379 [hep-ex]}}.

\bibitem{Aaij:2014qea}
{\bfseries LHCb} Collaboration, R.~Aaij {\em et~al.}, ``{Measurement of
  $\psi(2S)$ polarisation in $pp$ collisions at $\sqrt{s}=7$ TeV}'',
  \href{http://dx.doi.org/10.1140/epjc/s10052-014-2872-9}{{\em Eur. Phys. J. C}
  {\bfseries 74} (2014) 2872}, \href{http://arxiv.org/abs/1403.1339}{{\ttfamily
  arXiv:1403.1339 [hep-ex]}}.

\bibitem{Aaij:2017egv}
{\bfseries LHCb} Collaboration, R.~Aaij {\em et~al.}, ``{Measurement of the
  $\Upsilon$ polarizations in $pp$ collisions at $\sqrt{s}=7$ and 8 TeV}'',
  \href{http://dx.doi.org/10.1007/JHEP12(2017)110}{{\em JHEP} {\bfseries 12}
  (2017) 110}, \href{http://arxiv.org/abs/1709.01301}{{\ttfamily
  arXiv:1709.01301 [hep-ex]}}.

\bibitem{Lange:2001uf}
D.~Lange, ``{The EvtGen particle decay simulation package}'',
  \href{http://dx.doi.org/10.1016/S0168-9002(01)00089-4}{{\em Nucl.\ Instrum.\
  Meth.\ A} {\bfseries 462} (2001) 152--155}.

\bibitem{Barberio:1990ms}
E.~Barberio, B.~van Eijk, and Z.~Was, ``{PHOTOS: A Universal Monte Carlo for
  QED radiative corrections in decays}'',
  \href{http://dx.doi.org/10.1016/0010-4655(91)90012-A}{{\em Comput.\ Phys.\
  Commun.} {\bfseries 66} (1991) 115--128}.

\bibitem{Aaij:2015awa}
{\bfseries LHCb} Collaboration, R.~Aaij {\em et~al.}, ``{Forward production of
  $\Upsilon$ mesons in $pp$ collisions at $\sqrt{s}=7$ and 8TeV}'',
  \href{http://dx.doi.org/10.1007/JHEP11(2015)103}{{\em JHEP} {\bfseries 11}
  (2015) 103}, \href{http://arxiv.org/abs/1509.02372}{{\ttfamily
  arXiv:1509.02372 [hep-ex]}}.

\bibitem{Aaij:2018pfp}
{\bfseries LHCb} Collaboration, R.~Aaij {\em et~al.}, ``{Measurement of
  $\Upsilon$ production in $pp$ collisions at $\sqrt{s}$ = 13 TeV}'',
  \href{http://dx.doi.org/10.1007/JHEP07(2018)134}{{\em JHEP} {\bfseries 07}
  (2018) 134}, \href{http://arxiv.org/abs/1804.09214}{{\ttfamily
  arXiv:1804.09214 [hep-ex]}}. [Erratum: JHEP 05, 076 (2019)].

\bibitem{Bierlich:2022pfr}
C.~Bierlich {\em et~al.}, ``{A comprehensive guide to the physics and usage of
  PYTHIA 8.3}'', \href{http://arxiv.org/abs/2203.11601}{{\ttfamily
  arXiv:2203.11601 [hep-ph]}}.

\bibitem{Acharya:2019iur}
{\bfseries ALICE} Collaboration, S.~Acharya {\em et~al.}, ``{Studies of
  J/$\psi$ production at forward rapidity in Pb-Pb collisions at
  $\sqrt{s_{\rm{NN}}}$ = 5.02 TeV}'',
  \href{http://dx.doi.org/10.1007/JHEP02(2020)041}{{\em JHEP} {\bfseries 02}
  (2020) 041}, \href{http://arxiv.org/abs/1909.03158}{{\ttfamily
  arXiv:1909.03158 [nucl-ex]}}.

\bibitem{Butenschoen:2010rq}
M.~Butenschön and B.~A. Kniehl, ``{Reconciling $J/\psi$ production at HERA,
  RHIC, Tevatron, and LHC with NRQCD factorization at next-to-leading order}'',
  \href{http://dx.doi.org/10.1103/PhysRevLett.106.022003}{{\em Phys. Rev.
  Lett.} {\bfseries 106} (2011) 022003},
  \href{http://arxiv.org/abs/1009.5662}{{\ttfamily arXiv:1009.5662 [hep-ph]}}.

\bibitem{Ma:2010yw}
Y.-Q. Ma, K.~Wang, and K.-T. Chao, ``{$J/\psi (\psi^\prime)$ production at the
  Tevatron and LHC at ${\cal O}(\alpha_s^4v^4)$ in nonrelativistic QCD}'',
  \href{http://dx.doi.org/10.1103/PhysRevLett.106.042002}{{\em Phys. Rev.
  Lett.} {\bfseries 106} (2011) 042002},
  \href{http://arxiv.org/abs/1009.3655}{{\ttfamily arXiv:1009.3655 [hep-ph]}}.

\bibitem{Ma:2014mri}
Y.-Q. Ma and R.~Venugopalan, ``{Comprehensive Description of J/$\psi$
  Production in Proton-Proton Collisions at Collider Energies}'',
  \href{http://dx.doi.org/10.1103/PhysRevLett.113.192301}{{\em Phys. Rev.
  Lett.} {\bfseries 113} (2014) 192301},
  \href{http://arxiv.org/abs/1408.4075}{{\ttfamily arXiv:1408.4075 [hep-ph]}}.

\bibitem{Cheung:2018tvq}
V.~Cheung and R.~Vogt, ``{Production and polarization of prompt $J/\psi$ in the
  improved color evaporation model using the $k_T$-factorization approach}'',
  \href{http://dx.doi.org/10.1103/PhysRevD.98.114029}{{\em Phys. Rev. D}
  {\bfseries 98} (2018) 114029},
  \href{http://arxiv.org/abs/1808.02909}{{\ttfamily arXiv:1808.02909
  [hep-ph]}}.

\bibitem{Lansberg:2020rft}
J.-P. Lansberg, H.-S. Shao, N.~Yamanaka, Y.-J. Zhang, and C.~No\^us,
  ``{Complete NLO QCD study of single- and double-quarkonium hadroproduction in
  the colour-evaporation model at the Tevatron and the LHC}'',
  \href{http://dx.doi.org/10.1016/j.physletb.2020.135559}{{\em Phys. Lett. B}
  {\bfseries 807} (2020) 135559},
  \href{http://arxiv.org/abs/2004.14345}{{\ttfamily arXiv:2004.14345
  [hep-ph]}}.

\bibitem{Cacciari:2012ny}
M.~Cacciari, S.~Frixione, N.~Houdeau, M.~L. Mangano, P.~Nason, and G.~Ridolfi,
  ``{Theoretical predictions for charm and bottom production at the LHC}'',
  \href{http://dx.doi.org/10.1007/JHEP10(2012)137}{{\em JHEP} {\bfseries 10}
  (2012) 137}, \href{http://arxiv.org/abs/1205.6344}{{\ttfamily arXiv:1205.6344
  [hep-ph]}}.

\bibitem{Sirunyan:2018nsz}
{\bfseries CMS} Collaboration, A.~M. Sirunyan {\em et~al.}, ``{Measurement of
  nuclear modification factors of $\Upsilon$(1S), $\Upsilon$(2S), and
  $\Upsilon$(3S) mesons in PbPb collisions at $\sqrt{s_{_\mathrm{NN}}} =$ 5.02
  TeV}'', \href{http://dx.doi.org/10.1016/j.physletb.2019.01.006}{{\em Phys.
  Lett. B} {\bfseries 790} (2019) 270--293},
  \href{http://arxiv.org/abs/1805.09215}{{\ttfamily arXiv:1805.09215
  [hep-ex]}}.

\bibitem{Cheung:2018upe}
V.~Cheung and R.~Vogt, ``{Production and polarization of prompt
  $\varUpsilon$($n$S) in the improved color evaporation model using the
  $k_T$-factorization approach}'',
  \href{http://dx.doi.org/10.1103/PhysRevD.99.034007}{{\em Phys. Rev. D}
  {\bfseries 99} (2019) 034007},
  \href{http://arxiv.org/abs/1811.11570}{{\ttfamily arXiv:1811.11570
  [hep-ph]}}.

\end{thebibliography}\endgroup
%\input {bibliography.tex}  

%%%%%%%%%%%%%%%%%%%%%%%%%%%%%%%%
% Appendices: yours (if any) + authorlist
%%%%%%%%%%%%%%%%%%%%%%%%%%%%%%%%
\newpage
\appendix

\section{Appendix}

\begin{figure}[!htb]
\includegraphics[width=0.49\linewidth]{{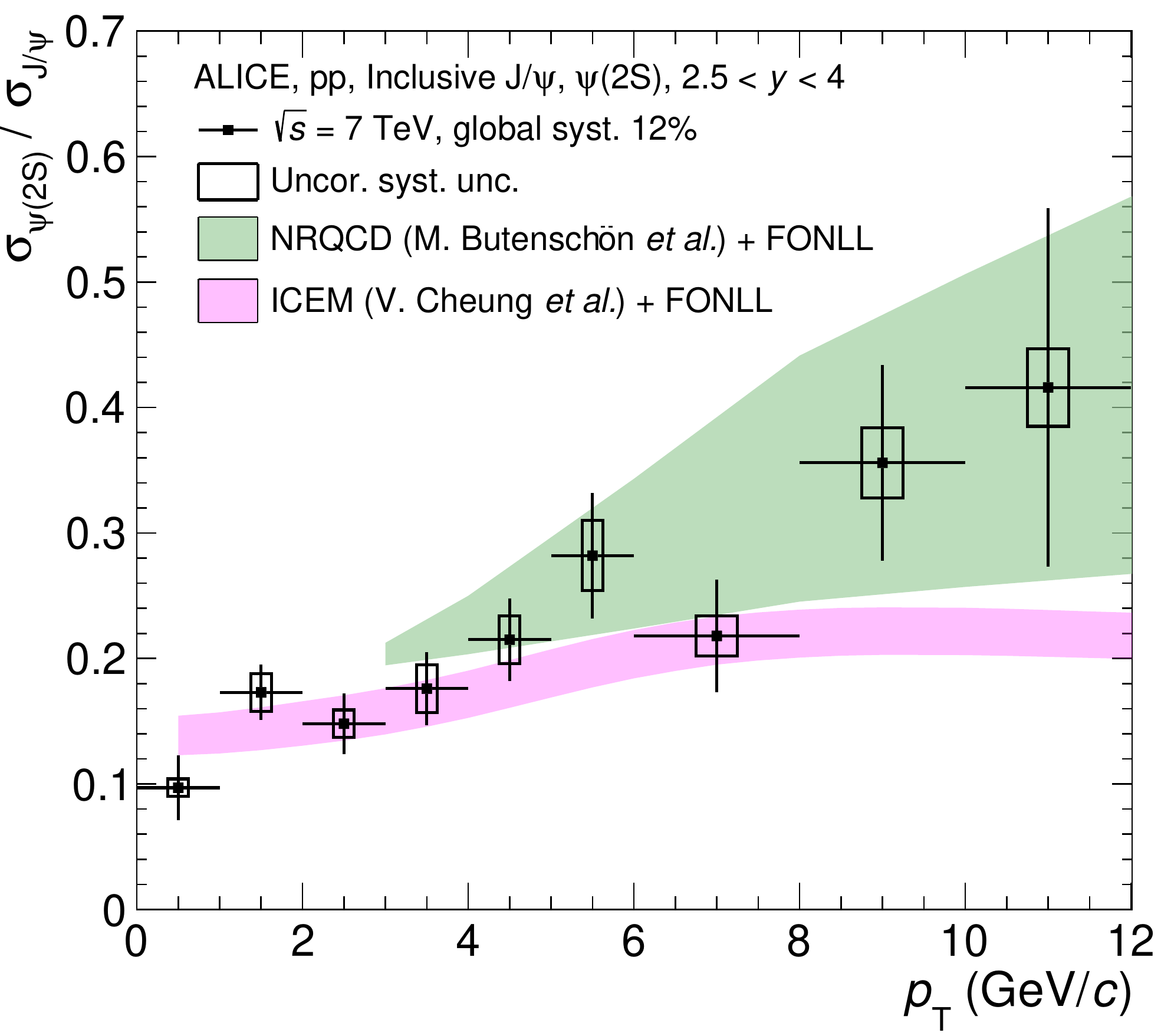}}
\includegraphics[width=0.49\linewidth]{{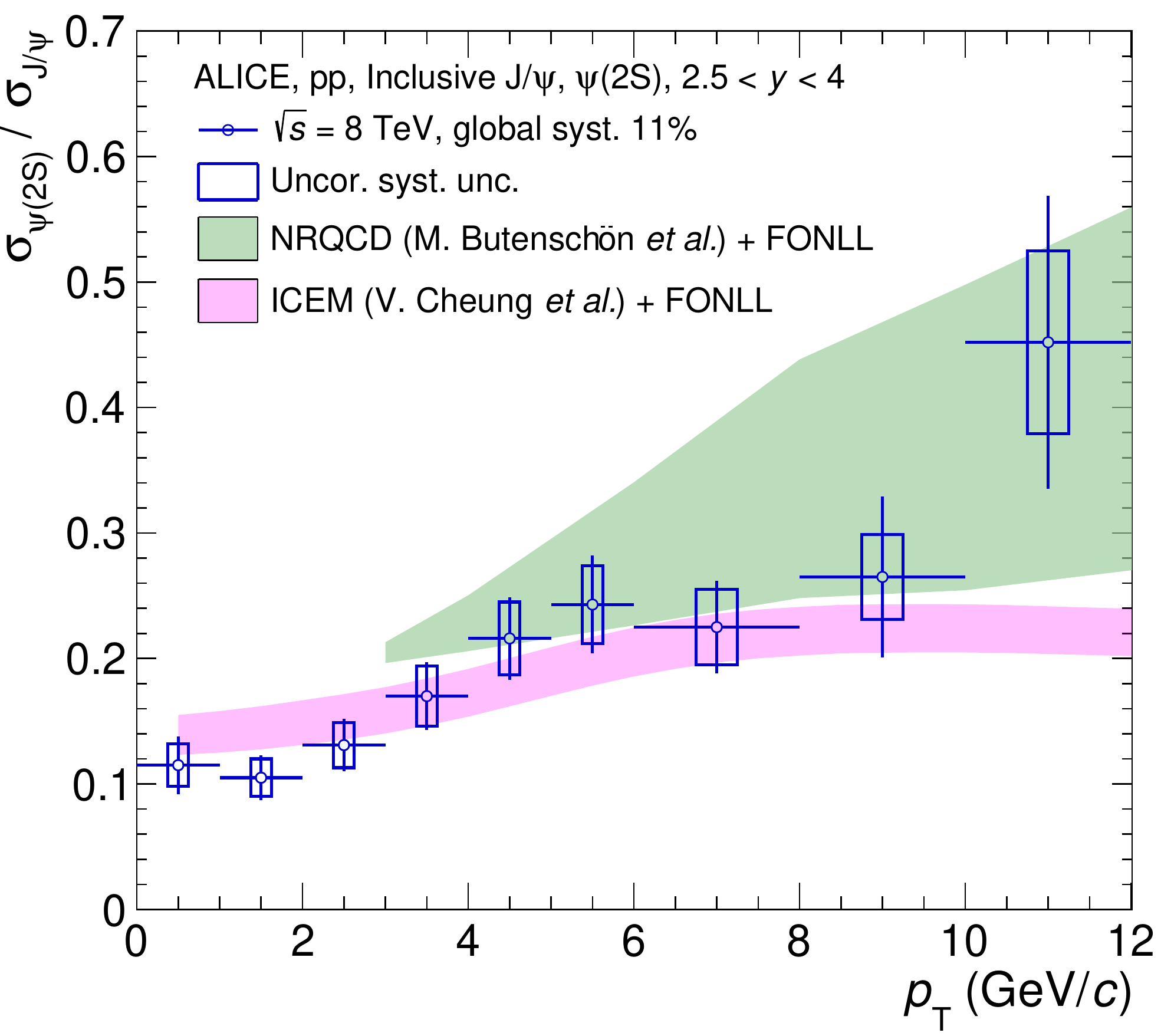}}
\includegraphics[width=0.49\linewidth]{{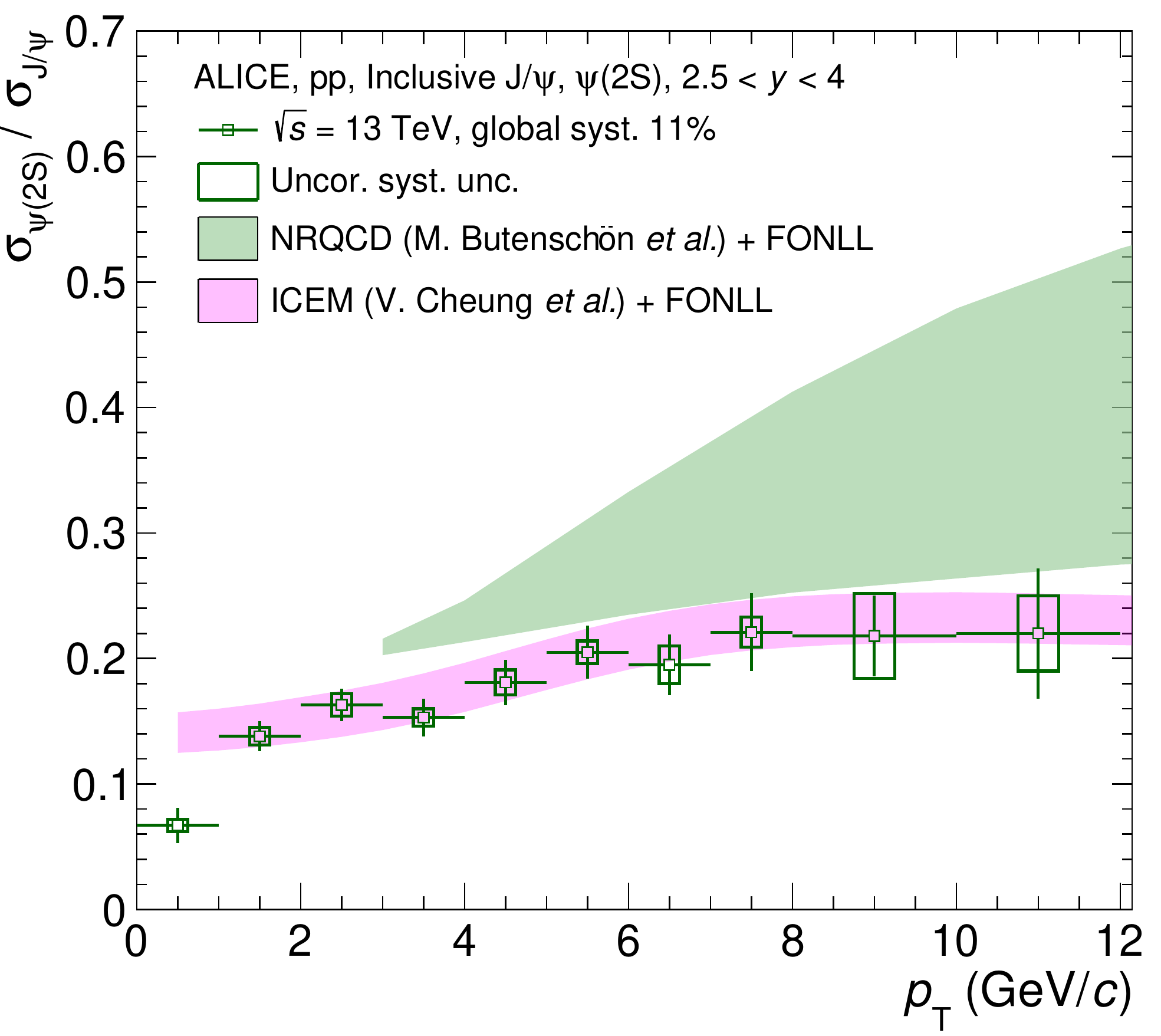}}
\caption{Inclusive \psitwos-to-\jpsi cross section ratio as a function of \pt, at forward \rapidity, in pp collisions at $\s$~=~7~\cite{Abelev:2014qha} (top left), 8~\cite{Adam:2015rta} (top right), and 13 TeV~\cite{Acharya:2017hjh} (bottom left). The data are compared with NRQCD theoretical calculations from Butenschön~\etal~+ FONLL~\cite{Butenschoen:2010rq,Cacciari:2012ny} and with theoretical calculations from ICEM + FONLL~\cite{Cheung:2018tvq,Cacciari:2012ny}.}
\label{fig:psitwosoverjpsienergycomp_appendix}
\end{figure}

The \psitwos-to-\jpsi cross section ratio is displayed as a function of \pt for \pp collisions at $\s$ = 7~\cite{Abelev:2014qha}, 8~\cite{Adam:2015rta} and 13 TeV~\cite{Acharya:2017hjh} in the top left, top right, and bottom left panel of Fig.~\ref{fig:psitwosoverjpsienergycomp_appendix}, respectively. It is compared to the NRQCD model from  Butenschön~\etal~\cite{Butenschoen:2010rq} and to the ICEM~\cite{Cheung:2018tvq} model, as in Fig.~\ref{fig:psitwosoverjpsienergycomp} right for the results obtained at $\s$ = 5.02 TeV.

\begin{figure}[!htbp]
\includegraphics[width=0.49\linewidth]{{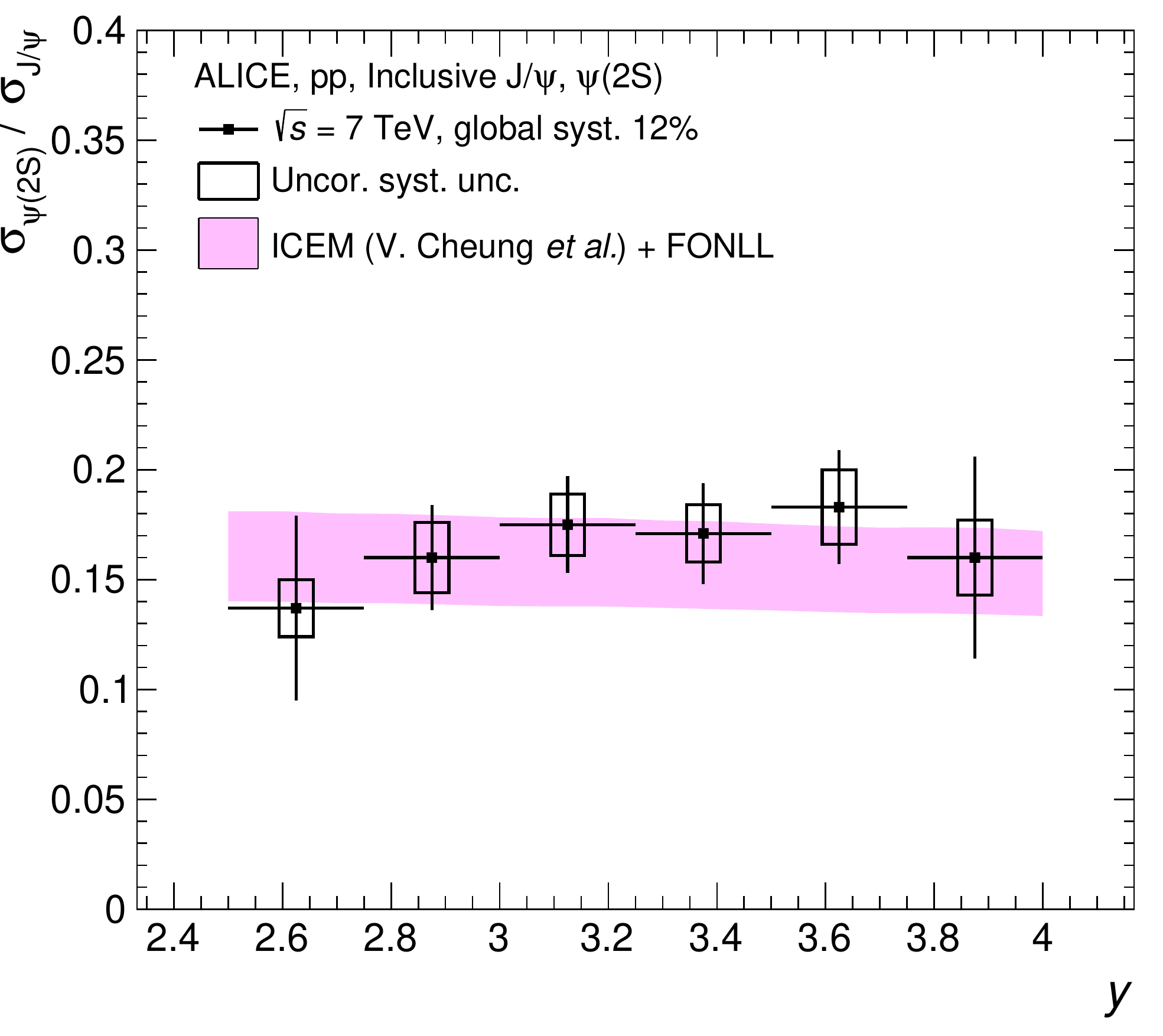}}
\includegraphics[width=0.49\linewidth]{{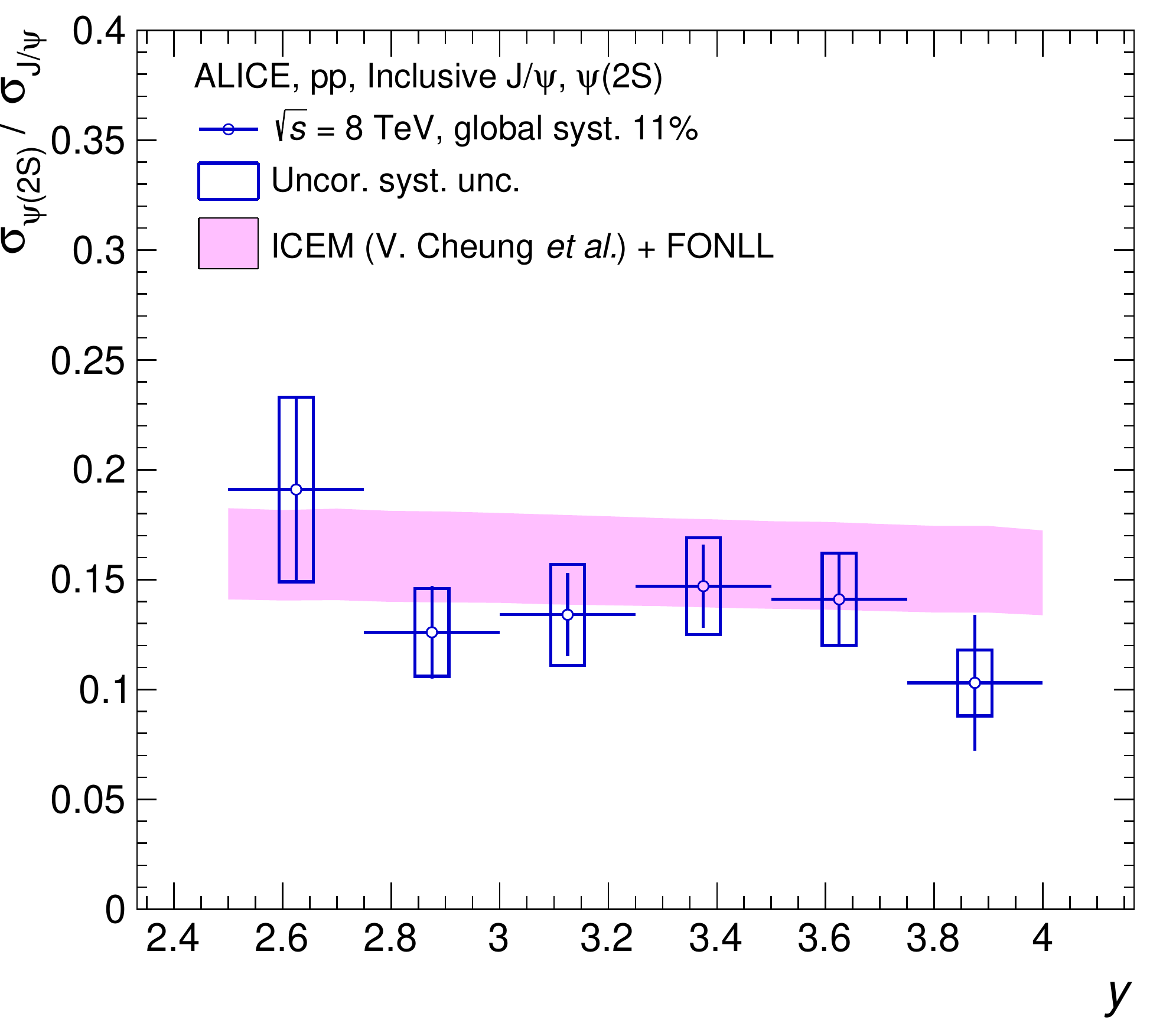}}
\includegraphics[width=0.49\linewidth]{{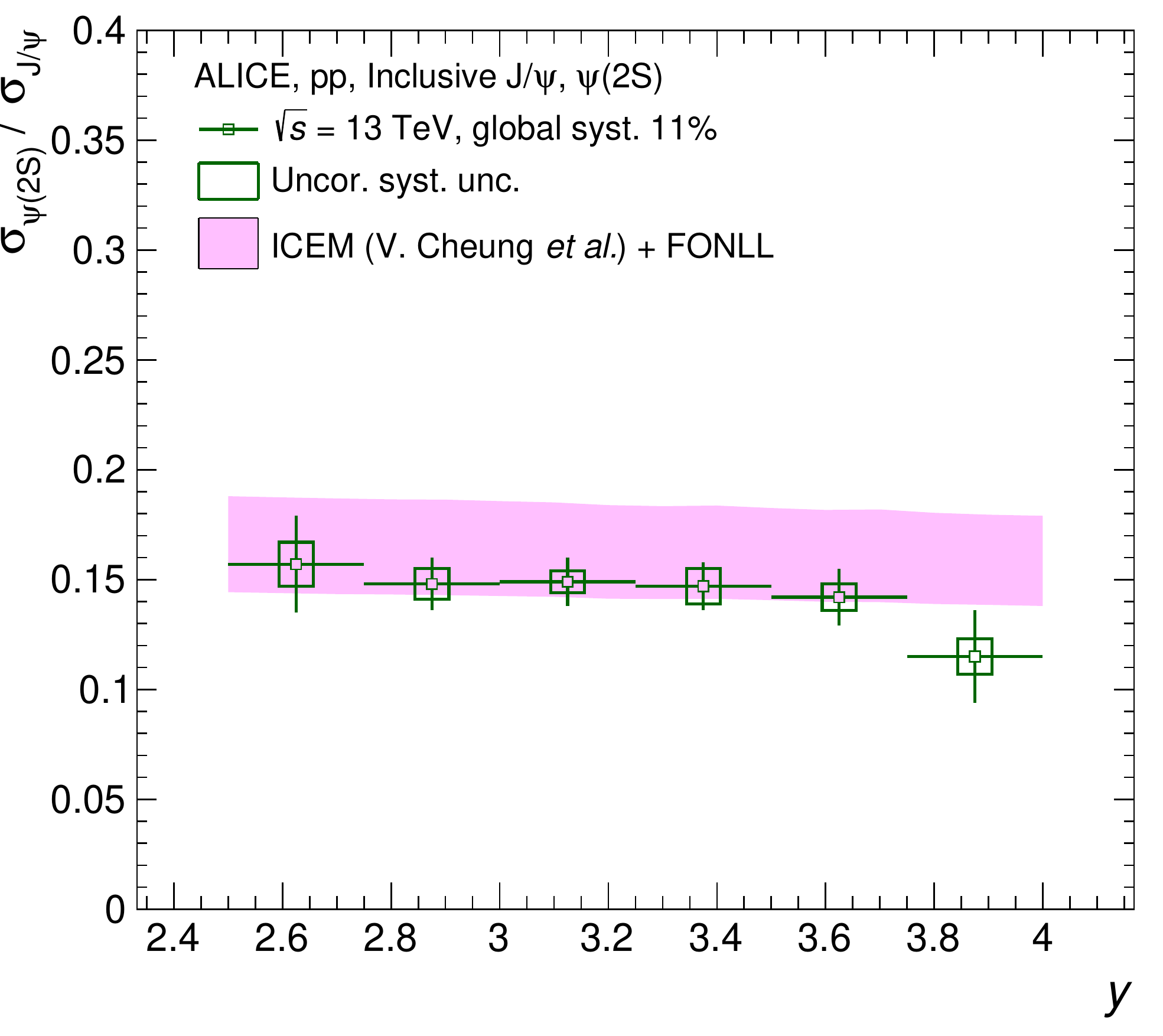}}

\caption{Inclusive \psitwos-to-\jpsi cross section ratio as a function of \rapidity in pp collisions at $\s$ = 7~\cite{Abelev:2014qha} (top left), 8~\cite{Adam:2015rta} (top right), and 13 TeV~\cite{Acharya:2017hjh} (bottom left).  
The data are compared with theoretical calculations from ICEM + FONLL~\cite{Cheung:2018tvq,Cacciari:2012ny}.}
\label{fig:psitwosoverjpsienergycomp2_appendix}
\end{figure}

The \psitwos-to-\jpsi cross section ratio is displayed as a function of \rapidity for \pp collisions at $\s$ = 7~\cite{Abelev:2014qha}, 8~\cite{Adam:2015rta} and 13 TeV~\cite{Acharya:2017hjh} in the top left, top right, and bottom left panel of Fig.~\ref{fig:psitwosoverjpsienergycomp2_appendix}, respectively. It is compared with the ICEM~\cite{Cheung:2018tvq} calculation, as in Fig.~\ref{fig:psitwosoverjpsienergycomp2} right for the results obtained at $\s$~=~5.02~TeV. 

\newpage

%
%\input{} % put your appendices here (if any)
%

%%%%% Authorlist - please do not touch: handled by EB chairs 
\section{The ALICE Collaboration}
\label{app:collab}
% ALICE Collaboration author list for 2021-09-21
\small
\begin{flushleft} 

S.~Acharya$^{\rm 142}$, 
D.~Adamov\'{a}$^{\rm 97}$, 
A.~Adler$^{\rm 75}$, 
J.~Adolfsson$^{\rm 82}$, 
G.~Aglieri Rinella$^{\rm 34}$, 
M.~Agnello$^{\rm 30}$, 
N.~Agrawal$^{\rm 54}$, 
Z.~Ahammed$^{\rm 142}$, 
S.~Ahmad$^{\rm 16}$, 
S.U.~Ahn$^{\rm 77}$
% ...

\section*{Affiliation notes}

$^{\rm I}$ Deceased\\
$^{\rm II}$ Also at: Italian National Agency for New Technologies, Energy and Sustainable Economic Development (ENEA), Bologna, Italy\\
$^{\rm III}$ Also at: Dipartimento DET del Politecnico di Torino, Turin, Italy\\
$^{\rm IV}$ Also at: M.V. Lomonosov Moscow State University, D.V. Skobeltsyn Institute of Nuclear, Physics, Moscow, Russia\\
$^{\rm V}$ Also at: Department of Applied Physics, Aligarh Muslim University, Aligarh, India
\\
$^{\rm VI}$ Also at: Institute of Theoretical Physics, University of Wroclaw, Poland\\
$^{\rm VII}$ Also at: University of Kansas, Lawrence, Kansas, United States\\

\section*{Collaboration Institutes}

$^{1}$ A.I. Alikhanyan National Science Laboratory (Yerevan Physics Institute) Foundation, Yerevan, Armenia\\
$^{2}$ AGH University of Science and Technology, Cracow, Poland\\
$^{3}$ Bogolyubov Institute for Theoretical Physics, National Academy of Sciences of Ukraine, Kiev, Ukraine\\
$^{4}$ Bose Institute, Department of Physics  and Centre for Astroparticle Physics and Space Science (CAPSS), Kolkata, India\\
$^{5}$ Budker Institute for Nuclear Physics, Novosibirsk, Russia\\
$^{6}$ California Polytechnic State University, San Luis Obispo, California, United States\\
$^{7}$ Central China Normal University, Wuhan, China\\
$^{8}$ Centro de Aplicaciones Tecnol\'{o}gicas y Desarrollo Nuclear (CEADEN), Havana, Cuba\\
$^{9}$ Centro de Investigaci\'{o}n y de Estudios Avanzados (CINVESTAV), Mexico City and M\'{e}rida, Mexico\\
$^{10}$ Chicago State University, Chicago, Illinois, United States\\
$^{11}$ China Institute of Atomic Energy, Beijing, China\\
$^{12}$ Chungbuk National University, Cheongju, Republic of Korea\\
$^{13}$ Comenius University Bratislava, Faculty of Mathematics, Physics and Informatics, Bratislava, Slovakia\\
$^{14}$ COMSATS University Islamabad, Islamabad, Pakistan\\
$^{15}$ Creighton University, Omaha, Nebraska, United States\\
$^{16}$ Department of Physics, Aligarh Muslim University, Aligarh, India\\
$^{17}$ Department of Physics, Pusan National University, Pusan, Republic of Korea\\
$^{18}$ Department of Physics, Sejong University, Seoul, Republic of Korea\\
$^{19}$ Department of Physics, University of California, Berkeley, California, United States\\
$^{20}$ Department of Physics, University of Oslo, Oslo, Norway\\
$^{21}$ Department of Physics and Technology, University of Bergen, Bergen, Norway\\
$^{22}$ Dipartimento di Fisica dell'Universit\`{a} and Sezione INFN, Cagliari, Italy\\
$^{23}$ Dipartimento di Fisica dell'Universit\`{a} and Sezione INFN, Trieste, Italy\\
$^{24}$ Dipartimento di Fisica dell'Universit\`{a} and Sezione INFN, Turin, Italy\\
$^{25}$ Dipartimento di Fisica e Astronomia dell'Universit\`{a} and Sezione INFN, Bologna, Italy\\
$^{26}$ Dipartimento di Fisica e Astronomia dell'Universit\`{a} and Sezione INFN, Catania, Italy\\
$^{27}$ Dipartimento di Fisica e Astronomia dell'Universit\`{a} and Sezione INFN, Padova, Italy\\
$^{28}$ Dipartimento di Fisica e Nucleare e Teorica, Universit\`{a} di Pavia, Pavia, Italy\\
$^{29}$ Dipartimento di Fisica `E.R.~Caianiello' dell'Universit\`{a} and Gruppo Collegato INFN, Salerno, Italy\\
$^{30}$ Dipartimento DISAT del Politecnico and Sezione INFN, Turin, Italy\\
$^{31}$ Dipartimento di Scienze e Innovazione Tecnologica dell'Universit\`{a} del Piemonte Orientale and INFN Sezione di Torino, Alessandria, Italy\\
$^{32}$ Dipartimento di Scienze MIFT, Universit\`{a} di Messina, Messina, Italy\\
$^{33}$ Dipartimento Interateneo di Fisica `M.~Merlin' and Sezione INFN, Bari, Italy\\
$^{34}$ European Organization for Nuclear Research (CERN), Geneva, Switzerland\\
$^{35}$ Faculty of Electrical Engineering, Mechanical Engineering and Naval Architecture, University of Split, Split, Croatia\\
$^{36}$ Faculty of Engineering and Science, Western Norway University of Applied Sciences, Bergen, Norway\\
$^{37}$ Faculty of Nuclear Sciences and Physical Engineering, Czech Technical University in Prague, Prague, Czech Republic\\
$^{38}$ Faculty of Science, P.J.~\v{S}af\'{a}rik University, Ko\v{s}ice, Slovakia\\
$^{39}$ Frankfurt Institute for Advanced Studies, Johann Wolfgang Goethe-Universit\"{a}t Frankfurt, Frankfurt, Germany\\
$^{40}$ Fudan University, Shanghai, China\\
$^{41}$ Gangneung-Wonju National University, Gangneung, Republic of Korea\\
$^{42}$ Gauhati University, Department of Physics, Guwahati, India\\
$^{43}$ Helmholtz-Institut f\"{u}r Strahlen- und Kernphysik, Rheinische Friedrich-Wilhelms-Universit\"{a}t Bonn, Bonn, Germany\\
$^{44}$ Helsinki Institute of Physics (HIP), Helsinki, Finland\\
$^{45}$ High Energy Physics Group,  Universidad Aut\'{o}noma de Puebla, Puebla, Mexico\\
$^{46}$ Hiroshima University, Hiroshima, Japan\\
$^{47}$ Hochschule Worms, Zentrum  f\"{u}r Technologietransfer und Telekommunikation (ZTT), Worms, Germany\\
$^{48}$ Horia Hulubei National Institute of Physics and Nuclear Engineering, Bucharest, Romania\\
$^{49}$ Indian Institute of Technology Bombay (IIT), Mumbai, India\\
$^{50}$ Indian Institute of Technology Indore, Indore, India\\
$^{51}$ Indonesian Institute of Sciences, Jakarta, Indonesia\\
$^{52}$ INFN, Laboratori Nazionali di Frascati, Frascati, Italy\\
$^{53}$ INFN, Sezione di Bari, Bari, Italy\\
$^{54}$ INFN, Sezione di Bologna, Bologna, Italy\\
$^{55}$ INFN, Sezione di Cagliari, Cagliari, Italy\\
$^{56}$ INFN, Sezione di Catania, Catania, Italy\\
$^{57}$ INFN, Sezione di Padova, Padova, Italy\\
$^{58}$ INFN, Sezione di Pavia, Pavia, Italy\\
$^{59}$ INFN, Sezione di Roma, Rome, Italy\\
$^{60}$ INFN, Sezione di Torino, Turin, Italy\\
$^{61}$ INFN, Sezione di Trieste, Trieste, Italy\\
$^{62}$ Inha University, Incheon, Republic of Korea\\
$^{63}$ Institute for Gravitational and Subatomic Physics (GRASP), Utrecht University/Nikhef, Utrecht, Netherlands\\
$^{64}$ Institute for Nuclear Research, Academy of Sciences, Moscow, Russia\\
$^{65}$ Institute of Experimental Physics, Slovak Academy of Sciences, Ko\v{s}ice, Slovakia\\
$^{66}$ Institute of Physics, Homi Bhabha National Institute, Bhubaneswar, India\\
$^{67}$ Institute of Physics of the Czech Academy of Sciences, Prague, Czech Republic\\
$^{68}$ Institute of Space Science (ISS), Bucharest, Romania\\
$^{69}$ Institut f\"{u}r Kernphysik, Johann Wolfgang Goethe-Universit\"{a}t Frankfurt, Frankfurt, Germany\\
$^{70}$ Instituto de Ciencias Nucleares, Universidad Nacional Aut\'{o}noma de M\'{e}xico, Mexico City, Mexico\\
$^{71}$ Instituto de F\'{i}sica, Universidade Federal do Rio Grande do Sul (UFRGS), Porto Alegre, Brazil\\
$^{72}$ Instituto de F\'{\i}sica, Universidad Nacional Aut\'{o}noma de M\'{e}xico, Mexico City, Mexico\\
$^{73}$ iThemba LABS, National Research Foundation, Somerset West, South Africa\\
$^{74}$ Jeonbuk National University, Jeonju, Republic of Korea\\
$^{75}$ Johann-Wolfgang-Goethe Universit\"{a}t Frankfurt Institut f\"{u}r Informatik, Fachbereich Informatik und Mathematik, Frankfurt, Germany\\
$^{76}$ Joint Institute for Nuclear Research (JINR), Dubna, Russia\\
$^{77}$ Korea Institute of Science and Technology Information, Daejeon, Republic of Korea\\
$^{78}$ KTO Karatay University, Konya, Turkey\\
$^{79}$ Laboratoire de Physique des 2 Infinis, Ir\`{e}ne Joliot-Curie, Orsay, France\\
$^{80}$ Laboratoire de Physique Subatomique et de Cosmologie, Universit\'{e} Grenoble-Alpes, CNRS-IN2P3, Grenoble, France\\
$^{81}$ Lawrence Berkeley National Laboratory, Berkeley, California, United States\\
$^{82}$ Lund University Department of Physics, Division of Particle Physics, Lund, Sweden\\
$^{83}$ Moscow Institute for Physics and Technology, Moscow, Russia\\
$^{84}$ Nagasaki Institute of Applied Science, Nagasaki, Japan\\
$^{85}$ Nara Women{'}s University (NWU), Nara, Japan\\
$^{86}$ National and Kapodistrian University of Athens, School of Science, Department of Physics , Athens, Greece\\
$^{87}$ National Centre for Nuclear Research, Warsaw, Poland\\
$^{88}$ National Institute of Science Education and Research, Homi Bhabha National Institute, Jatni, India\\
$^{89}$ National Nuclear Research Center, Baku, Azerbaijan\\
$^{90}$ National Research Centre Kurchatov Institute, Moscow, Russia\\
$^{91}$ Niels Bohr Institute, University of Copenhagen, Copenhagen, Denmark\\
$^{92}$ Nikhef, National institute for subatomic physics, Amsterdam, Netherlands\\
$^{93}$ NRC Kurchatov Institute IHEP, Protvino, Russia\\
$^{94}$ NRC \guillemotleft Kurchatov\guillemotright  Institute - ITEP, Moscow, Russia\\
$^{95}$ NRNU Moscow Engineering Physics Institute, Moscow, Russia\\
$^{96}$ Nuclear Physics Group, STFC Daresbury Laboratory, Daresbury, United Kingdom\\
$^{97}$ Nuclear Physics Institute of the Czech Academy of Sciences, \v{R}e\v{z} u Prahy, Czech Republic\\
$^{98}$ Oak Ridge National Laboratory, Oak Ridge, Tennessee, United States\\
$^{99}$ Ohio State University, Columbus, Ohio, United States\\
$^{100}$ Petersburg Nuclear Physics Institute, Gatchina, Russia\\
$^{101}$ Physics department, Faculty of science, University of Zagreb, Zagreb, Croatia\\
$^{102}$ Physics Department, Panjab University, Chandigarh, India\\
$^{103}$ Physics Department, University of Jammu, Jammu, India\\
$^{104}$ Physics Department, University of Rajasthan, Jaipur, India\\
$^{105}$ Physikalisches Institut, Eberhard-Karls-Universit\"{a}t T\"{u}bingen, T\"{u}bingen, Germany\\
$^{106}$ Physikalisches Institut, Ruprecht-Karls-Universit\"{a}t Heidelberg, Heidelberg, Germany\\
$^{107}$ Physik Department, Technische Universit\"{a}t M\"{u}nchen, Munich, Germany\\
$^{108}$ Politecnico di Bari and Sezione INFN, Bari, Italy\\
$^{109}$ Research Division and ExtreMe Matter Institute EMMI, GSI Helmholtzzentrum f\"ur Schwerionenforschung GmbH, Darmstadt, Germany\\
$^{110}$ Russian Federal Nuclear Center (VNIIEF), Sarov, Russia\\
$^{111}$ Saha Institute of Nuclear Physics, Homi Bhabha National Institute, Kolkata, India\\
$^{112}$ School of Physics and Astronomy, University of Birmingham, Birmingham, United Kingdom\\
$^{113}$ Secci\'{o}n F\'{\i}sica, Departamento de Ciencias, Pontificia Universidad Cat\'{o}lica del Per\'{u}, Lima, Peru\\
$^{114}$ St. Petersburg State University, St. Petersburg, Russia\\
$^{115}$ Stefan Meyer Institut f\"{u}r Subatomare Physik (SMI), Vienna, Austria\\
$^{116}$ SUBATECH, IMT Atlantique, Universit\'{e} de Nantes, CNRS-IN2P3, Nantes, France\\
$^{117}$ Suranaree University of Technology, Nakhon Ratchasima, Thailand\\
$^{118}$ Technical University of Ko\v{s}ice, Ko\v{s}ice, Slovakia\\
$^{119}$ The Henryk Niewodniczanski Institute of Nuclear Physics, Polish Academy of Sciences, Cracow, Poland\\
$^{120}$ The University of Texas at Austin, Austin, Texas, United States\\
$^{121}$ Universidad Aut\'{o}noma de Sinaloa, Culiac\'{a}n, Mexico\\
$^{122}$ Universidade de S\~{a}o Paulo (USP), S\~{a}o Paulo, Brazil\\
$^{123}$ Universidade Estadual de Campinas (UNICAMP), Campinas, Brazil\\
$^{124}$ Universidade Federal do ABC, Santo Andre, Brazil\\
$^{125}$ University of Cape Town, Cape Town, South Africa\\
$^{126}$ University of Houston, Houston, Texas, United States\\
$^{127}$ University of Jyv\"{a}skyl\"{a}, Jyv\"{a}skyl\"{a}, Finland\\
$^{128}$ University of Kansas, Lawrence, Kansas, United States\\
$^{129}$ University of Liverpool, Liverpool, United Kingdom\\
$^{130}$ University of Science and Technology of China, Hefei, China\\
$^{131}$ University of South-Eastern Norway, Tonsberg, Norway\\
$^{132}$ University of Tennessee, Knoxville, Tennessee, United States\\
$^{133}$ University of the Witwatersrand, Johannesburg, South Africa\\
$^{134}$ University of Tokyo, Tokyo, Japan\\
$^{135}$ University of Tsukuba, Tsukuba, Japan\\
$^{136}$ Universit\'{e} Clermont Auvergne, CNRS/IN2P3, LPC, Clermont-Ferrand, France\\
$^{137}$ Universit\'{e} de Lyon, CNRS/IN2P3, Institut de Physique des 2 Infinis de Lyon, Lyon, France\\
$^{138}$ Universit\'{e} de Strasbourg, CNRS, IPHC UMR 7178, F-67000 Strasbourg, France, Strasbourg, France\\
$^{139}$ Universit\'{e} Paris-Saclay Centre d'Etudes de Saclay (CEA), IRFU, D\'{e}partment de Physique Nucl\'{e}aire (DPhN), Saclay, France\\
$^{140}$ Universit\`{a} degli Studi di Foggia, Foggia, Italy\\
$^{141}$ Universit\`{a} di Brescia, Brescia, Italy\\
$^{142}$ Variable Energy Cyclotron Centre, Homi Bhabha National Institute, Kolkata, India\\
$^{143}$ Warsaw University of Technology, Warsaw, Poland\\
$^{144}$ Wayne State University, Detroit, Michigan, United States\\
$^{145}$ Westf\"{a}lische Wilhelms-Universit\"{a}t M\"{u}nster, Institut f\"{u}r Kernphysik, M\"{u}nster, Germany\\
$^{146}$ Wigner Research Centre for Physics, Budapest, Hungary\\
$^{147}$ Yale University, New Haven, Connecticut, United States\\
$^{148}$ Yonsei University, Seoul, Republic of Korea\\

\end{flushleft} 
  
\end{document}